\documentclass[12pt, preprint]{emulateapj}

\usepackage{natbib}
\usepackage{amsmath}

\shortauthors{Shi et al.}

\begin{document}

\title{A Joint Model of the X-ray And Infrared Extragalactic Backgrounds: I. Model Construction And First Results}

\author{Yong  Shi\altaffilmark{1},  George Helou\altaffilmark{1},  Lee Armus\altaffilmark{1},  
Sabrina Stierwalt\altaffilmark{1}, Daniel Dale\altaffilmark{2}  } 

\altaffiltext{1}{Infrared  Processing and Analysis   Center,   California    Institute   of   Technology,   1200 E. California Blvd, Pasadena, CA 91125}
\altaffiltext{2}{Department of Physics and Astronomy, University of Wyoming, Laramie, WY 82071, USA}

\begin{abstract}

We present an extragalactic  population model of the cosmic background
light to interpret the rich  high-quality survey data in the X-ray and
IR  bands.   The model  incorporates  star-formation and  supermassive
black  hole  (SMBH)  accretion  in  a  co-evolution  scenario  to  fit
simultaneously   617   data   points   of  number   counts,   redshift
distributions  and  local  luminosity  functions (LFs)  with  19  free
parameters. The model  has four main components, the  total IR LF, the
SMBH accretion energy fraction in  the IR band, the star-formation SED
and  the unobscured  SMBH SED  extinguished with  a HI  column density
distribution. As a result of the observational uncertainties about the
star-formation  and SMBH  SEDs,  we present  several  variants of  the
model.  The best-fit reduced $\chi^{2}$ reaches as small as 2.7-2.9 of
which a significant amount ($>$0.8) is contributed by cosmic variances
or caveats  associated with data.   Compared to previous  models, the
unique result of  this model is to constrain  the SMBH energy fraction
in the  IR band that is found  to increase with the  IR luminosity but
decrease with redshift up to $z$ $\sim$ 1.5; this result is separately
verified  using  aromatic feature  equivalent  width  data. The  joint
modelling of X-ray and mid-IR  data allows for improved constraints on
the  obscured AGN,  especially the  Compton-thick AGN  population. All
variants  of  the  model  require  that  Compton-thick  AGN  fractions
decrease with the SMBH luminosity but increase with redshift while the
type-1 AGN fraction has the reverse trend.

\end{abstract}


\section{Introduction}

The  extragalactic background light  (EBL) represents  the accumulated
radiation generated over cosmic history after the Big Bang. The EBL in
the X-ray,  or cosmic X-ray background  (CXB), is now known  to be the
relic  emission of  cosmic  supermassive black  hole (SMBH)  accretion
\citep[e.g.][]{Comastri95},  while   the  cosmic  infrared  background
(CIRB) arises  mainly from the integrated emission  of cosmic obscured
star  formation activities  \citep[e.g.][]{CE01}.   Extragalactic deep
surveys in the X-ray and IR bands have been key tools, identifying the
discrete sources  responsible for the  CXB and CIB,  respectively, and
reconstructing  the  time-lines  of  SMBH  accretion  and  dusty  star
formation. Cosmic energy generation in  both X-ray and IR bands show a
rapid increase  back to redshift $z$  $\sim$ 1-2, with  the density of
luminous   sources   generally    peaking   at   a   higher   redshift
\citep[e.g.][]{Cowie96,    LeFloch05,    Ueda03,   Barger05}.     This
down-sizing   character   of  evolution   is   thought   to  imply   a
mass-dependent  formation scenario  for both  SMBH and  galaxies, with
more massive systems formed  at earlier times.  This parallel behavior
implies  a  statistical  co-evolution   of  SMBH  accretion  and  star
formation, with strong evidence  for physical co-evolution provided by
the tight relationships  between the masses of the  stellar bulges and
central BH’s in galaxies \citep{Kormendy95, Magorrian98, Ferrarese00}.

While  rare in the  local universe,  IR luminous  dusty sources  are a
major    presence   at    high    redshift   \citep[$z\gtrsim$    0.7,
e.g.][]{LeFloch05,  Magnelli11}.   The  quantification  of  fractional
contributions by accretion and star formation to their luminosities is key
to understanding co-evolution and its relationship to mechanisms, such
as gas  accretion or galaxy interactions and  mergers, responsible for
building today’s galaxies and  BHs. Studies of local luminous infrared
galaxies (LIRGs; 10$^{11}$ L$_{\odot}$ $<$ $L_{\rm TIR}$ $<$ 10$^{12}$
L$_{\odot}$) and  ultra luminous infrared  galaxies (ULIRGs; 10$^{12}$
L$_{\odot}$ $<$ $L_{\rm TIR}$  $<$ 10$^{13}$ L$_{\odot}$) have yielded
major  insights.   Early  on,  optical  spectroscopy  established  the
importance  of active galactic  nuclei (AGN)  in powering  IR luminous
galaxies, and the  increase of this importance with  the IR luminosity
\citep{Veilleux95,  Sanders96,  Armus07,  Veilleux99}.  X-ray  surveys
revealed these objects to be X-ray  faint, and many of them are highly
extinguished even  in the hard X-ray band  or completely Compton-thick
\citep{Risaliti00, Iwasawa11}.  Observations in the mid-IR with much
lower extinction based on many available AGN/star-formation diagnostics
found that accretion contributes to the total luminosity about 10-20\%
in  LIRGs  and  20-50\%  in ULIRGs  \citep[e.g.][]{Genzel98,  Armus07,
Veilleux09,  Petric11}.   However,  the  situation  beyond  the  local
universe is  much more uncertain.  Studies  of sub-millimeter galaxies
(SMGs)  at $z$  $\sim$  2 show  that  the number  fraction of  objects
hosting AGN  or the SMBH luminosity  fraction are lower  than in local
objects  with  similar  luminosities \citep{Alexander05a,  Valiante07,
Pope08}.   On  the  other  hand,  the  mid-IR-selected  hyper-luminous
infrared   galaxies    (HLIRGs)   show   significant    AGN   activity
\citep[e.g.][]{Wu12}.  For LIRGs and ULIRGs, the selection method also
plays  an important  role  in the  derived  SMBH luminosity  fraction,
ranging from about 5\% to 30\% \citep{Yan07, Fu10, Fadda10}.

A complete census  of the AGN population across  a wide redshift range
is necessary for a full understanding of co-evolution, but is hampered
by the  large column densities often  obscuring the AGN.   Hard ($>$ 2
keV)  X-rays  arising from  hot  gas close  to  the  central SMBH  are
efficient  at identifying  AGN,  because the  photon  can penetrate  a
significant amount of gas (e.g.   70\% photons at 5keV can get through
$N_{\rm HI}$=10$^{23}$  cm$^{-2}$) to escape the  galaxy.  However, a
poorly known  number of  Compton-thick  AGN (defined  as  $N_{\rm HI}$  $>$
10$^{24}$ cm$^{-2}$ in this  paper) escape detection because even hard
X-rays are  suppressed to undetectable levels.   Current techniques to
probe  these  objects combine  X-ray  data  with  photometry at  other
wavelengths    especially   the    mid-IR    \citep{Lacy04,   Stern05,
Alonso-Herrero06,  Daddi07,  Donley08,  Alexander08, Fiore09,  Luo11}.
The dusty  torus surrounding a  SMBH absorbs the  UV/optical radiation
from the central accretion  disk and the reprocessed radiation emerges
in the IR and peaks at wavelengths significantly shorter than emission
from star  formation regions.  The  apparent gas-to-dust ratio  of the
AGN  is  significantly  larger  than   the  value  in  the  Milky  Way
\citep[e.g.][]{Risaliti00, Maiolino01,  Shi06}, leading to  low mid-IR
optical depths  even for Compton-thick AGN.   Still, broad-band mid-IR
AGN   selection   techniques  suffer   from   both  contamination   by
star-forming  galaxies   and  incompleteness  in   AGN  identification
\citep[e.g.][]{Donley08, Petric11}.

Galaxy population  models have been  used to represent  information in
deep  surveys  and EBL,  typically  focusing  on  either AGN  and  CXB
\citep[e.g.][]{Setti89,   Madau94,  Comastri95,   Ueda03,  Treister05,
Gilli01,  Gilli07} or  on star  formation and  CIRB \citep{Beichman91,
CE01,   Xu01,  Lagache03,  Lagache04,   Rowan-Robinson09,  LeBorgne09,
Valiante09,  Franceschini10, Gruppioni11, Bethermin11}.   However, the
rich collection  of multi-wavelength data accumulated  during the last
decade make  coherent models  taking into account  simultaneously star
formation and  SMBH accretion not only possible,  but necessary.  This
need  is demonstrated  by several  lines of  evidence that  Spitzer 24
$\mu$m   sources   have  significant   AGN   contributions  to   their
mid-infrared luminosity.   At the high flux end  ($f_{24{\mu}m}$ $>$ 5
mJy),  the  Spitzer  5MUSES  legacy  program  (PI:  George  Helou),  a
mid-infrared spectroscopic survey of 330 galaxies selected essentially
by their  24 $\mu$m  flux, found 30\%  of its  spectra have a  low 6.2
$\mu$m   aromatic  equivalent  width   indicative  of   AGN  dominance
\citep{Wu10,  Wu11}.   At   lower  fluxes,  \citet{Donley08}  employed
multiple  broad-band  photometric methods  to  look  for sources  with
AGN-dominated mid-IR emission, and found that the fraction varies from
30\% at $f_{24{\mu}m}=$ 1 mJy to around 10\% at $f_{24{\mu}m}$=0.2 mJy
and remains at 10\% down  to  the  survey  limit ($f_{24{\mu}m}$=0.08  mJy).  These
fractions  are  generally consistent  with  the Spitzer  spectroscopic
results  of  relatively small  samples  \citep{Yan07, Fu10,  Fadda10}.
Further  demonstration  of  this  need  comes  from  Chandra  surveys,
currently  reaching deep enough  to explore  X-ray emission  from star
formation.  The  fraction of galaxies powered by  star formation rises
steeply with decreasing X-ray flux, reaching around 50\% at the survey
limit (10$^{-17}$ erg/s/cm$^{2}$  at 0.5-2 keV) \citep{Bauer04}. Since
these  fainter fluxes  are  also populated  by  heavily obscured  AGN,
taking into  account the star  formation contribution is  critical for
deriving a correct AGN census.   Attempts to combine X-ray and IR data
to  improve the  constraints on  the obscured  AGN have  produced some
interesting  results  \citep{Ballantyne06,  Han12}, indicating  higher
obscured fractions at higher redshift.

This paper introduces a model of galaxy populations across cosmic time
that fits  simultaneously X-ray and  IR survey data and reproduces  the X-ray
and  IR  extragalactic   background.   This  approach  allows
estimations of  quantities inaccessible with  either IR or  X-ray data
alone, such  as the SMBH energy  fraction as a function  of the IR
luminosity out  to $z$ $\sim$ 3.  In this first paper,  we present the
model’s  construction  and the  basic  outputs,  namely  the total  IR
Luminosity  Function  (LF),  the  distributions  of  SMBH  energy
fraction  and HI  column density.   Our philosophy  is  to incorporate
known observational results in the model instead of leaving everything
as  free parameters, but  also present  several different  variants to
reflect  the uncertainties  on the  adopted observational  results. In
addition to this paper that focuses on the model construction, we will
discuss  in other two  papers the  implications   on the
Compton-thick AGN abundance and the cosmic link between obscured
star formation and SMBH accretion, respectively.
   The  overall  structure of  this  paper  is  summarized below.   We
briefly   summarize  previous   models  in   \S~\ref{background}.   In
\S~\ref{model_component}, we describe the structure of our model.  The
general   procedure   to  fit   the   observed   data   is  given   in
\S~\ref{numerical_approach}.  The datasets used for the fit are listed
in  \S~\ref{data_for_fit}.  The results  of the  fit and  the best-fit
parameters are shown  in \S~\ref{fit_result}.  The fundamental outputs
of the model are  discussed in \S~\ref{basic_output}.  Through out the
paper, we divide galaxies into three types based on the X-ray luminosities:
quasi-stellar objects (QSOs)  with log$L_{\rm 2-10keV}$ $>$ 10$^{44}$
erg/s, Seyfert galaxies with  10$^{42}$ erg/s $<$ log$L_{\rm 2-10keV}$
$<$ 10$^{44}$ erg/s, and BH quiescent galaxies (BHQGs) with log$L_{\rm
2-10keV}$ $<$  10$^{42}$ erg/s. AGN include QSOs and Seyferts. Based  on the HI column  density, the
AGN are  divided into type-1  ($N_{\rm HI}$ $<$  10$^{22}$ cm$^{-2}$);
Compton-thin  type-2   (10$^{22}$  cm$^{-2}$  $<$   $N_{\rm  HI}$  $<$
10$^{24}$  cm$^{-2}$) and Compton-thick  (($N_{\rm HI}$  $>$ 10$^{24}$
cm$^{-2}$).  The total  IR luminosity is defined to be  from 8 to 1000
$\mu$m.   The galaxies  are also  defined  according to  the total  IR
luminosity  as  IR  quiescent   galaxies  (IRQGs;  $L_{\rm  TIR}$  $<$
10$^{11}$ L$_{\odot}$), LIRGs (10$^{11}$ L$_{\odot}$ $<$ $L_{\rm TIR}$
$<$ 10$^{12}$ L$_{\odot}$),  ULIRGs (10$^{12}$ L$_{\odot}$ $<$ $L_{\rm
TIR}$  $<$  10$^{13}$  L$_{\odot}$)  and  HLIRGs  ($L_{\rm  TIR}$  $>$
10$^{13}$  L$_{\odot}$).   We adopt  a  cosmology  with $H_{0}$=70  km
s$^{-1}$ Mpc$^{-1}$, $\Omega_{\rm m}$=0.3 and $\Omega_{\Lambda}$=0.7.

\section{Background: The Synthesis Model Of Extra-galactic Background Light}\label{background}

Before we describe our EBL  model, we here summarize previous CXB and
CIRB models.  In general,  a synthesis model first parametrizes the
LF at a wavelength and its redshift evolution.  With assumed SEDs, the
model can then predict a  LF at any observed monochromatic wavelength,
which will be compared to  the data to derive the best-fit parameters.
The approach can, in principle, fit all the survey data 
regardless of  field size, survey depth and  wavelength coverage. 

\subsection{Cosmic X-ray Synthesis Model}

The CXB  was discovered almost half a  century ago \citep{Giacconi62},
and is now understood to  be dominated by the integrated emission from
gas  being heated and  accreted onto  SMBHs at  the centers  of active
galaxies.  The pure black-body spectrum of the CMB places strong upper
limits on the contributions to  the CXB from diffuse hot intergalactic
gas \citep[$10^{-4}$, ][]{Wright94}.  Deep extra-galactic surveys have
now  resolved a  significant fraction  ($\sim$70\%) of  the background
light  at   $<$  10  keV  into   discrete  sources  \citep{Hasinger98,
Mushotzky00,  Alexander03,  Worsley05},  with  the  resolved  fraction
decreasing with the increasing  energy bands.  The exact fraction also
depends  on  the absolute  normalization  of  the  CXB spectrum  whose
dispersion  reaches  $\sim$30\%  at  low  energies ($<$  10  keV)  and
$\sim$10\%     above     10     keV    among     different     studies
\citep[e.g.][]{Marshall80, Gruber99, Vecchi99, Revnivtsev03, Ajello08,
Churazov07, Moretti09, Turler10}.

The CXB  spectrum peaks around 20-30  keV and drops  toward both lower
and higher energies, and its slope  at $<$ 20 keV is harder than those
of Seyfert-1 galaxies.  In order to reproduce this  slope and turnover
around  20-30 keV,  the general  strategy is  to invoke  the SED  of a
type-1  AGN along  with a  varying HI  column  density \citep{Setti89,
Madau94,  Comastri95},  consistent with  the  unified  scheme for  AGN
\citep{Antonucci93} where Seyfert-2  galaxies are obscured variants of
Seyfert-1 galaxies and exhibit  harder X-ray spectra. Today CXB models
not only  reproduce the  CXB spectrum (1-100  keV) but also  the X-ray
number counts, redshift distributions, and spectral properties in both
soft and  hard bands \citep[e.g.][]{Treister05,  Gilli07, Treister09}.
It  is generally  accepted that  the  majority of  X-ray AGN  activity
occurs in  the obscured phase and  that the evolution of  X-ray AGN is
luminosity  dependent, with  the  density of Seyfert-like AGN 
peaking around $z$=1 and quasars peaking at $z$=2-3.
However,  there  are  still  important uncertainties  in  the  derived
parameters, such as the type-1/type-2 AGN ratio and its variation with
luminosity and redshift, and the amount of Compton-thick objects. Such
difficulties are at  least partly caused by the  fact that the current
X-ray surveys at $<$ 10  keV are not sensitive to heavily-extinguished
AGN,  especially  Compton-thick  objects,  and overall  number  counts
remain limited.

\subsection{Cosmic Infrared Synthesis Model}

The CIRB was first directly measured in the 1990s by COBE \citep[for a
review,  see][]{Hauser01}.  The energy  of CIRB  is comparable  to the
cosmic UV/optical energy \citep{Dole06},  while the ratio in the local
universe is only one third \citep{Driver08}.  This implies an increase
in dusty sources  with increasing redshift.  The slope  of the CIRB in
the  far-IR/submm  range  is  flatter  than that  of  local  starburst
galaxies, which provides further  evidence for the strong evolution of
IR sources whose peak emission is redshifted toward longer wavelength,
flattening the CIRB spectrum at the longer wavelength.

Unlike the CXB  model where the HI column density  is considered to be
the  main factor  driving the  variation of  the SED,  the  CIRB model
typically varies the SED mainly  as a function of IR luminosity.  They
either  adopt   a  fixed  SED  at  a   given  luminosity  \citep{CE01,
LeBorgne09}, or  further consider  the scatter of  the SED at  a given
luminosity  \citep{Chapman03,  Valiante09},  or  assume  different  IR
populations   that  exhibit  different   SEDs  and   occupy  different
luminosity ranges \citep{Xu01, Lagache03, Lagache04, Rowan-Robinson09,
Franceschini10, Gruppioni11,  Bethermin11}.  The SMBH  radiation, with
typically  a much warmer  IR SED  compared to  that of  star formation
regions is energetically important in mid-IR surveys: the contribution
of  AGN   to  the  mid-IR  background  reaches
noticeable fractions, e.g. 10-20\% in the 1-20 $\mu$m range by \citet{Silva04}
or  2-10\%   in   the   3-24   $\mu$m  by \citet{Treister06}; 
classifications (AGN vs star-formation dominated) of 24 $\mu$m sources
indicate  that  the  fraction  of  objects whose  mid-IR  emission  is
dominated  by AGN  ranges from  $\sim$40\% at  f$_{24{\mu}m}$ $>$  5  mJy and
drops down with decreasing fluxes but remains around 10\% down to
the  faintest  end  f$_{24{\mu}m}$ $\sim$ 0.08  mJy \citep{Donley08,  Fu10,  Wu10,  Choi11}.
However, the 24 $\mu$m number counts can be reproduced equally well by
models  with AGN  \citep{Valiante09, Franceschini10,  Gruppioni11} and
without AGN \citep{Bethermin11, LeBorgne09}, reflecting the limitation
of the  CIRB synthesis models  in decomposing SMBH  and star-formation
emission. Models without  AGN can re-produce the effect  of AGN in the
mid-IR  simply  by  modifying  the  star-formation mid-IR  SED  or  by
adjusting the slope of LFs and their redshift evolution.

\section{The Construction Of A Joint Model Of X-ray And IR Backgrounds}\label{model_component}

Our  model has four  basic components,  each of  which contains  a few
components.  The four components  are a luminosity function defined in
the total  IR band  (8-1000 $\mu$m), the  SMBH energy fraction  in the
total  IR  band, the  luminosity-dependent  SED  for the  star-forming
component,  and  the type-1  SMBH  SED  extinguished  by a  HI  column
density.  The first component describes the number density of galaxies
at  a given redshift  and total  IR luminosity,  while the  second one
decomposes  this  luminosity into  star-formation  and  SMBH total  IR
emission.  The last two then generate star-formation and SMBH emission
at any  observed band from their corresponding  total IR luminosities.
The general  rule for setting the  total number of  free parameters is
that  the model starts  with the  minimum set  of free  parameters and
additional ones are added if the reduced $\chi^{2}$ decreases by about
0.5. The goal  is to fit model predictions  to number counts, redshift
distributions  and  local LFs  data,  which  offers  hundreds of  data
points. As discussed  in the following of this section,  we will run a
total of four  variants of the model to  reflect current uncertainties
on the star-forming and SMBH SEDs.

\subsection{Infrared Luminosity Function}\label{model_component_LF}

We represent  the galaxy total IR LF as a double power-law distribution with ten free parameters:
\begin{eqnarray}
 \Phi(L_{\rm TIR}, z) &=& \frac{dN}{{d{\rm log}L}{\times}dV} \nonumber \\
&=& \frac{\Phi_{*}}{(L/L_{*})^{\gamma_{1}}+(L/L_{*})^{\gamma_{2}}} [{\rm Mpc}^{-3} {\rm log}L_{\odot}^{-1}]
\end{eqnarray}
We adopt a simple independent luminosity and density
evolution  (ILDE). The  evolution  is described as a polynominal
function  of  redshift:
\begin{equation}\label{eq_evl_lstar}
{\rm log}L_{*}(z)   =
{\rm log}L_{*, 0}    + \eta{\times}k_{1, l} + \eta^{2}{\times}k_{2, l}+\eta^{3}{\times}k_{3, l} 
\end{equation}
\begin{equation}\label{eq_evl_denstar}
{\rm log}\Phi_{*}(z) = 
{\rm log}\Phi_{*, 0}  + \eta{\times}k_{1, d} + \eta^{2}{\times}k_{2, d}+\eta^{3}{\times}k_{3, d}
\end{equation} where  $\eta = {\rm log}(1+z)$.  For  this component of
the  model,  there are  ten  free  parameters  including four  in  the
luminosity evolution  (${\rm log}L_{*, 0}$, $k_{1, l}$,  $k_{2, l}$ \&
$k_{3, l}$),  four in the  density evolution (${\rm  log}\Phi_{*, 0}$,
$k_{1, d}$,  $k_{2, d}$ \&  $k_{3, d}$) and two  slopes ($\gamma_{1}$ \&
$\gamma_{2}$).

\subsection{The SMBH Energy Fraction}\label{model_component_BHfrac}

\begin{figure}
\epsscale{1.0}
\plotone{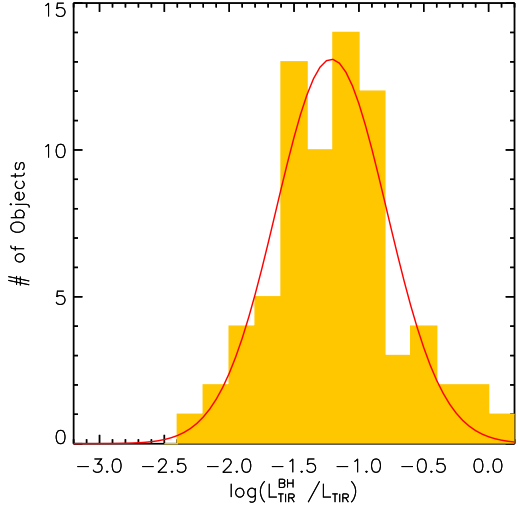}
\caption{\label{ew_1JyULIRGs} The distribution of the SMBH energy fraction in the total IR band of 74 1-Jy
ULIRGs based on the aromatic feature EW measurements by \citet{Veilleux09}. The filled histogram shows the 
derived SMBH energy fraction distribution. The solid line is the Gaussian
fit to the filled histogram. For the detail, see \S~\ref{model_component_BHfrac}.  }
\end{figure}

Individual objects  in the model  are assumed to be  experiencing both
star formation and SMBH  accretion activities, so that $L_{\rm TIR}$
= $L_{\rm TIR}^{\rm BH}$ + $L_{\rm TIR}^{\rm SF}$.  For the formula of
the SMBH energy fraction, we  assume a bounded Gaussian distribution for
the logarithm of  the SMBH energy fraction in the  total IR band $f^{\rm
BH}_{\rm TIR}=\frac{L_{\rm TIR}^{\rm  BH}}{L_{\rm TIR}}$, as motivated
by works of \citet{Valiante09}:
\begin{equation}
P^{\rm BH}({\rm log}f^{\rm BH}_{\rm TIR}) =\frac{dN}{d{\rm log}f^{\rm BH}_{\rm TIR}}\propto {\rm exp}(-\frac{ ({\rm log}f^{\rm BH}_{\rm TIR} - {\rm log}f^{\rm BH}_{\rm *, TIR})^{2} }{2(\sigma^{\rm BH})^{2}})
\end{equation}
\begin{equation}
{\int_{\rm min}^{\rm max}}P^{\rm BH}({\rm log}f^{\rm BH}_{\rm TIR}) d{\rm log}f^{\rm BH}_{\rm TIR} = 1
\end{equation}
where the boundary of  log$f^{\rm BH}_{\rm TIR}$ is set to have the minimum and maximum at -6.0 and 0.0, respectively (for details,
see  \S~\ref{numerical_approach}). 

This Gaussian distribution is assumed to be redshift and luminosity dependent. The mean value of the 
distribution (${\rm log}f^{\rm BH}_{*}$) is  a function of both total IR luminosity and redshift:
\begin{equation}{\label{fBH}}
{\rm log}f^{\rm BH}_{*} =  {\rm log}f^{\rm BH}_{*, z} + p_{f^{\rm BH}}{\times}{\rm log}\frac{L_{\rm TIR}}{L_{*, f^{\rm BH}}}
\end{equation}
where
\begin{equation}
 {\rm log}f^{\rm BH}_{*, z} = {\rm log}f^{\rm BH}_{*, 0} + \eta{\times}k_{1, d}^{\rm BH} + \eta^{2}{\times}k_{2, d}^{\rm BH} 
\end{equation}
The width of the distribution is only luminosity dependent:
\begin{equation}
  \sigma^{\rm BH} = \sigma_{0}^{\rm BH} + p_{\sigma}{\times}{\rm log}\frac{L_{\rm TIR}}{L_{*, f^{\rm BH}}}
\end{equation}

At  $z$$=$0  and $L_{\rm  TIR}$=$L_{*,  f^{\rm  BH}}$,  the SMBH  energy
fraction is fixed at $f^{\rm BH}_{*, 0}$, which
is based on the sample of  local 1 Jy ULIRGs \citep{Kim98}.  The advantage of ULIRGs
compared to lower luminosity objects  is that the SMBH contributions are
important so  that the distribution of  the SMBH energy  fraction can be
derived to a high accuracy.  For a sample of 74 ULIRGs, we derived the
SMBH IR luminosity from the observed 6.2 $\mu$m equivalent width and 6.0
$\mu$m continuum luminosity as measured by \citet{Veilleux09}:
\begin{eqnarray}\label{eqn_LAGN_ew}
 L_{6{\mu}m}^{\rm BH}={\rm max}[(1.0-EW_{6.2{\mu}m}/EW_{\rm star-formation}){\times} \nonumber \\
L_{6{\mu}m}{\times}exp(\tau_{6{\mu}m}), 0]
\end{eqnarray}  where  $EW_{6.2{\mu}m}$  is  the observed  6.2  $\mu$m
equivalent width,  $L_{6{\mu}m}$ is the observed  6.0 $\mu$m continuum
luminosity, $EW_{\rm star-formation}$ is  the equivalent width of pure
star  formation and  taken to  be 4.2  $\mu$m  \citep{Veilleux09}, and
$\tau_{6{\mu}m}$ is the effective optical depth at 6 $\mu$m and set to
be  10\% of  the  measured 9.7  $\mu$m  silicate depth  using the  SMC
extinction curve \citep{Pei92}. Given  the star-forming EWs still have
a small scatter \citep[e.g.][]{Wu10,  Stierwalt12}, we set the maximum
value of  the derived SMBH radiation  in Equation~\ref{eqn_LAGN_ew} to
be zero.   Note that the aromatic  feature EW relies on  the method to
measure  it, e.g.,  sometimes  the  wing emission  of  the feature  is
included through  extrapolating the analytic  profile. As long  as the
same  method is  adopted,  the  SMBH energy  fraction  from the  above
equation should not  differ significantly.  The $L_{6{\mu}m}^{\rm BH}$
is then corrected  to the total SMBH IR luminosity by  a factor of 1.6
using  our  own  SMBH  SED.   Combined  with  the  measured  total  IR
luminosity,  the final  distribution of  the SMBH  energy  fraction is
shown  in  Figure~\ref{ew_1JyULIRGs}.  Our  assumption  of a  Gaussian
distribution is more or less  correct given the derived shape. The fit
with  a  Gaussian  gives  log$f^{\rm BH}_{*,  0}$=-1.21.   The  median
luminosity  of this  sample gives  log$L_{*, f^{\rm  BH}}$=12.3.  Note
that  the above  values do  not  depend significantly  on the  adopted
$EW_{\rm  star-formation}$  since  $EW_{6.2{\mu}m}$  is small  in  the
majority  of  the  objects.   However,   we  did  not  use  the  above
distribution  to  fix $\sigma_{0}^{\rm  BH}$  but  set  it as  a  free
parameter, because the  width measured in Figure~\ref{ew_1JyULIRGs} is
broadened  by the intrinsic  scatter in  $EW_{\rm SF}$  and luminosity
correction  $L_{\rm  TIR}/L_{6{\mu}m}$.   We  notice that  the  median
fraction derived here is  significantly lower than the energy fraction
of roughly 35-40\% as  derived by \citet{Veilleux09}.  The main reason
is that their numbers are given  in the bolometric band while ours are
integrated from 8-1000 $\mu$m.   While the emission is similar between
these two bands  for ULIRGs, a factor of $\sim$ 5  needs to be applied
for the SMBH radiation.

There  are five  free parameters  in total,  including  two describing
redshift evolution  of the mean fraction ($k_{1,  d}^{\rm BH}$ \& $k_{2,
d}^{\rm BH}$),  one characterizing  luminosity dependence of  the mean
fraction  ($p_{f^{\rm BH}}$),  one for  the  scatter ($\sigma_{0}^{\rm
BH}$)  at  the total  IR  luminosity  of  10$^{12.3}$ L$_{\odot}$  and
another for its luminosity dependence ($p_{\sigma}$).

\subsection{Star Formation SED Template}

For  the  IR  SED  of  the star-formation  emission,  we  adopted  the
luminosity-dependent SED  shapes and associated  dispersions.  This set
of SEDs is  well characterized in the local  universe; the uncertainty
lies  at high $z$  where the  difference is  already observed  but not
fully quantified.

\subsubsection{Local Infrared SED Templates}\label{SF_IR_SED}

 It is now established that  the IR luminosity is not the main driver
of the IR  SED shape, i.e., at a given luminosity,  the IR color shows
a large  scatter  \citep{Chapman03,  Hwang10,  Rujopakarn11a,  Elbaz11}.
Fortunately, this scatter has been well quantified. We thus adopted the
observed  trend between  IR luminosity  and  color ($\frac{f_{\nu}(\rm
IRAS-60{\mu}m)}{f_{\nu}(\rm  IRAS-100{\mu}m)}$) as well  as the associated
scatter as established by \citet{Chapman03} for local  IRAS galaxies.  The
probability  distribution  of IR  color  at  a  given star-forming  IR
luminosity is described by a Gaussian function:
\begin{equation}
P^{\rm SF{\textendash}color} = \frac{dN}{dC}= \frac{1}{\sigma_{c}\sqrt{2\pi}} {\rm exp}(-\frac{(C-C_{0})^2}{2(\sigma_{c})^2})
\end{equation}
where $C$=log($\frac{f_{\nu}(60{\mu}m)}{f_{\nu}(100{\mu}m)}$), $\sigma_{C}$=0.065, and
\begin{equation}{\label{color_lum}}
 C_{0} = C_{*} + {\gamma}{\times}{\rm log}(1+\frac{L_{\rm TIR}^{'}}{L_{*, C}}) - {\delta}{\times}{\rm log}(1+\frac{L_{*, C}}{L_{\rm TIR}^{'}})
\end{equation} $C_{*}$ = $-$0.35, $\gamma$=0.16, $\delta$=0.02, $L_{*,
C}$=5.0$\times10^{10}$  L$_{\odot}$,  and  $L_{\rm  TIR}^{'}$  is  the
star-forming total IR luminosity from  3 $\mu$m to 1100 $\mu$m.  For a
given   color    of   $\frac{f_{\nu}(\rm   IRAS-60{\mu}m)}{f_{\nu}(\rm
IRAS-100{\mu}m)}$,  we  used  the  infrared SED  template  library  of
\citet[hereafter DH02]{DH02},  which sorts the  IR SED by  that color.
The template  was built empirically  first from IRAS/ISO  3-100 $\mu$m
observations \citep{Dale01}  and then extended up  to submm wavelengths
based  on  SCUBA  and  ISO  data  \citep{DH02}.   {\it  Spitzer}  data
(3-160$\mu$m) confirms its  ability to cover the overall  range of SED
variations  and describe  the  median trend  between different  colors
below 200 $\mu$m  \citep[e.g.][]{Dale07, Wu10}.  Recent {\it Herschel}
far-IR     observations     further     demonstrated     its     power
\citep{Rowan-Robinson10,     Boselli10,     Dale12}.     Nevertheless,
observations  suggest that  this  set of  templates  likely lacks  the
dynamic range in the SED  variations above rest-frame 200 $\mu$m.  For
example,  DH02  templates  do  not  produce  objects  with  rest-frame
$f_{\nu}$(Herschel-500$\mu$m)/$f_{\nu}$(Spitzer-24$\mu$m)  $>$  6.5 as
observed  in \citet{Rowan-Robinson10} although  they bracket  well the
spread in $f_{\nu}$(Spitzer-70$\mu$m)/$f_{\nu}$(Spitzer-24$\mu$m). The
templates also  under-represent the diversity of slopes  in the mid-IR
\citep{Wu10}.  It should be  emphasized that all template libraries in
the literature  are based on  broad-band photometry, mainly  using ISO
and  IRAS filters.  However, the  detailed  SED shape  at fine  scales
within a  filter can actually affect model  predictions for broad-band
surveys as the different parts  of spectra are redshifted into a fixed
wavelength filter.

\subsubsection{IR SEDs At High-$z$}\label{IR_SED_HIGHZ_SF}

The sensitive infrared surveys carried out by Spitzer and Herschel are
revealing  the SED  shapes of  high  redshift luminous  sources to  be
significantly  different from  local galaxies  at the  same luminosity
\citep[e.g.][]{Papovich07,  Rigby08, Symeonidis09,  Muzzin10, Hwang10,
Rujopakarn11a,  Rujopakarn11b,  Elbaz11}.   Many  high-$z$  LIRGs  and
ULIRGs show IR SEDs similar  to those of local less luminous galaxies,
i.e.  high-z LIRGs  and  ULIRGs  are colder  compared  to their  local
counterparts; IR  luminosities do not correspond  to unique properties
of  star-forming regions.   To account  for this  fact, we  evolve the
$L_{*, C}$ in Equation~\ref{color_lum} through
\begin{equation}\label{EV_T_LTIR}
L_{*,  C}(z) = L_{*,  C}(z=0)(1+z)^{p_{c}}
\end{equation}

It is,  however, still difficult  to constrain $p_{c}$ because  of the
limited number of available  broad-band photometric bands, the dynamic
range of the IR luminosities at a fixed redshift and the observational
selection  biases.    It  is  possible  that  the   evolution  in  the
star-formation   SED   only  takes   place   at   $z$  $\gtrsim$   1.5
\citep{Hwang10,  Elbaz10}. But  it is  still not  certain  whether the
basic shape  of Equation  \ref{color_lum}, i.e., a  two-power-law mean
trend  and  a  Gaussian  distribution  of the  scatter,  is  valid  at
high-$z$.   For  example, \citet{Elbaz11}  have  described the  high-z
$L_{\rm TIR}$/$L_{8{\mu}m}$  color as  a Gaussian distribution  plus a
higher-value tail that  is independent of the IR  luminosity.  In this
paper, we assumed that the form of the local relationship holds at the
high  z  and  explored  two  possibilities of  $p_{c}$,  zero  for  no
evolution  and  2.0  for  rapid  evolution to  bracket  the  range  as
discussed  in  the  literature \citep[e.g.][]{Symeonidis09,  Muzzin10,
Hwang10, Magnelli09, Chapin11}. Notice that  we did ran the model with
various $p_{c}$  values from  0.0 to  3.0 but failed  to found  that a
non-zero $p_{c}$ value can give the  better fit compared to the no SED
evolution case. Because of this, the $p_{c}$=2.0 is used to illustrate
the effect of the SED evolution in the model's fitting.

\subsubsection{X-ray SED}

\begin{figure}
\epsscale{1.0}
\plotone{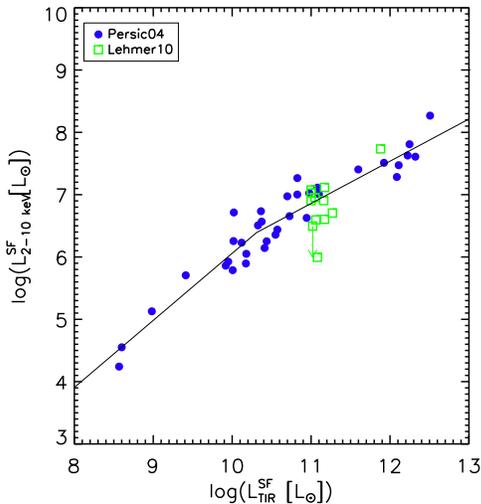}
\caption{\label{lx_ir_sf} The relationship between IR and X-ray emission of star-forming galaxies
based on the sample of \citet{Persic04} and \citet{Lehmer10} (see Equation \ref{eqn_lx_ir_sf}). }
\end{figure}

Deep X-ray  surveys, especially in  the soft band, can  probe emission
from star-formation  regions.  We thus related  the star-forming X-ray
emission to the star formation IR luminosity and constructed the X-ray
spectrum to  go with  each IR SED  template.  In galaxies  with active
star formation,  the X-ray emission  is dominated by  short-lived high
mass  X-ray binary  (HMXB)  and is  known  to be  related  to the  SFR
\citep[e.g.][]{Grimm03,  Persic04, Lehmer10}.   In  quiescent objects,
the low-mass X-ray binary  (LMXB) emission becomes important and X-ray
luminosity  is no  longer  a reliable  SFR  tracer.  Nevertheless,  IR
emission also has a contribution from dust heated by long-lived stars,
so it is reasonable to expect  a relation between IR and X-ray even in
quiescent    objects.    We    thus    compiled   measurements    from
\citet{Persic04}  and \citet{Lehmer10}  but excluding  AGN  and AGN+SF
composite objects.   For a  few common objects  in both  studies, data
from the  latter are used.   The total IR luminosities  unavailable in
the former study  are measured by our own based on  the IRAS four band
fluxes  through  the  \citet{Sanders96}  equation.  The  final  sample
covers  the   IR  luminosity  from   10$^{8}$  up  to   $>$  10$^{12}$
L$_{\odot}$, as  shown in Figure~\ref{lx_ir_sf}.  A two  power law fit
to the data gives:
\begin{equation}\label{eqn_lx_ir_sf}
{\rm log}L_{\rm 2-10keV}^{\rm SF} = 
\begin{cases}
 -4.70 + 1.08\times{\rm log}L_{\rm TIR}^{\rm SF} & \text{  for }  {\rm log}\frac{L_{\rm TIR}^{\rm SF}}{L_{\odot}} \leq 10.31\\
 -0.59 + 0.68\times{\rm log}L_{\rm TIR}^{\rm SF} & \text{  for }  {\rm log}\frac{L_{\rm TIR}^{\rm SF}}{L_{\odot}} > 10.31\\
\end{cases}
\end{equation}  
where  all  luminosities  are  in  units  of  solar  luminosity.   The
1-$\sigma$ scatters for trends at the low and high luminosity ends are
0.28 and  0.27 dex, respectively.  The fit does  not use the  only one
upperlimit  but  the result  does  not  change  if including  it.   As
discussed by \citet{Lehmer10}, the  scatter between X-ray emission and
SFR  can be  reduced if  including  the stellar  mass quantity  which,
however,  is   not  feasible  for   our  model  that  does   not  have
optical/near-IR components to infer stellar masses. Independent  of the X-ray luminosity, the
spectrum from 0.2  to 100 keV is fixed to be  that of the star-forming
galaxy M82 based on  the BeppoSAX observation from \citet{Cappi99} and
Swift data from \citet{Baumgartner11}.   We further assume no redshift
evolution  for  the   $L_{\rm  X-ray}^{\rm  SF}-L_{\rm  IR}^{\rm  SF}$
relationship and  X-ray spectrum.  The  lack of redshift  evolution in
the  $L_{\rm  X-ray}^{\rm  SF}-L_{\rm  IR}^{\rm SF}$  relationship  is
tentatively suggested by some works \citep{Grimm03, Symeonidis11}.

\subsection{SMBH SED Template}\label{BH_SED_TEMPLATE}

The grand unified model of AGN \citep{Antonucci93} states that all AGN
are intrinsically the same and  the emerging SED is only determined by
the line of sight with respect  to the dusty torus; the type-2 objects
are the  obscured version of  the type-1 objects. Adopting
this scenario, we first constructed  the intrinsic AGN SED from type-1
data-set and  then modified it with extinction  as characterized by
the  HI  column density    ($N_{\rm  HI}$)  to obtain  the
extinction-dependent SED. A reflected component in the X-ray is always added:
\begin{eqnarray}
f_{\nu}^{\rm AGN{\textendash}observed} = f_{\nu}^{\rm AGN{\textendash}intrinsic}{\times} \nonumber \\
    {\rm exp}(-\tau(\nu, N_{\rm HI})) + {\rm reflected-component}
\end{eqnarray} 
where $f_{\nu}^{\rm AGN{\textendash}observed}$  is the observed SED of
AGN,  $f_{\nu}^{\rm AGN{\textendash}intrinsic}$  is the  intrinsic SED
before extinction and $\tau(\nu, N_{\rm HI})$ is the extinction curve. 

The significant  difference from  the star-formation SED  templates is
that  we  only  used  the  mean  SMBH  SED  but  did  not  consider  its
dispersion. This  is partly because to keep  computations manageable but
also because of the lack  of studies that address fully the dispersion
of the AGN SED from X-ray all the way to submm as a function of the BH
luminosity \citep[e.g.][]{Hao12}.

\subsubsection{The IR/submm SED  of Unobscured AGN}


\begin{figure}
\epsscale{1.0}
\plotone{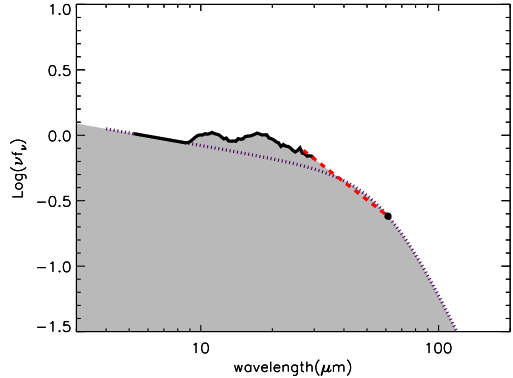}
\caption{\label{pg_spec} The  IR spectrum of  type-1 AGN based
on the  {\it Spitzer}  IRS
spectrum and  MIPS photometry of  PG  quasars (the upper  bound of the grey  area). First, the
median IR SED is measured for  the whole PG sample. The star formation
contribution is subtracted based  on the measured aromatic features of
this median spectrum. The black  solid line and filled circles are the
IRS spectrum  and MIPS 70  $\mu$m photometry after  subtractions. Note
that the observed MIPS 160  $\mu$m is dominated by star formation. The
purple dotted line describes  the underlying continuum of the silicate
features. It  is a  fit to  the 5-8$\mu$m IRS  spectrum and  70 $\mu$m
photometry with  two power laws where  the index at  long wavelength is
fixed to be 4 to mimic the modified black-body emissivity of 1.0 (${\nu}f_{\nu}$
$\propto$ $\nu^{4}$). A single  power law (red dashed line) connecting
the IRS  30 $\mu$m emission and  MIPS 70 $\mu$m photometry  is used to
model the silicate emission above the continuum roughly between 30 $\mu$m and 38 $\mu$m. }
\end{figure}


\begin{figure}
\epsscale{1.0}
\plotone{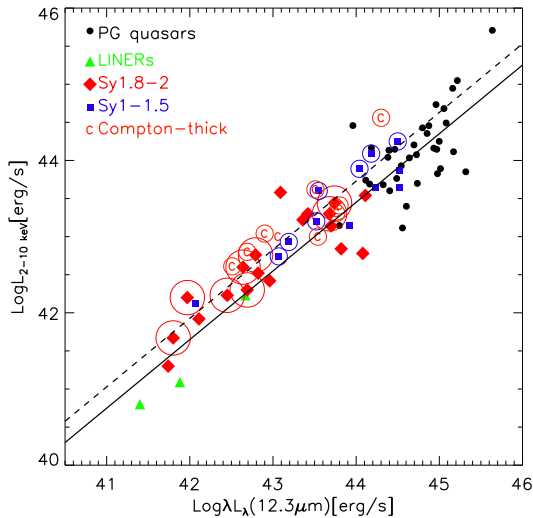}
\caption{\label{lx_lir} The relationship between 2-10 keV unobscured X-ray and
12.3 $\mu$m luminosities. The dashed line is the best fit for all Seyfert and 
LINER galaxies with well-resolved images at 12.3 $\mu$m \citep[circled symbols; ][]{Gandhi09}. The solid line
is the fit adopted in this paper that is exclusively for PG quasars.
It is derived by fixing the slope as that for Seyfert and LINER ones but recalculating the 
normalization to produce the median trend of PG quasars. 
The offset between the dashed and solid lines is due to the strong silicate features in the PG quasars.}
\end{figure}

The  distinct IR  continuum feature  of the  unobscured AGN  is strong
emission of  dust  heated  to near the  sublimation temperature
($\sim$ 1000 K) by UV/optical radiation from the accretion disk.  This
feature  serves  as the  basis  of various  mid-IR AGN  selection
techniques  \citep{Lacy04,  Stern05,  Armus07, Donley08,  Wu11}.   The
shape  of this  feature  has  been well  characterized  with the 
emission peaking  at    3-5   $\mu$m   and   the    1-30   $\mu$m   slope
$\alpha$=1$\pm$0.3  ($f_{\nu}$  $\propto$  $\nu^{-\alpha}$), although the
 slope estimate depends on the definition of the photometric bands and their placement with
respect  to the  silicate  feature besides  the intrinsic  dispersion.
Generally,  the broad-band  photometric  observations indicate  slopes
between 1.0 and 1.5 \citep[e.g.][]{Neugebauer87,  Haas03}. The Spitzer
IRS observations of a large  sample of local AGN reveal smaller slopes
for the  continuum underlying the silicate feature.   For example, the
whole sample of 87 PG  quasars shows a relatively universal slope with
median and  standard deviations of $\alpha_{5-30{\mu}m}$=0.75$\pm$0.35
($f_{\nu}$  $\propto$ $\nu^{-\alpha}$)  \citep{Shi07}
\citep[also see][]{Netzer07, Maiolino07}.  The Seyfert 1
objects     in    the     complete    12     $\mu$m     sample    have
$\alpha_{15-30{\mu}m}$=0.85$\pm$0.61  \citep{Wu09} where  the slightly
steeper  slope and  the larger  dispersion  are likely  caused by  the
larger star  formation contamination in  these low luminosity  AGN. No
obvious evidence has been seen for  the dependence of the slope on the
BH luminosity in the above studies.

The most prominent  spectral feature of the unobscured  AGN SED is the
silicate emission  features at 9.7  and 18 $\mu$m as established
unambiguously  by Spitzer  observations  \citep[e.g.][]{Hao05, Shi06}.
It is  important to include this  feature in the SED  template for our
model as it can affect the broad-band photometry. The  strength  of  this 
 feature depends weakly on the  SMBH  luminosity \citep{Maiolino07}, with only a factor
of 2 increase for a SMBH luminosity  range of 10$^{4}$. We thus adopt the same
feature strength independent of the SMBH luminosity.

The most uncertain part of the  SMBH IR SED lies above $\sim$ 30 $\mu$m
where the star-formation emission  becomes important or even dominates 
for  quasars.   One solution  is  to  spatially  separate the  nuclear
emission  from   star  formation   regions  in  host   galaxies;  such
observations     are     only     feasible     below     30     $\mu$m
\citep[e.g.][]{Gandhi09}.   Another   challenging  problem  with  this
strategy  is  that we  have  no  idea if  the  dusty  torus itself  is
experiencing star  formation and how  important that is.   The current
approach is  to scale  the star formation  with the  observed aromatic
feature  and  then subtract  it  off \citep[e.g.][]{Shi07,  Netzer07}.
This leads  to the  conclusion that above  $\sim$ 100 $\mu$m,  all the
observed emission  typically comes from  star formation, and  that SED
part then relies  on the extrapolation from the  short wavelength.  In
spite  of significant  caveats associated  with the  AGN SED  above 30
$\mu$m,  its effect  on our  model is  limited.  This  is  because the
majority (90\%) of the AGN IR emission (3-1000$\mu$m) emerges below 30
$\mu$m; their far-IR contribution is negligible to the observed far-IR
radiation.

In practice,  we constructed  the unobscured IR  SED based  on Spitzer
observations of the  whole PG sample that is selected  based on the UV
excess   and  are   thus  type-1   objects  with   little  obscuration
\citep{Green86}.  To construct the intrinsic SED, we first derived the
median  composite spectrum of  Spitzer IRS  spectra (5-30  $\mu$m) and
MIPS photometry (24,  70 and 160 $\mu$m) of all  the PG objects.  With
the composite  spectrum, we measured the 11.3  $\mu$m aromatic feature
and  scaled it  to derive  the  star-forming emission  at the  Spitzer
wavelength based  on the  DH02 template.  The  star-formation emission
contributes less than 15\% at  $<$ 30 $\mu$m and reaches $\sim$50\% in
the 70 $\mu$m filter while  dominating the 160 $\mu$m filter emission.
A  continuous  IR  SED  template  is then  constructed  based  on  the
star-formation subtracted IRS spectrum and photometry in the 70 $\mu$m
filter where all the aromatic features are also carefully removed.  To
fill   the  gap   in  the   SED  beyond   30  $\mu$m,   as   shown  in
Figure~\ref{pg_spec}, the underlying continuum of the silicate feature
is fitted  with a  double power  law (purple dotted  line) to  the 5-8
$\mu$m IRS part and MIPS 70$\mu$m photometry where the power law index
at  long wavelength  part  is fixed  to  be 4  to  mimic the  modified
black-body   with   emissivity   of  1.0   (${\nu}f_{\nu}$   $\propto$
$\nu^{4}$). A  single power law  connecting the IRS  30$\mu$m emission
and MIPS  70$\mu$m photometry (red dashed  line) is used  to model the
silicate emission between  30 and 38 $\mu$m.  The  resulting IR SED is
quite consistent with  the photometry based SED derived  from the SDSS
quasars   by   \citet{Richards06},   except   for  the   much   better
characterized silicate features in our template.

Extracting the intrinsic IR SED for a large sample of AGN with a range
of  host  galaxy  contamination,  extinctions,  etc,  would  be  quite
difficult. Moreover, the spatial  and spectral resolution data are not
available to  allow for a  result comparable to that  for star-forming
galaxies with well characterized mean  SED and associated scatter as a
function of the luminosity  \citep{DH02, Chapman03}. We thus adopt the
mean PG  quasar IR SED with  no luminosity dependence  or scatter. The
recent work  by \citet{Mullaney11} claimed a  luminosity dependence of
the AGN intrinsic mid-IR/far-IR ratio  in a small AGN sample. However,
their low-  and high-luminosity sub-samples  have completely different
extinctions, which could easily  induce such a dependence. The studies
of  X-ray AGN  in the  COSMOS field  show no  apparent  luminosity and
redshift dependence for their SED \citep{Hao12}.

\subsubsection{X-ray SED As A Function Of $N_{\rm HI}$}\label{Xray_SED_HI_AGN}

For the X-ray SED as a function of the HI column density, we basically
adopted the work of \citet{Gilli07}.  Since no scatter is included for
the IR part  in our model, we did not  consider the scatter in
the X-ray spectral slope either. While \citet{Gilli07} argued that the
scatter  in   the  X-ray  spectra   is  important  to   constrain  the
Compton-thick AGN population, this will  not be the case for our model
due to  the fundamental difference  in the way that  the Compton-thick
AGN   abundance  is   constrained.   Their   CXB  model   derived  the
Compton-thick AGN by subtracting  the Compton-thin AGN contribution to
the CXB spectrum around 30 keV.  The extrapolation of Compton-thin AGN
contribution from $<$ 10 keV to 30 keV certainly depends critically on
the  scatter  of  the  power-law  slope.   Our  model  constrains  the
Compton-thick AGN through mid-IR data which depends on the SMBH IR/X-ray
relationship.  The  additional difference between their  work and this
study   is  that  we   included  the   reflected  Component   for  the
quasars \citep[e.g.][]{Piconcelli10}.   A summary is given
here:

(1) The unobscured SMBH SED is composed of three components, a primary
power-law with a  cut-off at high ($>$ 100  keV) energies, a reflected
component  and iron 6.4  keV emission  line.  The  power-law component
arises  directly from  the hot  corona above  the accretion  disk. Its
shape   is   characterized  by   a   photon   index  of   $\Gamma$=1.9
\citep{Nandra94,  Reeves00, Piconcelli05,  Mainieri07}  and a  cut-off
energy  of 300  keV \citep{Molina06,  Dadina08}. For  the  latter, the
uncertainty is  still important.  This component can  be reflected off
the  far side  of the  accretion disk  or dusty  torus,  hardening the
emerging radiation.  The reflected  component depends on  the incident
spectrum  slope and  the  observed inclination.   It  is modelled  for
$\Gamma$=1.9 and the  relative normalization is fixed to  be 1.3 which
is the average for type-1 lines of sight assuming a torus half-opening
angle of  45$^{\circ}$. The  iron line is  assumed to have  a Gaussian
profile with width of 0.4 keV and equivalent width of 280 eV.

(2) For the obscured Compton-thin  AGN, the intrinsic power-law is the
same  as that  for  the  unobscured one  but  the reflected  component
normalization is reduced to 0.88 given different viewing angles toward
obscured nuclei.  The extinction curve is the photoelectric absorption
at solar abundances as characterized by \citet{Morrison83}.  In total,
we  adopts log$\frac{N_{\rm HI}}{\rm  cm^{-2}}$=[21.5, 22.5,  23.5] to
sample the Compton-thin X-ray SEDs.

(3) For  Compton-thick AGN with $N_{\rm HI}$  $<$ 10$^{25}$ cm$^{-2}$,
the intrinsic  power-law is  the same as  that for the  unobscured one
while  the normalization  of the  reflected component  is fixed  at
0.37.   The extinction  curve needs  to consider  the non-relativistic
Compton scattering as modelled through the code of \citet{Yaqoob97}.

(4) For  Compton-thick AGN with $N_{\rm HI}$  $>$ 10$^{25}$ cm$^{-2}$,
all transmitted  photons are down-scattered by Compton  recoil and only
the reflected Component is visible.

The relative normalization  of the X-ray spectrum with  respect to the
IR part is  derived through the modified version  of the tight
relationship between the 2-10 keV  and 12.3 $\mu$m luminosities for 22
Seyfert   galaxies   in    \citet{Gandhi09}.    These   galaxies   are
well-resolved at  12.3 $\mu$m  so that star  formation can  be removed
reliably.  Galaxies in  that work are a mix of Seyfert  1, Seyfert 2 and
Compton-thick AGN, while what we  need here is the relationship only
for type-1 AGN, specifically PG quasars.  At 12.3 $\mu$m, the silicate
emission feature  can be important.  As  shown in Figure~\ref{lx_lir},
we thus modify the  \citet{Gandhi09} relationship by keeping the slope
but modifying the normalization to produce the median X-ray/12.3$\mu$m
ratio  of 34  PG quasars whose X-ray and  IR data  are derived  from
\citet{Cappi06} and \citet{Shi07},  respectively.  For these PG objects,
the contribution  from star formation  to the 12.3 $\mu$m  emission is
negligible  as evidenced  by extreme  low aromatic  feature equivalent
width (EW$_{\rm 11.3 {\mu}mPAH}$  $\lesssim$ 0.05 $\mu$m). The derived
relationship is:
\begin{equation}\label{eqn_xray_ir}
 {\rm log} \frac{L_{\rm 2-10keV}}{\rm 10^{43} erg/s} = 0.90{\rm log}\frac{L_{12.3{\mu}m}}{\rm 10^{43} erg/s} - 0.45
\end{equation}
where $L_{\rm  2-10keV}$ and $L_{12.3{\mu}m}$ are  unobscured 2-10 keV
and 12.3 $\mu$m luminosities,  respectively.  To evaluate the
effect  of this  scatter  on  the model,  we  explored two  additional
variants  of the  model with  the  normalizations 0.2  dex higher  and
lower, respectively, where 0.2 dex is the 3-$\sigma$ dispersion of the mean
ratio of the PG objects.

\subsubsection{HI Column Density Distribution}\label{model_HI}

The HI column density distribution is key  to understanding
AGN physics  and the cosmic  X-ray background. Its  proper characterization
needs unbiased sample selections  and secure measurements of HI column
densities. In the local universe, optically-selected samples offer
the least bias compared to UV/X-ray (biased against heavily-extinguished objects
and  low-luminosity   AGNs)  and  IR  (biased against   unobscured  objects  and
low-luminosity AGNs).  \citet{Maiolino95a} have attempted  to compile
all Seyfert galaxies  at $m_{B}$ $<$ 13.4. The Seyfert  2 to Seyfert 1
ratio is  found to be  as high as  4:1. This ratio is  consistent with
that  found  in the  Palomar  nuclear  spectroscopic  survey of  local
galaxies  at  $m_{B}$  $<$  12.5  \citep{Ho97}.   Both  studies  dealt
carefully with  host galaxy  light dilution and  the result  should be
quite reliable.  Both studies probed very low-luminosity  AGN with bolometric
luminosities  around 10$^{9}$  L$_{\odot}$.  \citet{Hao05}  instead found
1:1 for Seyfert-2/Seyfert-1 ratio around $L_{\rm bol}$ $\sim$ 10$^{9}$
L$_{\odot}$ and decreasing  toward higher luminosities.  The discrepancy
can be partly due to the  luminosity dependence but also that the SDSS
spectra in \citet{Hao05} with  large  aperture   and  relatively  low  quality  suffer
significant  host  dilution.  By  building  a  Seyfert  2 sample  from
\citet{Maiolino95a} and \citet{Ho97}, \citet{Risaliti99} measured
the  HI column  density  distribution.  Combining this  with works  of
\citet{Cappi06}   for   Seyfert  1   galaxies,   we
characterized  the  column density  distribution  with three
bins:
\begin{equation}
P^{\rm N_{HI} } = \frac{dN}{d{\rm log}N_{\rm HI}} =
\begin{cases}
0.5f_{\rm type-1} \text{ for } {\rm log}\frac{N_{\rm HI}}{\rm cm^{-2}} \in [0, 22] \\
0.5(1-f_{\rm type-1})(1-f_{\rm CT}) \text{ for } {\rm log}\frac{N_{\rm HI}}{\rm cm^{-2}} \in [22, 24]  \\
0.5(1-f_{\rm type-1})f_{\rm CT}  \text{ for } {\rm log}\frac{N_{\rm HI}}{\rm cm^{-2}} \in [24, 26]  \\
\end{cases}
\end{equation}
where $P^{\rm  N_{HI} }$  is the probability  per logarithm of  the HI
column  density,  the  unobscured  AGN fraction  $f_{\rm  type-1}$  is
defined  as  the  number  of objects  with  log$\frac{N_{\rm  HI}}{\rm
cm^{-2}}$ $<$ 22 to the total,  while  $f_{\rm CT}$
is  defined  to  be  the  fraction of  CT  (log$\frac{N_{\rm  HI}}{\rm
cm^{-2}}$  $>$ 24)  among the  obscured AGN  (22  $<$ log$\frac{N_{\rm
HI}}{\rm cm^{-2}}$ $<$ 26).

Beyond  the local  universe,  it  is increasingly  difficult  to  directly
measure  the HI column  density distribution.   Our model  thus allows
both luminosity and redshift evolution of $f_{\rm type-1}$ and $f_{\rm
CT}$:
\begin{eqnarray}
f_{\rm type-1} = f_{0, \rm type-1}{\times}\frac{1}{(1+z)^{\beta_{z, {\rm type-1}}}} \nonumber \\
  {\times}{\rm exp}( -({\rm log}L_{\rm TIR}^{\rm BH}-9.0){\times}\beta_{l, {\rm type-1}} )
\end{eqnarray}
\begin{eqnarray}
f_{\rm CT} = f_{0, \rm CT}{\times}\frac{1}{(1+z)^{\beta_{z, {\rm CT}}}} \nonumber \\
     {\times}{\rm exp}( -({\rm log}L_{\rm TIR}^{\rm BH}-9.0){\times}\beta_{l, {\rm CT}} )
\end{eqnarray}
where $f_{0, \rm type-1}$ is fixed  to be 0.2 and $f_{0, \rm CT}$
is fixed to  be 0.5 at $L_{\rm TIR}^{\rm  BH}$ = 10$^{9}$ L$_{\odot}$,
in order to approximate the result of \citet{Maiolino95a}, \citet{Ho97} and
\citet{Risaliti99}. We also impose a maximum and minimum fractions of 95\% and 5\%, respectively, for each
column range (type-1, Compton-thin type-2 \& Compton-thick). There are four free parameters in total to describe the HI 
column density distribution.

\subsubsection{The  UV/optical Part Of Unobscured SED}

The composite UV/optical/near-IR  SED is taken from \citet{Richards06}
based on the  SDSS type-1 quasars.  We combined this  part with X-ray emission
through the X-ray/optical luminosity relationship as established by \citet{steffen06}:
\begin{equation}
 \alpha_{\rm ox}=-0.107{\rm log}(\frac{L_{\nu,2500\AA}}{\rm ergs/s/Hz})+1.739
\end{equation}
where $\alpha_{\rm ox}$=-0.384log($L_{\nu, 2500\AA}/L_{\nu, \rm 2keV}$).

\subsubsection{The  Extinction Curve At UV/optical/IR Wavelength}

The extinction curve  for the UV/optical/IR emission is  adopted to be
that of  the Small Magellan Cloud  from \citet{Pei92} which  is a good
approximation for reddened AGN  as discussed in \citet{Hopkins07}.  To
relate  the amount  of extinction  in the  optical/IR to  that  in the
X-ray, we  need to assume  a gas-to-dust ratio represented  by $N_{\rm
HI}$/$A_{\rm V}$ in AGN.  Studies  show that in AGN, the extinction in
the optical  and IR is significantly  less than what  is expected from
the    gas-to-dust    ratio    in     the    Milky    Way    or    SMC
\citep[e.g.][]{Risaliti00,  Maiolino01, Shi06}.  Two  possible reasons
are that there is a large  amount of dust-free gas surrounding SMBH or
that the  X-ray and IR emitting  sources are spatially  offset so that
they  are  not  extinguished  exactly  by the  same  material.   We  here
increased  the gas-to-dust  ratio  by a  factor  of 100  to match  the
observed average  ratio between X-ray  HI column density  and silicate
absorption feature \citep{Shi06}.

\subsection{A Summary Of Model Components}

Our model first assumes total IR LFs that are described by a two power
law with four parameters (break  luminosity and density along with the
faint and  bright end  slopes). The  IR LF is  assumed to  evolve with
redshift  in  both break luminosity  and   density,  each  of  which  is  a
polynominal function  of redshift with three parameters.  Therefore the total
IR LF has a total of 10 parameters. The model then introduces the SMBH
energy fraction  in the total IR  band to decompose  the emission into
SMBH  and  star-forming components.  At  a  given  IR luminosity,  the
distribution  of  SMBH  energy  fractions  is assumed  to be a  Gaussian
function with both mean and scatter depending on the luminosity, while
the mean  is further assumed  to show redshift dependence.   There are
total five  free parameters including  two for redshift  dependence of
the mean and one for its luminosity dependence, one for the scatter at
the total  IR luminosity  of 10$^{12.3}$ L$_{\odot}$  and one  for its
luminosity dependence.   The mean fraction  at 10$^{12.3}$ L$_{\odot}$
is   fixed  based  on   local  ULIRG   objects.  For   each  power source
(star-forming and  SMBH), the model  then convert the emission  in the
total  IR  band  to  other  wavelengths  using  star-forming  and  AGN
SEDs.  For the former,  luminosity  dependent empirical  SEDs are
used.  For the  latter,  HI columns  are added  to  extinguish the
type-1 empirical  SEDs.  HI  columns are divided into  three ranges
representing  type-1, Compton-thin type  2 and  Compton thick,  which are
described  by two parameters  to give  the relative frequencies of 
occurrence (the sum of which needs to be normalized).  Each  parameter  has  a local value fixed at  the  SMBH
IR luminosity  of 10$^{9}$  L$_{\odot}$, and is assumed to follow a simple power law dependence on each of
 luminosity and redshift. Therefore  the HI  columns have four  parameters in  total. The
model in total has 19 free parameters.

\section{Numerical Approach}\label{numerical_approach}


\begin{figure}
\epsscale{0.8}
\plotone{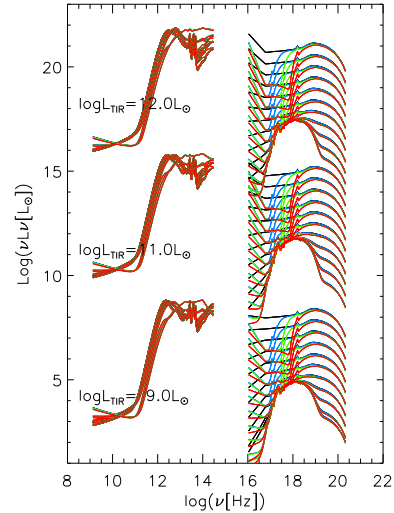}
\caption{\label{example_sed} Examples of SEDs at three given total IR luminosities. Note that
the model adopts a total of 2340 SEDs at a given total IR luminosity and
a redshift that covers the range of the SMBH energy fraction, the extinction and the scatter in the star-formation SEDs.
Here it shows SEDs at lower resolution grids over the above three parameters for clarity. Different colors are for
different extinctions.}
\end{figure}

Our approach to constraining the model is to fit both X-ray and IR survey data, with
the basic assumption that all  the galaxies are experiencing both star
formation and  nuclear accretion activities  with relative intensities
that vary with both  redshift and luminosity. The model  is composed of
  four components (see  \S~\ref{model_component} for details):
the total IR LF, the SMBH  energy fraction in the IR, the star-formation
SED  and SMBH SED.   In principle,  the combination  of these  four can
predict any  observed properties that  are related to  galaxy number
densities and  SEDs, such as number counts,  redshift distributions, LFs
etc.

The  model has  a  total of  19 free parameters.  To incorporate  the
uncertainties on the observational constraints that we adopted for the
model, we  present four  variants of the  model as  summarized in
Table~\ref{table_list_model},  varying the SMBH IR/X-ray  ratio and redshift  evolution in the
star-formation  SED.  The numerical  calculation starts with
the generation of  a multi-dimensional array of the  flux density at a
given  filter  for  an  object  at  the  given  total  IR  luminosity,
star-formation  color, redshift, SMBH  energy fraction  and HI  column
density, i.e.,  $f_{\nu}$(filter, $L_{\rm TIR}$, star-formation-color,
$z$, $f^{\rm  BH}$, $N_{\rm  HI}$), where the  numerical grid  of each
dimension is pre-defined and  thus $f_{\nu}$ is fixed.  Each $f_{\nu}$
has an associated total probability function (the product of four
model components):
\begin{equation}
P^{\rm tot} = \Phi(L_{\rm TIR}, z){\times}P^{\rm BH}{\times}P^{\rm SF{\textendash}color}{\times}P^{\rm N_{HI}}
\end{equation}
$P$$^{\rm tot}$ contains all free parameters whose best values will be obtained by fitting to the data.
The general procedure to derive $f_{\nu}$ and $P^{\rm tot}$ is as follows:

(a) $\Phi(L_{\rm TIR}, z)$ is a function of $L_{\rm TIR}$ and $z$. The
numerical  grid of  $L_{\rm  TIR}$ covers  the  luminosity range  from
10$^{7}$ L$_{\odot}$ to 10$^{14}$ L$_{\odot}$ with a resolution of 0.2
dex.   The  redshift  grid  starts  with z=0.001  and  increases  with
$\Delta$$z$/$z$=0.1 until $z$ $=$ 1.0 with a total of 63 bins.  At $z$
$\geq$ 1.0, $\Delta$$z$ is fixed to be 0.1 up to $z$=6.

(b)  The  given  $L_{\rm  TIR}$  is divided  into  the  star-formation
component $L_{\rm TIR}^{\rm SF}$  and SMBH component $L_{\rm TIR}^{\rm
BH}$ according to the SMBH  energy fraction ${\rm log}f^{\rm BH}$. The
${\rm log}f^{\rm BH}$  has a range from -6.0 to  0.0 with a resolution
of 0.2 dex. At ${\rm log}f^{\rm  BH}$ = -6.0, the SMBH contribution at
any wavelength  is negligible.   Also, such a  low value is  needed to
assure the possibility  of negligible SMBH radiation in  a IR luminous
object, i.e., $L_{\rm 2-10 keV}$ $<$ 10$^{42}$ erg/s.

(c) For  the given $L_{\rm  TIR}^{\rm SF}$, the star-formation  SED is
assigned according to all  possible star-formation color as defined by
log($\frac{f_{\nu}({\rm                   IRAS}-60{\mu}m)}{f_{\nu}({\rm
IRAS}-100{\mu}m)}$).     The   star-formation   color    varies   from
$C_{0}$-3$\sigma_{\rm c}$ to $C_{0}$+3$\sigma_{\rm  c}$ with a step of
0.5$\sigma_{\rm c}$ where $C_{0}$ is the function of $L_{\rm TIR}^{\rm
SF}$  (see  \S~\ref{SF_IR_SED}).   Each  star-formation  color  has  a
probability  of  $P^{\rm  SF{\textendash}color}$.  With  the  assigned
star-formation SED, the $L_{\nu}^{\rm  SF}(\rm filter)$ is obtained by
convolving the SED with the filter curve.  Note that in the X-ray, the
flux is the direct integral of the SED between the edges of the band.

(d) For a given $L_{\rm TIR}^{\rm BH}$, the SMBH SED is assigned according to
the HI column density. The  $N_{\rm HI}$  grid  is defined
to be log$N_{\rm HI}/$cm$^{-2}$ = [0, 21.5, 22.5, 23.5, 24.5, 25.5],  each of
which has a probability of  $P^{\rm N_{HI}}$.  The assigned SED is then used
to derive the $L_{\nu}^{\rm BH}(\rm  filter)$.

(f) The final  luminosity in a filter is  the sum of $L_{\nu}^{\rm
SF}(\rm filter)$ and $L_{\nu}^{\rm  BH}(\rm filter)$.  This is divided
by   the  distance  modulus   to  give   the  final   $f_{\nu}(\rm filter)$.  

At a given  total IR luminosity and redshift,  the above grids produce
2340  SEDs  covering  the  range  of the  SMBH  energy  fraction,  the
extinction and  the scatters in the star-formation  SED, with examples
shown in Figure~\ref{example_sed}. In  total, there are 10 million SEDs
 to cover the IR luminosity, redshift, the BH energy fraction,
the  star   formation  SED  and  the  HI   columns.   After  producing
$f_{\nu}$(filter,  $L_{\rm TIR}$,  star-formation-color,  $z$, $f^{\rm
BH}$, $N_{\rm HI}$), the free  parameters in $P^{\rm tot}$ are derived
by  fitting   to  the  observations.    The  fitting  scheme   is  the
Levenberg-Marquardt  least-square  technique   with  the  IDL  program
MPFIT.pro \citep{Markwardt09}.

\section{Data For  the Fit}\label{data_for_fit}

Our model fits a total of 617  data points from the local LFs plus two
basic  types of  data  from  deep surveys,  namely  number counts  and
redshift distributions.   The basic information of  these data-sets is
listed  in  Table~\ref{data_set}.   The  main advantages  of  our  fit
strategy  compared  to those  fitting  directly  to  LFs at  different
redshifts  include: (1)  that  we  isolated the  number  counts as  an
independent dataset for  the fit. These are the  most direct result of
the  deep   survey  observations   and  contain  the   least  modeling
uncertainties; (2)  that we  avoided the uncertainties  and systematic
differences associated  with the K-corrections for the  LFs derived in
different studies.

While a large number of data  sets are available in the literature, we
generally  preferred those  with  good statistics.   We thus  included
either those  based on  the combined fields  or individual  results of
fields with either  large areas or high sensitivities.   We also avoid
double fitting datasets  of the same field but  from different authors
or papers.   Overall, the  number counts include  X-ray data  at 17-60
keV,  15-55 keV,  2-10  keV and  0.5-2  keV bands,  and IR/submm  data
ranging  from 24 $\mu$m  to 1200  $\mu$m.  The  redshift distributions
also  cover a  range of  wavelengths,  field areas  and limiting  flux
cuts. For  the same field,  the redshift distributions  with different
flux  cuts  offer  constraints  on  the model  LFs  in  the  different
luminosity and redshift ranges. The redshift completeness in data sets
we used is high, $>$ 90\%  either from spec-z or photo-z or both.  The
error bars include Poisson noise  and cosmic variances with the latter
estimated by  two point correlation function  from direct observations
or  the cosmological  model  of \citet{Trenti08}.   Except  for a  few
data-sets with high flux  cuts, photo-z measurements always constitute
an important  fraction especially  at $z$ $>$  1.  For  X-ray sources,
photo-z  errors can be  significant due  to the  faint optical/near-IR
counterparts  along with  the  power-law featureless  shape.  We  thus
added additional errors to redshift distributions for the photo-z part
through the simple bootstrapping method.


\subsection{Number Counts}

\setcounter{figure}{5}
\begin{figure*}
\epsscale{1.0}
\plotone{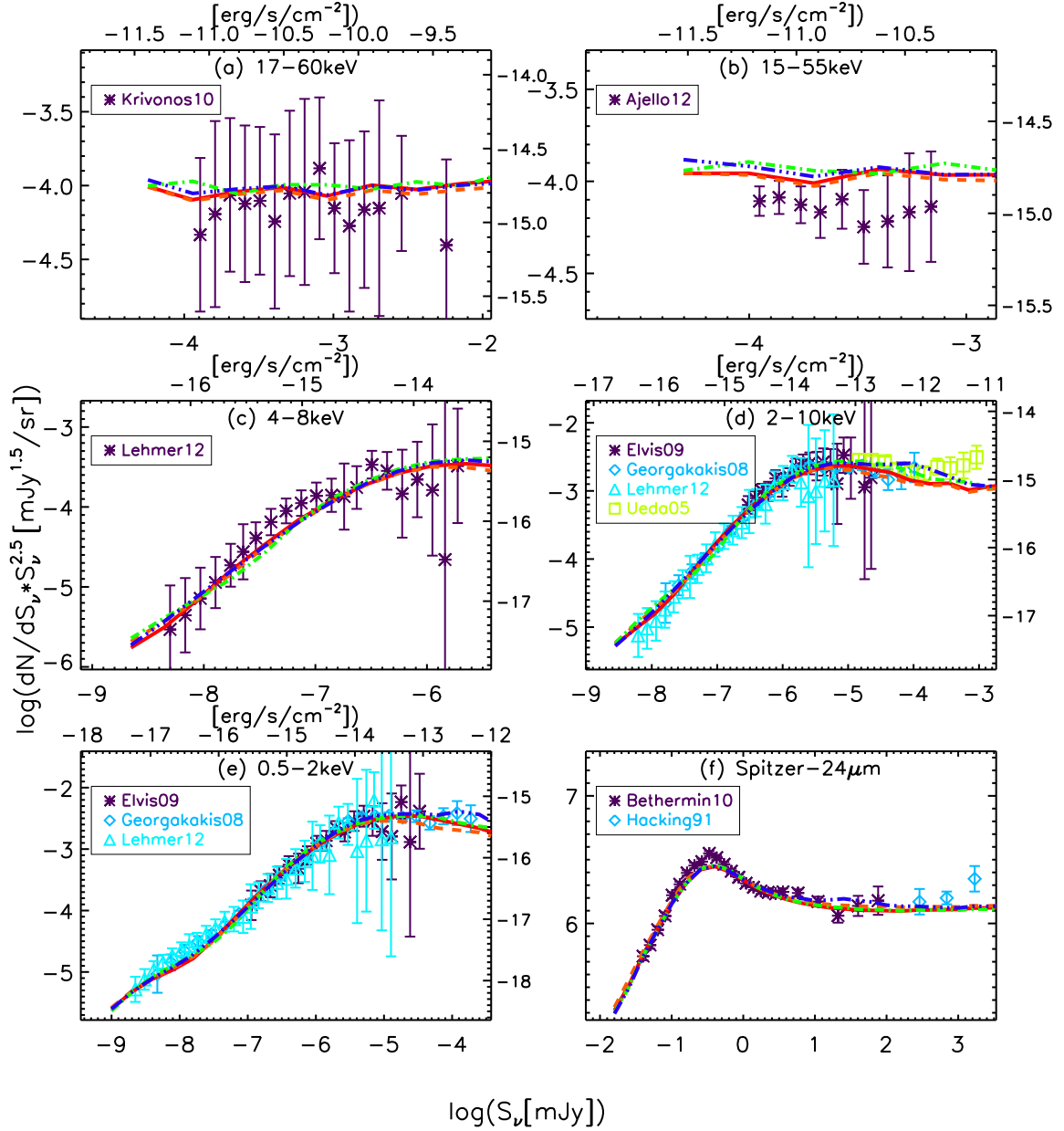}
\caption[]{\label{number_counts}  Fit dataset  I --  number  counts.
All symbols are the observations while those in the black are not used
for the fitting. The four variants of the model are shown with different line styles:
solid line for reference variant, dashed line for fast\_evol\_SED\_SF, dot-dashed line for
low\_IR2X\_BH and three-dot-dashed line for high\_IR2X\_BH.
The reference model is the one with the minimum
$\chi^{2}$. The fast\_evol\_SED\_SF variant assumes strong
redshift evolution of the star-forming SED. The low\_IR2X\_BH and high\_IR2X\_BH 
variants have X-ray luminosities at given IR luminosities for the SMBH radiation
0.2 dex lower and higher, respectively  (see Table~\ref{table_list_model}).  }
\end{figure*}

\setcounter{figure}{5}
\begin{figure*}
\epsscale{1.0}
\plotone{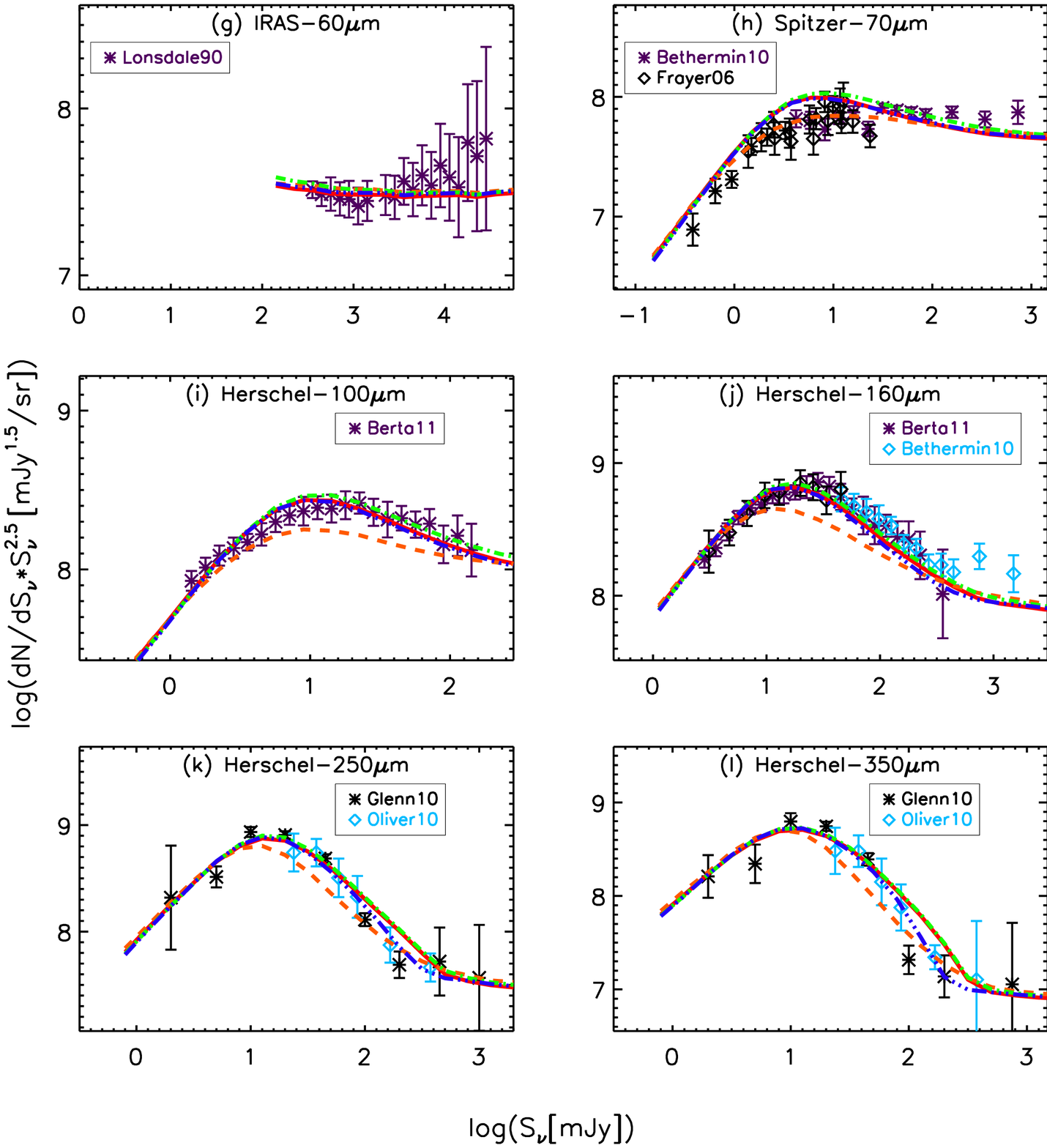}
\caption{Continued}
\end{figure*}

\setcounter{figure}{5}
\begin{figure*}
\epsscale{1.0}
\plotone{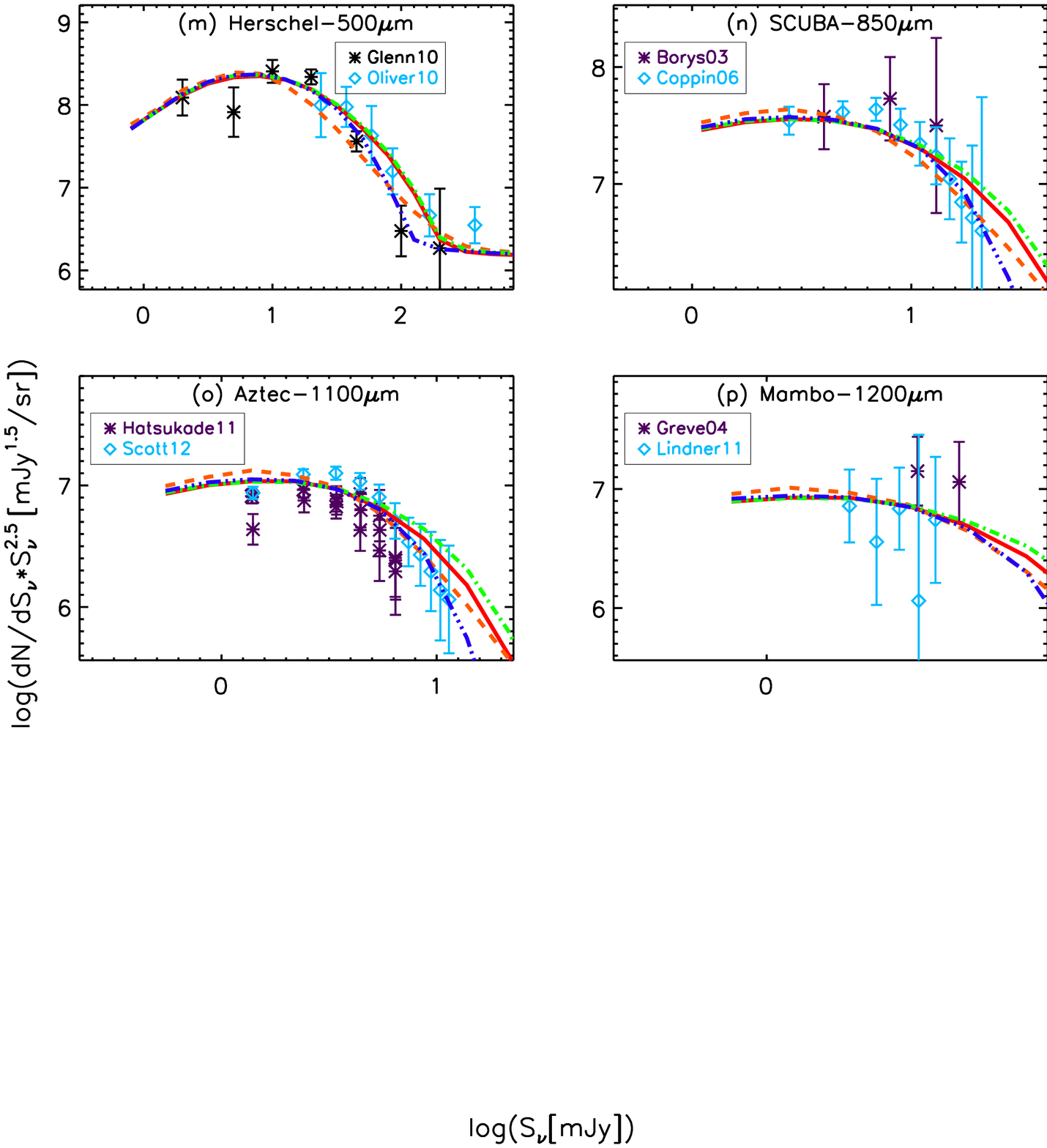}
\caption{Continued}
\end{figure*}

\subsubsection{X-ray Number Counts}

All sky  surveys in the  17-60 keV by INTEGRAL  \citep{Krivonos10} and
15-55  keV  by  Swift   \citep{Ajello12}  probe  bright  AGN  down  to
6-7$\times$10$^{-12}$ erg/s/cm$^{2}$, whose number counts offer useful
constraints on the  model's behavior in the local  universe.  The {\it
Chandra}  deep  survey  data   are  used  to  probe  high-z  universe.
\citet{Georgakakis08}  combined  several  surveys  including  CDF-N/S,
ECDF-S, EGS, EN1  and XBOOTES fields, to offer  number counts in 0.5-2
and 2-10  keV with  high S/N  across a large  flux range.   It reaches
1.0$\times$10$^{-17}$ and  1.0$\times$10$^{-16}$ erg/s/cm$^{2}$ in two
bands, respectively.  Although galactic  stars are not removed in this
work, their  contributions are small  \citep[$<$10\%;][]{Bauer04}. The
additional  data  in the  COSMOS  field  by  \citet{Elvis09} are  also
included for the fit. The  recent 4 Ms CDFS data from \citet{Lehmer12}
are further used to probe number counts at the faintest end, for which
4-8 keV  data along with 0.5-2 keV  and 2-10 keV are  all included. An
additional large  area ASCA  survey in the  2-10 keV is  also included
\citep{Ueda05}.   The  small  difference  ($\sim$1\%)  in  the  counts
between two  different galactic latitudes  of this survey  indicates a
small contribution from  Galactic contaminators.  The published errors
of  the number  counts in  above studies  have taken  into  account of
Poisson noises  and cosmic variances.  If a fixed Photon  index (e.g.,
1.4) is used to convert the count rate to the flux, we added an additional
uncertainty of 0.1 dex quadratically.

\subsubsection{{\it Spitzer} Number Counts}

We fitted to the {\it Spitzer}  number counts at  24, 70 and  160 $\mu$m
measured by  \citet{Bethermin10}, which are  generally consistent with
previous works \citep{Papovich04,  Shupe08, LeFloch09, Frayer09}.  The
study  incorporates  data  from  fields including  SWIRE,  COSMOS  and
FIDEL.    Besides  standard   data   reduction  and   photometry
measurements, the incompleteness and Eddington bias are also corrected
in  that work,  with further  removal of  stars. The  counts  reach 35
$\mu$Jy, 3.5  mJy and 40 mJy  at 24, 70 and  160 $\mu$m, respectively.
Their stacking data at two longer wavelengths are not used in the fits but
plotted for illustration.   At 24 $\mu$m, we added  the IRAS 25 $\mu$m
data   from   \citet{Hacking91}   by    assuming   a   flat   SED   in
${\nu}{f_{\nu}}$, which provides constraints in the local universe.

\subsubsection{IRAS Number Counts}

In addition  to the  IRAS 25  $\mu$m number counts,  the counts  at 60
$\mu$m   from  \citet{Lonsdale90}   are  also   included   to  provide
constraints  at the  brightest flux  end (i.e.   local  universe). The
error is  caused by the large  scale structure of  the local universe.
We estimated it roughly as the difference in the number counts between
South and North galactic cap.

\subsubsection{{\it Herschel} Number Counts}

Our  fit includes  the number  counts at  250, 350  and 500  $\mu$m of
detected sources in the HerMES field as estimated by \citet{Oliver10}.
The depth reaches around 15 mJy at three wavelengths.  Their published
uncertainties  account  for   Poisson  noises,  cosmic  variances  and
uncertainties  associated with  flux  measurements and  incompleteness
corrections.   Deeper counts obtained  through $P(D)$  (probability of
deflection) analysis  of the  same field \citep{Glenn10}  are included
only for illustration. At 100 and 160 $\mu$m, we adopted the result of
\citet{Berta11} from three fields  (COSMOS, Lockman Hole and GOODS-N).
The  data reach  down  to 2.0  mJy  at 100  $\mu$m and  5  mJy at  160
$\mu$m. The estimated  errors on the counts include  Poisson noise and
field-to-field variations.

\subsubsection{SCUBA/Aztec/Mambo Submm Number Counts}

The SCUBA  counts at 850$\mu$m are collected  from \citet{Borys03} and
\citet{Coppin06} with the  overall flux range from 2.5  to 25 mJy. The
Aztec  number  counts   at  1.1  mm  are  from   two  recent  studies,
\citet{Hatsukade11}  and \citet{Scott12}.   The combined  counts cover
the flux range from  about 0.5 to 10 mJy. The Mambo  1.2 mm counts are
from \citet{Greve04} and \citet{Lindner11}.

\subsection{Redshift Distributions}

\subsubsection{X-ray Redshift Distributions}

For   503  objects   in  the   {\it  Chandra}   CDF-N  2   Ms  catalog
\citep{Alexander03},  the  redshift  measurements  are  compiled  from
\citet{Barger05},  \citet{Donley07}  and  \citet{Trouille08}.  In  the
hard  X-ray  band, we  constructed  two  redshift distributions  above
limiting    fluxes     3$\times10^{-16}$    and    3$\times$10$^{-15}$
erg/s/cm$^{2}$, respectively, and in  the soft band (0.5-2 keV), above
two  limiting   fluxes  of  6$\times10^{-16}$   and  6$\times10^{-17}$
erg/s/cm$^{2}$. For  each band, the  first flux is near  the detection
limit and the second one is about 10 times brighter. The spectroscopic
redshift and all redshift (spec-z+photo-z) completeness is around 65\%
and 90\%, respectively.

For 462 sources in the  {\it Chandra} CDF-S 2Ms catalog \citep{Luo08},
the redshift measurements are available in \citet{Luo10}. Two redshift
distributions     in    the     hard-xray    are     measured    above
3.0$\times$10$^{-16}$    and   3.0$\times$10$^{-15}$   erg/s/cm$^{2}$,
respectively.    In   the   soft   band,  the   limiting   fluxes   of
5.0$\times10^{-17}$  and 5.0$\times10^{-16}$ erg/s/cm$^{2}$  are used.
The  spec-z and  all redshift  completeness is  around 55\%  and 97\%,
respectively.

With  deep  multiple optical/near-IR  follow-ups,  virtually all  1761
X-ray  AGN in the  COSMOS field  are identified  \citep{Civano12}. The
redshift distribution above 3$\times$10$^{-15}$ erg/s/cm$^{2}$ in 2-10
keV and  6$\times$10$^{-16}$ erg/s/cm$^{2}$ in  0.5-2 keV are  used for
the fit,  with spec-z and  all redshift completeness reaching  60\% and
98\%, respectively

The  extended   CDF-S  (ECDF-S)  contains  a  total   of  762  sources
\citep{Lehmer05}.  The redshift distribution in 2-10keV is constructed
at  $>$ 2.0$\times10^{-15}$  erg/s/cm$^{2}$  \citep{Silverman10}.  The
spec-z and  all redshift completeness is 64\%  and 95\%, respectively.
The  XMS  offers  318   X-ray  sources  at  bright  fluxes  ($\gtrsim$
$\times10^{-14}$  erg/s/cm$^{2}$ )  with  high optical  identification
(90\%)  \citep{Barcons07}. The  redshift  distribution is  constructed
above the 2-10 keV  flux of 5$\times10^{-14}$ erg/s/cm$^{2}$ with high
spectroscopic completeness (87\%).

In  the soft  band,  two  additional surveys  by  ROSAT are  included.
\citet{Henry06}    have   defined    a   statistical    sample   above
2$\times$10$^{-14}$ erg/s/cm$^{2}$ from  the ROSAT north ecliptic pole
survey.   The  spectroscopic  redshifts are highly  complete  ($>$95\%,
Gioia03).    We  have   measured  the   redshift   distribution  above
5$\times$10$^{-14}$  erg/s/cm$^{2}$ from this  sample. The  second one
above   3$\times$10$^{-14}$   erg/s/cm$^{2}$   is   constructed   from
ROSAT-RIXOS survey  \citep{Mason00}. Note  that for these  two surveys
with  high limiting  fluxes,  the galactic  stars  are identified  and
removed.

\subsubsection{ IR/submm Redshift Distributions}

In  the   COSMOS  field,  \citet{LeFloch09}   have  measured  redshift
distributions of 24 $\mu$m  sources. Here we included two measurements
with flux limits of 0.08 and  0.15 mJy, respectively. The one with the
highest  flux limit (0.3  mJy) is  not used  since the  bright sources
excluded from that study may affect the distribution at $z$ $<$ 1. The
photo-z completeness  reaches 94\%.   In  the
GOODS-N, \citet{Barger08} carried out a highly spec-z complete survey.
At $f_{24{\mu}m}$  $>$ 0.3 mJy, we measured  the redshift distribution
with spec-z completeness  of 90\%. From the {\it  Spitzer} 5MUSES data
(Helou  et al.   2011), we  measured the  redshift distribution  of 24
$\mu$m  sources   at  $f_{24{\mu}m}$  $>$  5  mJy   where  the  spec-z
completeness  reaches 98\%.  At  60$\mu$m in  the local  universe, the
spec-z distribution of IRAS RBGS sample is included \citep{Sanders03}.

\citet{Chapman05} carried out a spec-z  survey of submm sources at 850
$\mu$m and modelled the  redshift distribution for $f_{850{\mu}m}$ $>$
5 mJy. We fitted this  distribution in our study.  We further included
the redshift  distribution of sources  at Aztec 1100 $\mu$m as  measured by
\citet{Smolcic12} where the spec-z and redshift completeness are roughly
41\% and 100\%, respectively.

\subsection{Local LFs}

We   fitted   four  local   LFs,   including   local   15-55  keV   LF
\citep{Ajello12},  local  2-10 keV  LF  \citep{Ueda11},   the IRAS  12
$\mu$m LF \citep{Rush93}, and IRAS 25 $\mu$m LF \citep{Shupe98}.

\section{Data Fitting Results}\label{fit_result}

The fits to the number counts, redshift distribution and local LFs are
shown   in   Figure~\ref{number_counts},   Figure~\ref{red_dist}   and
Figure~\ref{local_lf},   respectively,   along   with   the   best-fit
parameters in Table~\ref{best_fit_parameter}  for four variants of the
model.  The model fits 617 data-points  in total, with 371, 195 and 51
points  from  number counts,  redshift  distributions  and local  LFs,
respectively. Overall our model reproduces the observations reasonably
well.  The  differences between  the  fits  and  data appear  at  some
data-points  but  are  generally   smaller  than  0.2  dex  or  within
2-$\sigma$.  Cosmic  variances or  photo-z errors are  significant for
some of these deviations.  The  reduced $\chi^{2}$ of four variants of
the model ranges from 2.7 to  2.9, with at least 0.75 contributed by
caveats associated with data themselves (see \S~\ref{sum_diff_mod} for
details).  In the  following, we go through each  version of the model
and point out  the differences between the fit and  the data and their
causes if possible.

\subsection{Fitting Number Counts}\label{res_number_counts}

For the reference model (the  solid line), the differences between the
data  and the  fit include  0.1 dex  constant over-fits  of the
15-55 keV number counts, 0.2-0.3  dex under-fits of the 2-10 keV
counts at the  bright end, the 0.2 dex  under-fits around 0.5-2
keV fluxes  from 1.5 to  4.0$\times$10$^{-17}$ erg/s/cm$^{2}$, 0.1-0.2
dex  under-fits around  the peak  24  $\mu$m count  in the  flux
density range of  0.2 to 0.5 mJy, 0.2  dex under- and over-fits
at the 70 $\mu$m bright  and faint end, respectively. Unlike the 15-55
keV counts, the 17-60 keV ones  are well produced by our model so that
the discrepancy seen in 15-55 keV  can be due to subtle differences in
the X-ray SED around  the range of 15 to 17 keV and  55 to 60 keV.  It
is  yet interesting  to notice  that, similar  to ours,  several other
models \citep{Gilli07, Treister09, Draper10} all overproduce the 15-55
keV counts by 0.1-0.2 dex, while  their adopted SEDs are not exactly the same.
The difference seen at the bright end of the 2-10 keV counts is likely
caused  by contamination of  galaxy clusters  in the  data themselves
\citep{Gilli99}.   In  0.5-2  keV  counts,  the data  clearly  show  a
transition  flux   (3$\times$10$^{-17}$  erg/s/cm$^{2}$)  below  which
normal galaxies other  than AGN start to dominate,  causing the counts
to go above the simple power law extrapolation from the higher fluxes \citep{Bauer04, Lehmer12},
while the model also produces a similar feature but at fluxes $\sim$3
times  lower.  For  the 24  $\mu$m peak  counts, although  the highest
observed  point is the  result of  the SWIRE  data near  the detection
limit  \citep{Bethermin10},  the  differences  are seen  over  several
adjacent data points.  Since this  peak is caused by aromatic features
moving into the 24 $\mu$m  filter around $z$=1, a possible explanation
is  that  $z$$\sim$1  star-forming  galaxies have  slightly  increased
aromatic  feature  EWs.   Differences  seen  in  70  $\mu$m  are  also
noticeable, for  which the subtle  caveat in star-forming SEDs  can be
the potential cause too.

For the fast\_evol\_SED\_SF  variant (dashed line), additional differences
between the fits and data,  besides those for the reference model, are
the significant under-estimates  of the 70, 100 and  160 $\mu$m counts
above     the      flux     density     of      the     peak     count
(Figure~\ref{number_counts}(h)-(j)). The  reason is that  the redshift
evolution in the star-formation SED results in a colder star-formation
color  $f_{60{\mu}m}/f_{100{\mu}m}$   that  decreases  the  rest-frame
30-100  $\mu$m  emission.  The  behavior  of  the  other two  variants
(low\_IR2X\_BH and  high\_IR2X\_BH) in fitting the  data are quite
similar to that of  the reference variant.

\subsection{Fitting Redshift Distributions}

\setcounter{figure}{6}
\begin{figure*}
\epsscale{1.0}
\plotone{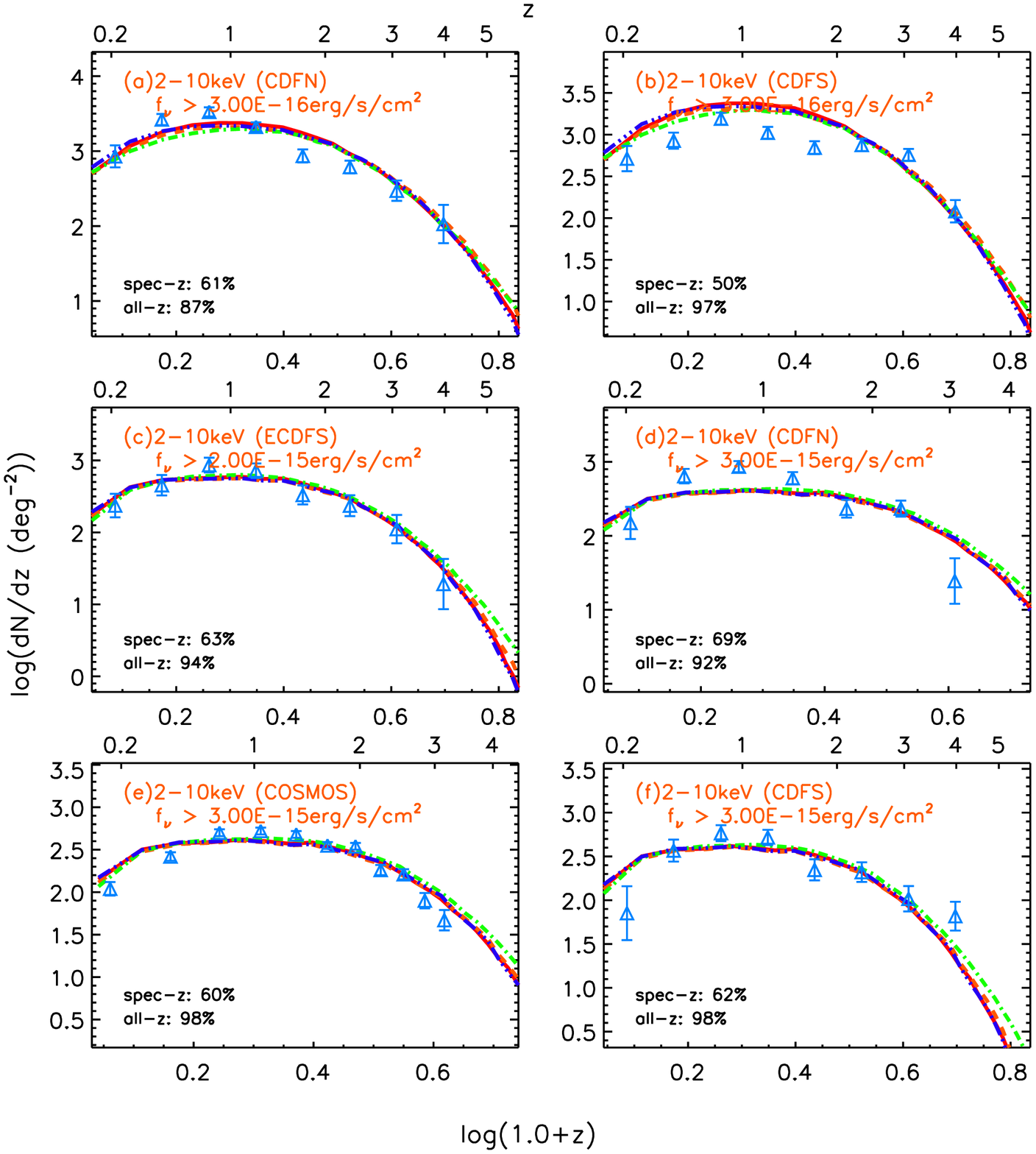}
\caption[]{\label{red_dist}  Fit dataset II -- redshift distribution. All symbols are the observations.
The four variants of the model are shown with different line styles:
solid line for reference variant, dashed line for fast\_evol\_SED\_SF, dot-dashed line for
low\_IR2X\_BH and three-dot-dashed line for high\_IR2X\_BH (see Table~\ref{table_list_model}).}
\end{figure*}

\setcounter{figure}{6}
\begin{figure*}
\epsscale{1.0}
\plotone{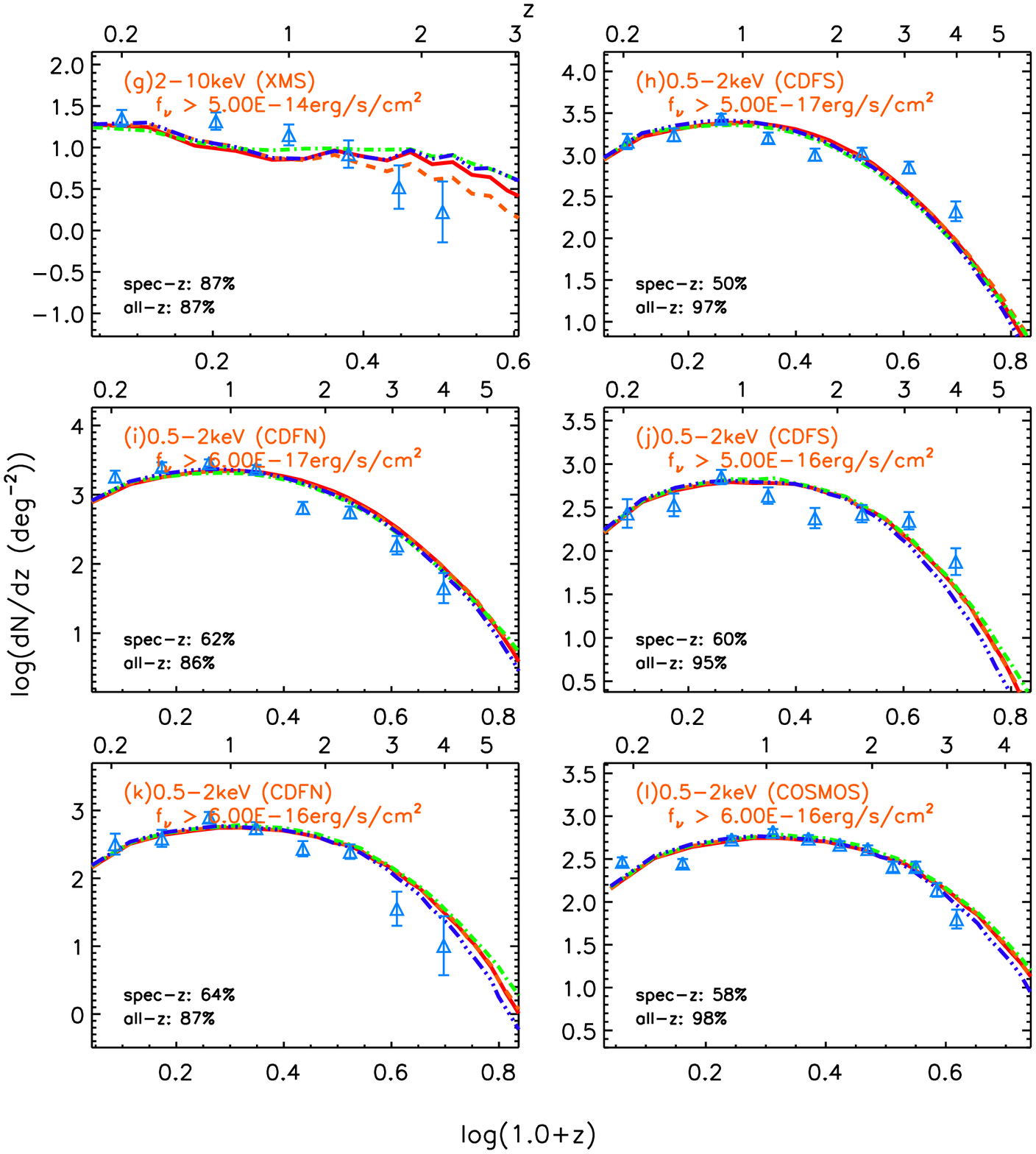}
\caption{\label{red_dist} Continued}
\end{figure*}

\setcounter{figure}{6}
\begin{figure*}
\epsscale{1.0}
\plotone{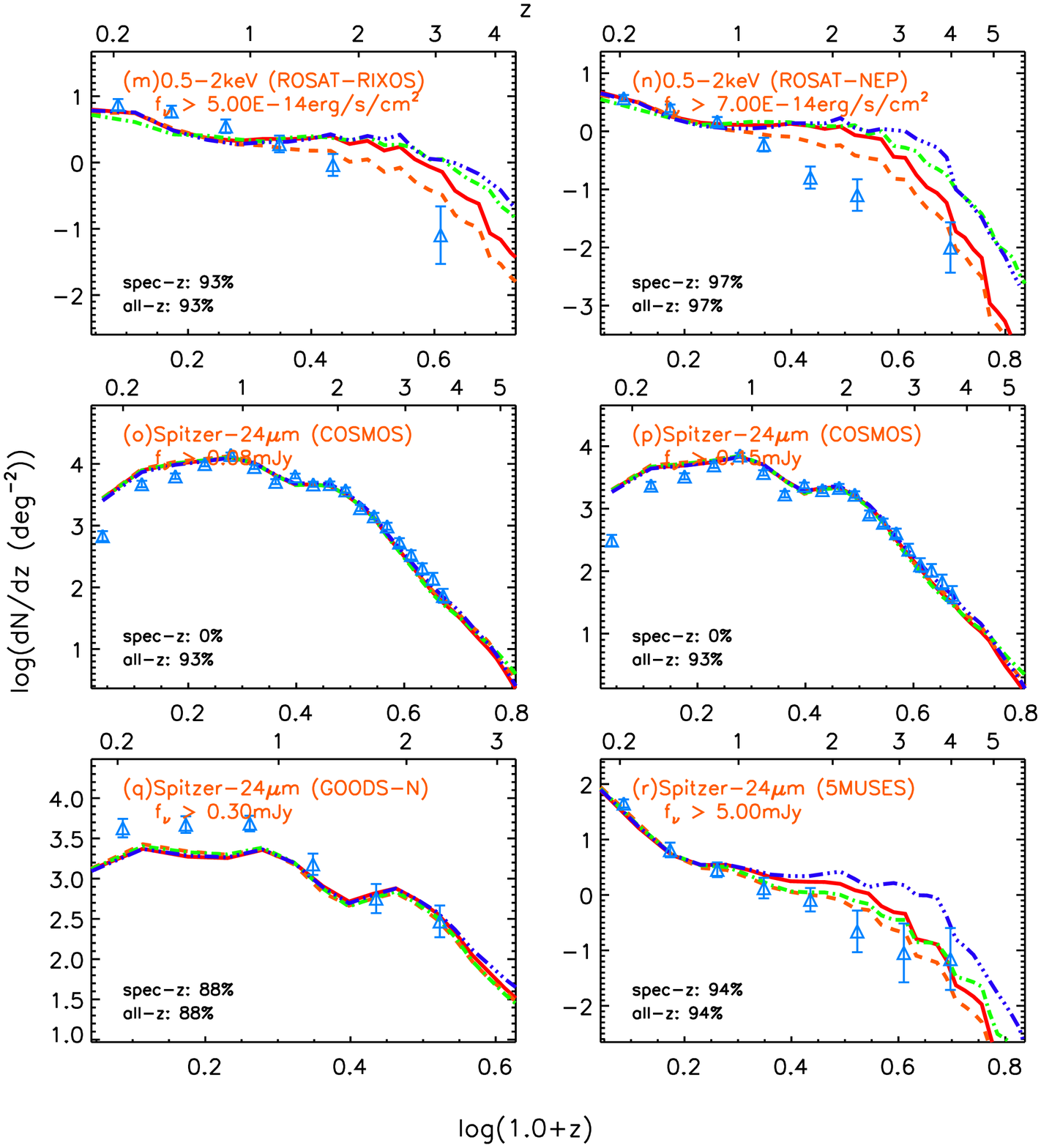}
\caption{Continued}
\end{figure*}

\setcounter{figure}{6}
\begin{figure*}
\epsscale{1.0}
\plotone{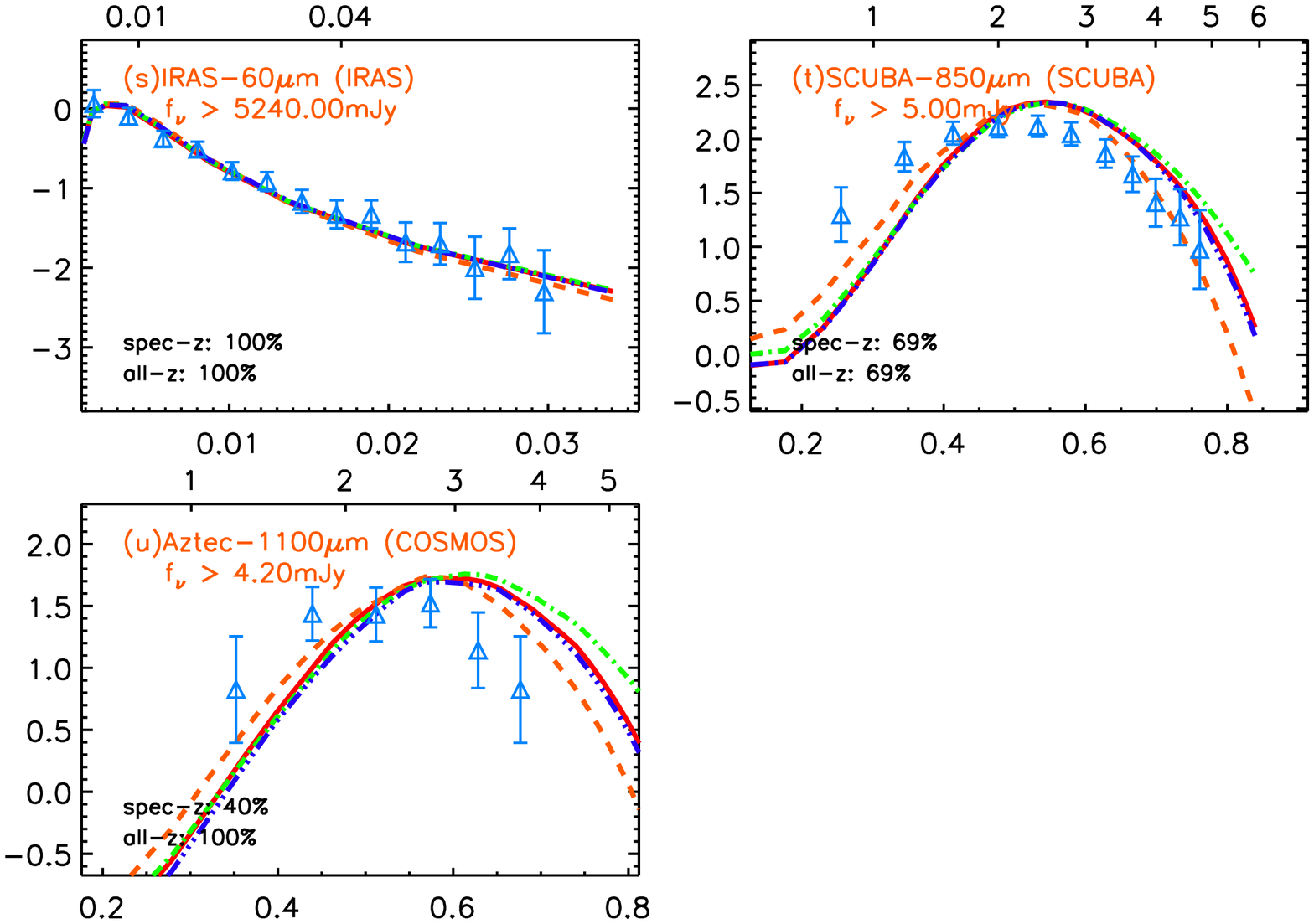}
\caption{Continued}
\end{figure*}

The  redshift distributions are  shown in  Figure~\ref{red_dist}.  The
distributions      at      2-10      keV      are      plotted      in
Figure~\ref{red_dist}(a)-(g). The data  are from five fields including
CDF-N, CDF-S,  ECDF-S, COSMOS  and XMS that  complement each  other by
probing from  small deep  field (0.12 square  degree) to  wide shallow
area   (3   square   degree).    The  limiting   fluxes   range   from
$>$3$\times$10$^{-16}$  up to $>$  5$\times$10$^{-14}$ erg/s/cm$^{2}$.
Compared to  the number  counts, the observational  uncertainties here
are larger due to the redshift measurement errors or cosmic variances.
For example, for the CDFN and CDFS (Figure~\ref{red_dist} (a) and (b),
respectively) with the same field size and limiting X-ray flux, we saw
obvious   differences    in   the    data.    In   the    soft   X-ray
(Figure~\ref{red_dist}(h)-(n)),      the     data      extend     from
$>$5$\times$10$^{-17}$ to  $>$7$\times$10$^{-14}$ erg/s/cm$^{2}$ while
the field  area ranges from 0.12  to 80 square  degree.  Again, cosmic
variances and photo-z errors  play roles in differences among redshift
distributions with  similar flux cuts, emphasizing  the limitations of
current available  data.  The redshift  distributions in the  IR bands
are  shown  in  Figure~\ref{red_dist}(o)-(u).  Although  the  redshift
measurements in general have  high completeness, the incompleteness is
always redshift  dependent, missing more objects at  higher $z$.  This
seems to be consistent with  that our model produces more objects than
the  observations   at  high  $z$   ends  in  several   panels  (e.g.,
Figure~\ref{red_dist}(m), (n) \& (r)).

All variants  of the model  perform in similar  ways.  The fit  to the
2-10 keV distributions are generally acceptable where the distribution
with the brightest cut  as shown in Figure~\ref{red_dist}(g) shows the
largest difference between  the fits and data.  In  the 0.5-2 keV, two
distributions  with the  brightest flux  cut (Figure~\ref{red_dist}(m)
and  (n)) have  worst fits.   Models over-produce  $z$ $>$  2 sources,
potentially  caused   by  the  fact  that   sources  without  redshift
measurements lie at high $z$.  At  24 $\mu$m, for two lowest flux cuts
(Figure~\ref{red_dist}(o)   \&  (p)),  all   variants  of   the  model
re-produce well two  peaks around z=1 and z=2  when different aromatic
features are redshifted into the 24 $\mu$m filter. The only difference
is  seen at  $z$  $<$ 0.5  where the  COSMOS  field does  not probe  a
significant       cosmic      volume.       For       the      GOODS-N
(Figure~\ref{red_dist}(q)), all  the fits are  systematically lower at
$z$ $<$ 0.7. Here the cosmic  variance or photo-z error should also be
important; the GOODS-N  distribution with a higher flux  cut but shows
higher densities at $z$ $<$  0.5 compared to two COSMOS distributions.
The match is excellent to the highest  flux cut at $z$ $<$ 2 (5 mJy in
24 $\mu$m,  Figure~\ref{red_dist}(r)) where  the data quality  is high
\citep[effectively 10 square degree  field size and 98\% spectroscopic
completeness,  ][]{Wu10}.   For  this  distribution, sources  with  no
redshift  measurements are  not included  but likely  lie at  high $z$
according  to  their  photo-z,   which  at  least  partly  causes  the
difference between  the model  and data at  $z$$>$2.  All the  fits to
redshifts  of local IRAS-60$\mu$m  sources brighter  than 5.24  Jy are
excellent. All the fits to  the SCUBA-850 $\mu$m and Aztec-1100 $\mu$m
are acceptable but produce higher mean redshifts.

\subsection{Fitting Local  LFs}\label{fit_result_local_lf}


\begin{figure*}
\epsscale{1.0}
\plotone{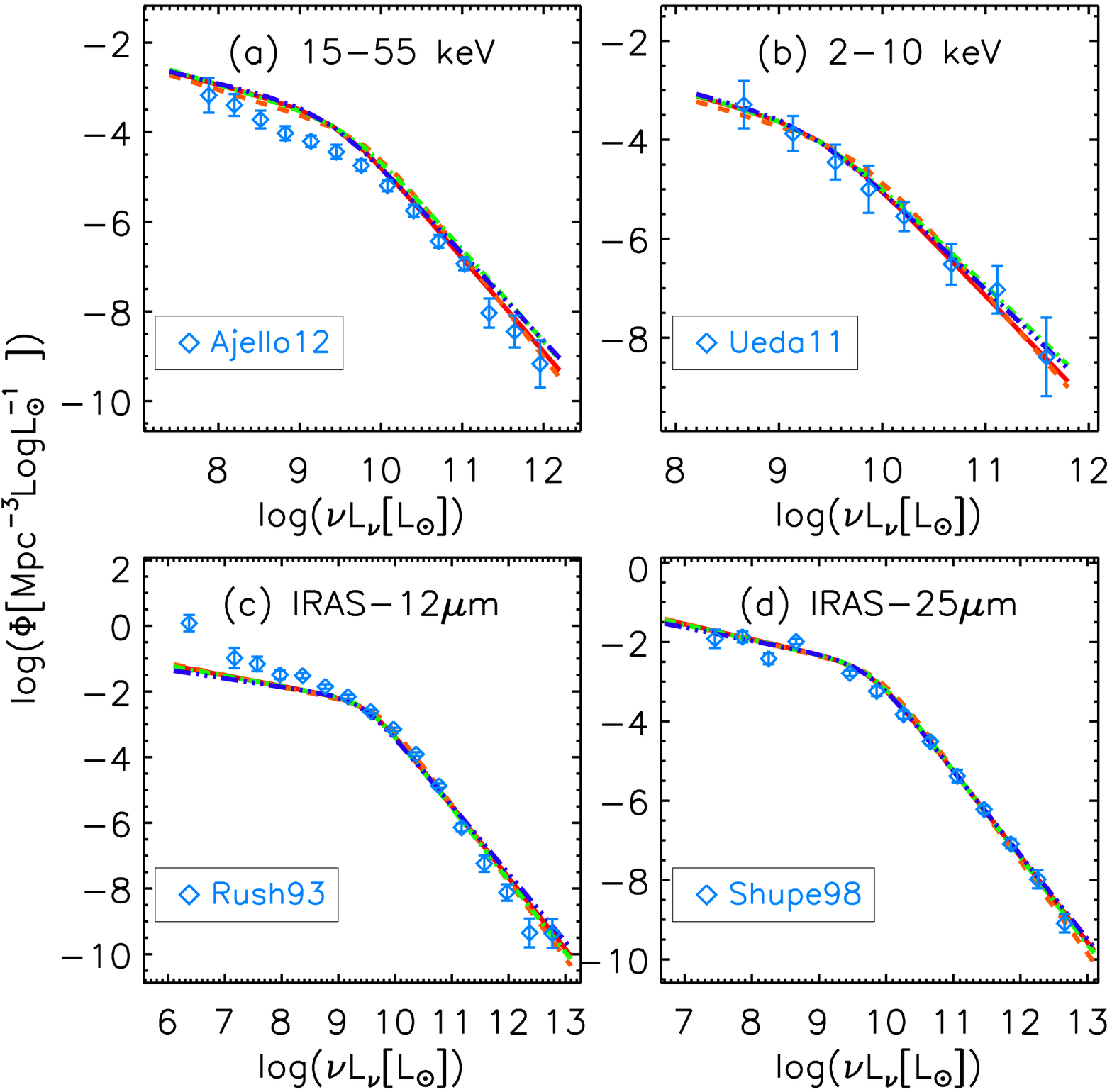}
\caption{\label{local_lf}  Fit dataset III -- local IRAS and X-ray luminosity functions.
All symbols are the observations.
The four variants of the model are shown with different line styles:
solid line for reference variant, dashed line for fast\_evol\_SED\_SF, dot-dashed line for
low\_IR2X\_BH and three-dot-dashed line for high\_IR2X\_BH (see Table~\ref{table_list_model}). }
\end{figure*}

For the 15-55 keV LF  (Figure~\ref{local_lf} (a)), all variants of the
model produce  0.3-0.5 dex higher spatial  densities than observations
below   the  break   luminosity.  The   fits  to   the  2-10   keV  LF
(Figure~\ref{local_lf}  (b))  offer acceptable  fits  given the  error
bars.  For the  IRAS 12 $\mu$m as shown  in Figure~\ref{local_lf} (c),
all variants of the  model produce slightly flatter bright-end slopes.
At the  faint end,  all variants under-estimate  the observed  LF. The
local  supercluster is  known to  boost the  spatial density  of faint
sources, an  effort which  is yet  not corrected in  the 12  $\mu$m LF
\citep{Rush93, Fang98, Shupe98}.  For the  IRAS 25 $\mu$m LFs as shown
in Figure~\ref{local_lf}(d), the fits are excellent.

\subsection{Systematics and Model Variants}\label{sum_diff_mod}

\begin{figure*}
\epsscale{0.8}
\plotone{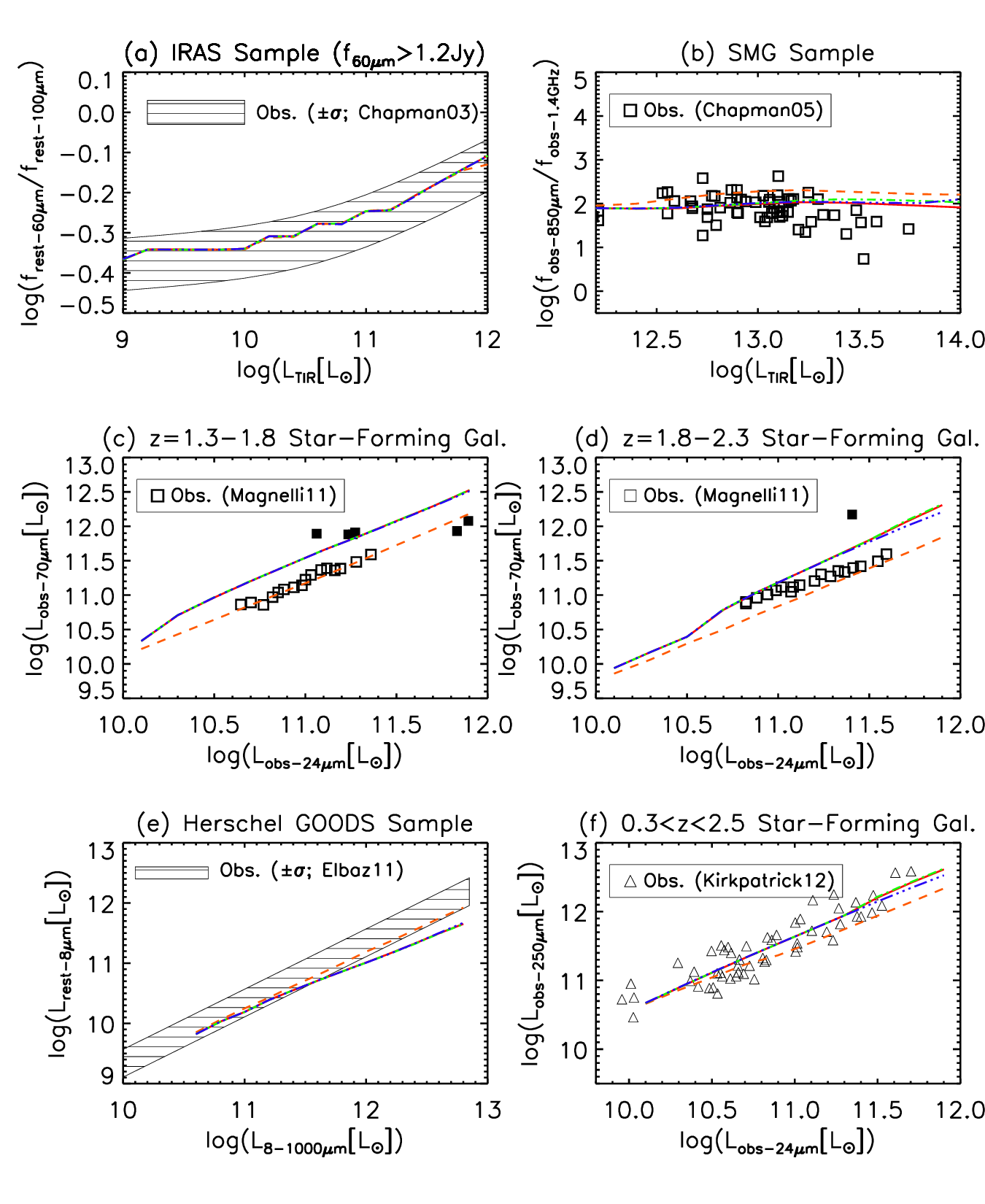}
\caption{\label{fig_color_ltir} The comparison  in the SED between the
model  and   observations.  (Equation~\ref{EV_T_LTIR}).  For every observed dataset,
we  apply similar  selection  functions  to our  model  to define  the
comparison    sample:   (a)    The   local    rest-frame    IR   color
(log($\frac{f_{60{\mu}m}}{f_{100{\mu}m}}$)) vs.  IR luminosity of IRAS
galaxies with  $f_{60{\mu}m}$ $>$  1.2 Jy \citep{Chapman03}.   (b) The
observed color between 850 $\mu$m and radio 1.4 GHz for the SMG sample
with $f_{850{\mu}m}$ $>$  3 mJy, $f_{\rm 1.4 GHz}$  $>$ 30 $\mu$Jy and
$z$ $>$ 1 \citep{Chapman05}. (c)  The observed 70 $\mu$m and 24 $\mu$m
luminosity relation  of star-forming galaxies  at 1.3 $<$ $z$  $<$ 1.8
where  open  squares  are  stacking  results and  filled  squares  are
individual detections \citep{Magnelli11}. Star-forming
galaxies in the  model are defined by having  the fractional SMBH luminosity
in the total IR band smaller than 20\%.   (d) The observed 70 $\mu$m
and 24 $\mu$m luminosity relation  of star-forming galaxies at 1.8 $<$
$z$ $<$ 2.3 where open squares are stacking results and filled squares
are individual detections \citep{Magnelli11}.  (e) The rest-frame IRAC
8 $\mu$m and total IR luminosity relation for Herschel GOODS sample as
defined  by  $f_{100{\mu}m}$ $>$  1.1  mJy  \citep{Elbaz11}. (f) The observed
250 $\mu$m and 24 $\mu$m luminosity relation of star-forming galaxies at 
0.3 $<$ $z$ $<$ 2.5 \citep{Kirkpatrick12}. The  four
variants of the model are shown with different line styles: solid line
for  reference variant,  dashed line  for  fast\_evol\_SED\_SF, dot-dashed
line    for   low\_IR2X\_BH    and    three-dot-dashed   line    for
high\_IR2X\_BH (see Table~\ref{table_list_model}). }
\end{figure*}

All    variants    of   the    model,    including   the    reference,
fast\_evol\_SED\_SF,  low\_IR2X\_BH and  high\_IR2X\_BH  as listed  in
Table~\ref{table_list_model},  offer  similar  performance in  fitting
number  counts,  redshift  distributions   and  local  LFs,  with  the
fast\_evol\_SED\_SF    variant    showing    additional    significant
under-estimates of the 70, 100  and 160 $\mu$m number counts above the
flux density  of the peak  count.  The reduced $\chi^{2}$  ranges from
2.7  to  2.9.   The   contribution  to  the  reduced  $\chi^{2}$  from
systematics associated with  data is estimated to be  at least 0.75 as
shown in the following.  For  the number counts, we assumed negligible
data  systematic   errors  so   that  the  difference   between  model
predictions and data are due to  the limitation of the model.  For the
local  12  $\mu$m  LF  (Figure~\ref{local_lf}(a)),  the  excess  faint
sources  due to  the local  super cluster  are not  corrected, causing
higher  densities  than  our  model's predictions.  We  quantitatively
estimate its contribution  to $\chi^{2}$ by summing the  square of the
difference  between  observations   and  predictions  divided  by  the
observed   errors  at   the  12   $\mu$m  luminosity   below  10$^{9}$
L$_{\odot}$.   Cosmic  variances or  photo-z  errors  are evident  for
redshift distributions between different fields with similar flux cuts
including at 2-10 keV, the  CDFN in Figure~\ref{red_dist} (a) vs. CDFS
in Figure~\ref{red_dist}  (b), the ECDFS  in Figure~\ref{red_dist} (c)
vs.     CDFN   in   Figure~\ref{red_dist}    (d),   the    COSMOS   in
Figure~\ref{red_dist} (e)  vs.  CDFS in  Figure~\ref{red_dist} (f); at
0.5-2  keV,  the  CDFS  in  Figure~\ref{red_dist}  (h)  vs.   CDFN  in
Figure~\ref{red_dist} (i),  the CDFN in  Figure~\ref{red_dist} (k) vs.
COSMOS in  Figure~\ref{red_dist} (l). In addition, in  24 $\mu$m bands
at $z$$<$0.3,  the COSMOS above 0.08 mJy  in Figure~\ref{red_dist} (o)
vs. the GOODS-N above  0.3 mJy in Figure~\ref{red_dist} (q).  Although
this  pair has  quite  different  flux cuts,  the  cosmic variance  or
photo-z errors must play dominant  roles because the GOODS-N with much
higher flux  cuts shows significantly  higher source densities  at $z$
$<$ 0.3  compared to the  COSMOS field.  Since  the best fit to  two Y
values at  fixed X  is the mean  of two  Y values, independent  of any
model assumption, we estimated  the minimum contribution to $\chi^{2}$
for each pair as listed above  by assuming the best-fit to be the mean
of two distributions in the  pair.  The sum of all above contributions
normalized by  the degree of  freedom gives 0.75, indicating  that the
true  reduced $\chi^{2}$  that our  four variants  of the  model would
achieve if data systematics was removed can be better than 2.0.

Comparison of the model variants  and their performance in fitting the
data does not favor strong  star-formation SED evolution; on the other
hand, direct  comparisons with  the observed high-z  SED color  do not
provide      definitive     evidence     about      SED     evolution.
Figure~\ref{fig_color_ltir}  shows such  comparisons where  we sampled
the  model  output to  mimic  the  observational  selection, which  is
crucial    to    eliminate    selection   biases    \citep{Magnelli09,
Chapin11}. However,  it is impossible to ascertain  that all selection
biases  have been properly  represented.  Fig.~\ref{fig_color_ltir}(a)
verifies that  the model re-produces the  local relationship (Equation
\ref{color_lum}).  The  difference in the  median color at a  given IR
luminosity  is around  0.02,  which is  reasonable  given the  limited
resolution (0.037) of the numerical calculation. Beyond z=0, we tested
the evolution  of the SED  shape by comparing  to five samples,  a SMG
sample   from  \citet{Chapman05},   two   star-forming  samples   from
\citet{Magnelli11}, the Herschel GOODS sample from \citet{Elbaz11} and
a star-forming sample from \citet{Kirkpatrick12}.  For the SMG sample,
the  radio emission  in our  model  is derived  through the  radio-FIR
relation    \citep[for   details,    see][]{DH02}.    As    shown   in
Fig.~\ref{fig_color_ltir}(b), although the scenario of no evolution in
the  star-formation  SED  offers   a  slightly  better  match  to  the
850$\mu$m/1.4GHz  ratio  of  the  SMG  sample, these  SMGs  are  radio
selected  for  follow-up observations  and  thus  subject to  possible
contamination       from      SMBH      radio       emission.       In
Fig.~\ref{fig_color_ltir}(c), the tight empirical relationship between
$L_{\rm   observed-24{\mu}m}$  and   $L_{\rm   observed-70{\mu}m}$  of
z=1.3-1.8  star-forming galaxies  is defined  by stacked  data  and it
favors strong  evolution in the  star-formation SED.  For  the stacked
color  of  $z=$1.8-2.3 galaxies,  the  variants  of  no SED  evolution
performs  better at  the low  luminosity  end while  the variant  with
strong  SED evolution  does better  at the  high luminosity  end.  The
well-defined  relationship between  rest-frame 8  $\mu$m and  total IR
luminosities     of      the     Herschel     GOODS      sample     in
Fig.~\ref{fig_color_ltir}(e)  is based  on  individual detections  but
from  different   redshifts.   Only   at  the  high   luminosity  end,
corresponding  to $z$$\sim$2,  the variant  with strong  SED evolution
produce     results     closer     to    the     observations.      In
Fig.~\ref{fig_color_ltir}(f),  the $L_{\rm obs-24{\mu}m}$  vs. $L_{\rm
obs-250{\mu}m}$  relationship is  based on  24 $\mu$m  sources ($>$0.1
mJy) with  Spitzer mid-IR  spectroscopic measurements at  the redshift
range of 0.3  to 2.5. We selected objects  with inferred AGN fractions
smaller than  20\% to  define the above  relationship. The  250 $\mu$m
detection rate  is quite high  (95\%) so that the  relationship should
not bias  against low 250$\mu$m/24$\mu$m sources.   The model variants
with no  SED evolution  match better the  median of  the distribution,
compared to the  variant with SED evolution.  At  the bright 24 $\mu$m
luminosity end  that corresponds to $z$=2, modelled  galaxies with SED
evolution are  too cold  to produce enough  observed frame  250 $\mu$m
(i.e., rest-frame  80 $\mu$m)  radiation.  We therefore  conclude that
although  the high-$z$ ULIRGs  are colder  than local  counterparts as
argued in the literature, their IR SEDs do not represent exactly those
of  local  normal  star-forming  galaxies  across the  whole  IR  band
\citep[e.g.][]{Kirkpatrick12}.


\begin{figure}
\epsscale{1.0}
\plotone{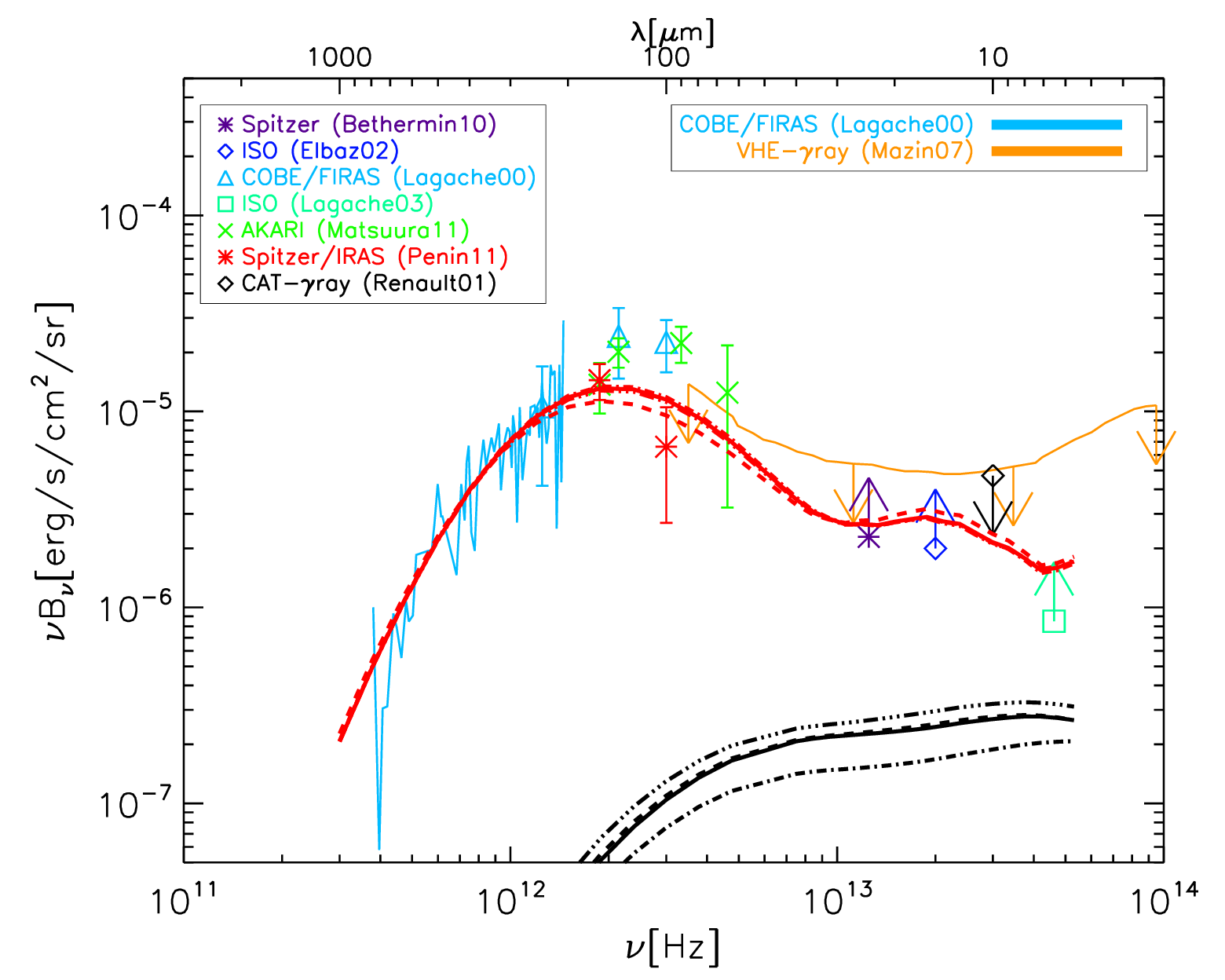}
\caption{\label{cosmic_ir_light} The cosmic IR background light. The red lines are model's
predictions on the total radiation while black lines are just for the emission from SMBH dusty tori: solid line
for  reference variant,  dashed line  for  fast\_evol\_SED\_SF, dot-dashed
line    for   low\_IR2X\_BH    and    three-dot-dashed   line    for
high\_IR2X\_BH (see Table~\ref{table_list_model}). All others are the observational constraints.
References: Bethermin10 -- \citet{Bethermin10};  Elbaz02 -- \citet{Elbaz02};  Lagache00 -- \citet{Lagache00}; 
Lagache03 -- \citet{Lagache03};  Matsuura11 -- \citet{Matsuura11};  Mazin07 -- \citet{Mazin07}; 
Penin11 -- \citet{Penin11};  Renault01 -- \citet{Renault01}. }
\end{figure}

\begin{figure}
\epsscale{1.0}
\plotone{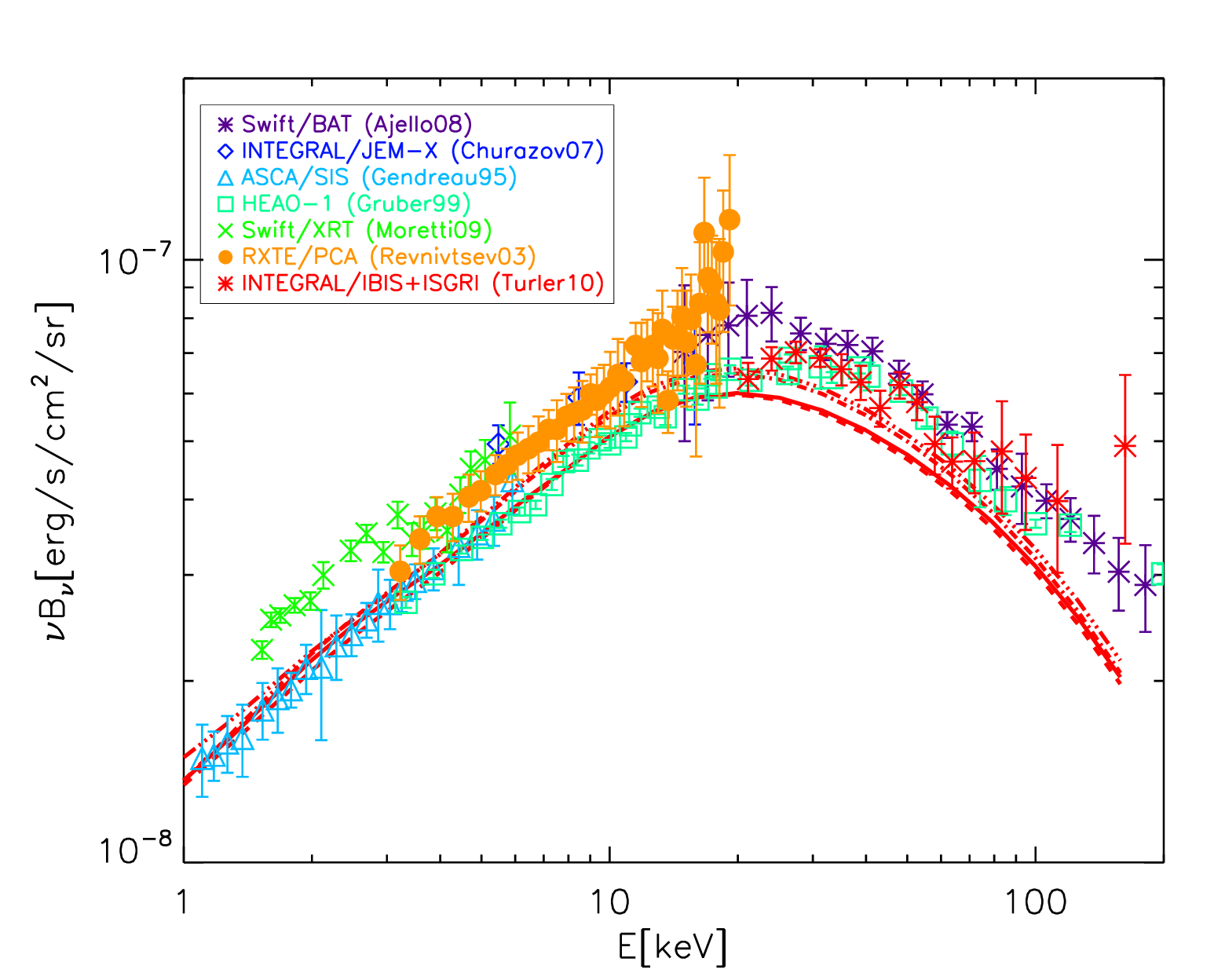}
\caption{\label{cosmic_xray_light} The cosmic X-ray background light. The red lines are model's
predictions: solid line
for  reference variant,  dashed line  for  fast\_evol\_SED\_SF, dot-dashed
line    for   low\_IR2X\_BH    and    three-dot-dashed   line    for
high\_IR2X\_BH (see Table~\ref{table_list_model}). All others are the observational constraints.
jello08 -- \citet{Ajello08};  Churazov07 -- \citet{Churazov07}; 
Gendreau95 -- \citet{Gendreau95};  Gruber99 -- \citet{Gruber99}; 
Moretti09 -- \citet{Moretti09};  Revnivtsev03 -- \citet{Revnivtsev03}; 
Turler10 -- \citet{Turler10}.
 }
\end{figure}

\section{The Basic Outputs Of The Model}\label{basic_output}

We discuss  here the model's  first results in  the form of  its three
basic outputs, i.e.,  the SMBH energy fraction, the  HI column density
and  the  total  IR  LFs  whose  best-fit  parameters  are  listed  in
Table~\ref{best_fit_parameter}.   The  first of  these  is the  unique
output compared  to previous models for  X-ray or IR  data alone.
As a final check  of our overall outputs, Figure~\ref{cosmic_ir_light}
and Figure~\ref{cosmic_xray_light}  show the comparisons  between data
and model's  predictions for CIRB and CXB  spectra, respectively.  All
variants of the  model re-produce the observed CIRB  data within error
bars.  It is  shown that  the emission  from the  SMBH dusty  torus is
around 15\%  at 5-10 $\mu$m  and decreases toward  longer wavelengths,
lower than \citet{Treister06}  but comparable to \citet{Silva04}.  All
variants of the model reproduce the CXB spectrum below the peak energy
but are about 10-20\% lower than observations at higher energies.

\subsection{The SMBH Energy Fraction}


\begin{figure}
\epsscale{1.0}
\plotone{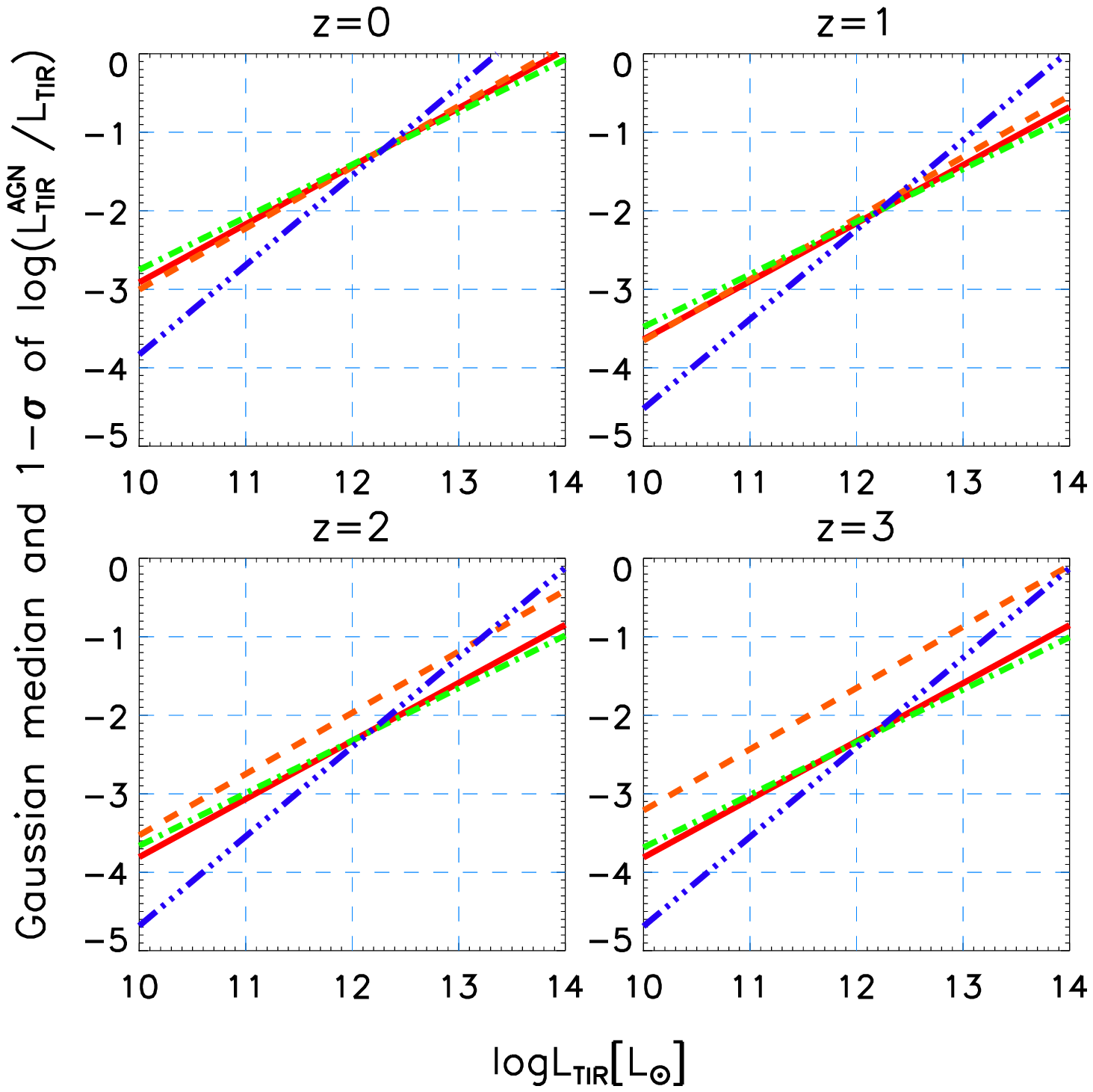}
\caption{\label{dist_BH_energy_frac}   The   median
 SMBH energy  fraction in the total IR band as a
function of the IR luminosity at four redshifts.The four variants of the model are shown with different line styles:
solid line for reference variant, dashed line for fast\_evol\_SED\_SF, dot-dashed line for
low\_IR2X\_BH and three-dot-dashed line for high\_IR2X\_BH (see Table~\ref{table_list_model}).  }
\end{figure}


\begin{figure}
\epsscale{1.0}
\plotone{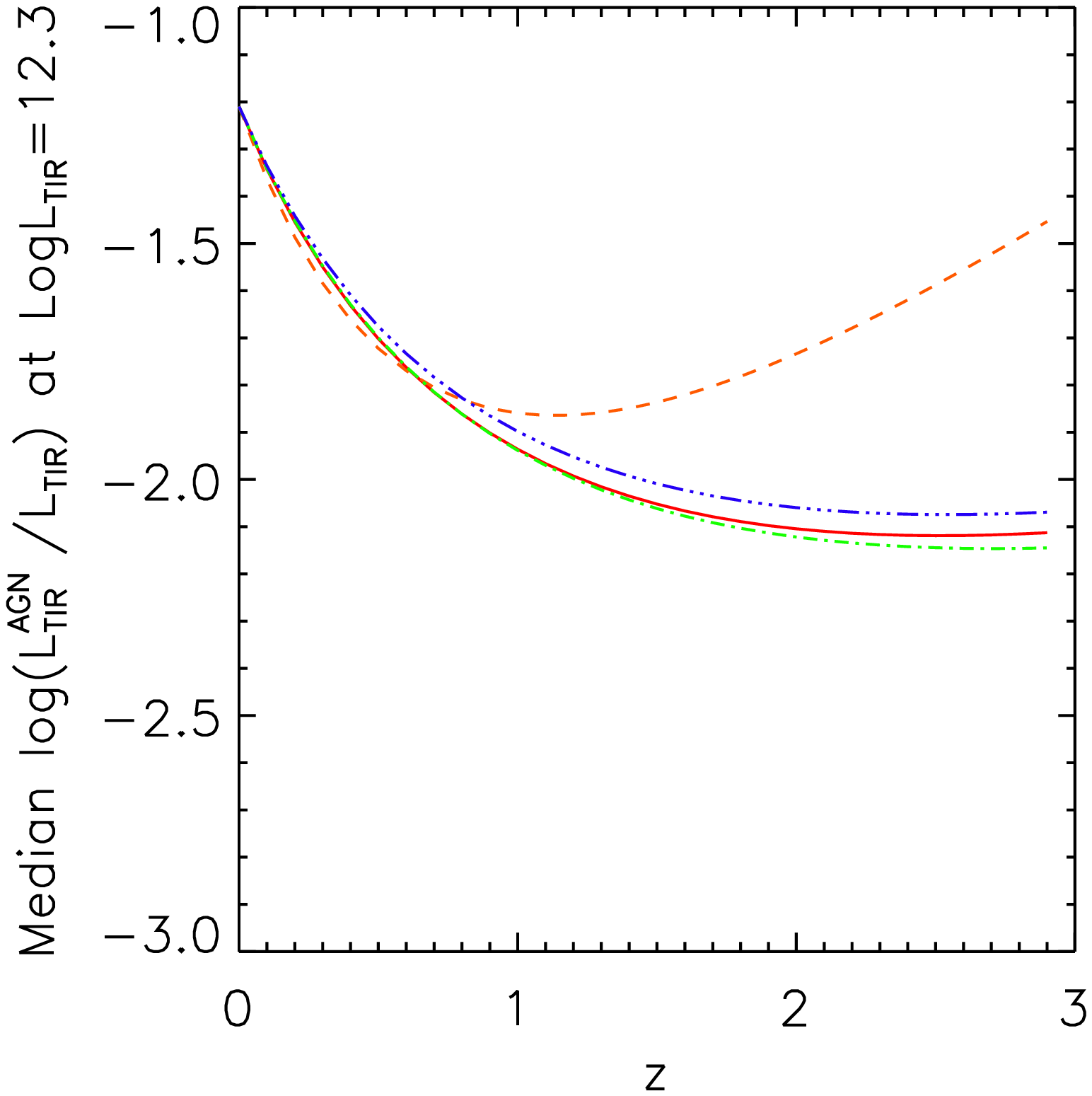}
\caption{\label{med_BH_frac_red}  The  redshift  evolution of  the  BH
energy  fraction  at  $L_{\rm  TIR}[L_{\odot}$]=12.3.  Different  line
styles are  for different  variants of the  model: 
solid line for reference variant, dashed line for fast\_evol\_SED\_SF, dot-dashed line for
low\_IR2X\_BH and three-dot-dashed line for high\_IR2X\_BH (see Table~\ref{table_list_model}).}
\end{figure}

As   listed    in   Table~\ref{best_fit_parameter}   and    shown   in
Figure~\ref{dist_BH_energy_frac},   and  Figure~\ref{med_BH_frac_red},
all variants of the model  indicate that the median SMBH energy fraction
in  the total  IR  band increases  with  the IR  luminosity, which  is
consistent   with    various   studies   in    the   local   universe
\citep[e.g.][]{Sanders96,  Genzel98, Armus07, Desai07,  Veilleux09, Petric11} and
at high-z \citep{Kartaltepe10},  while the
Gaussian  width  of the  distribution  increases  with the  decreasing
luminosity.  The overall trend of the median SMBH fraction with redshift
shows a minimum  around $z$$\sim$1-2.  To test the  robustness of this
redshift  trend,  we first  ran  a  model  without redshift  evolution
($k_{1,  d}^{\rm BH}\equiv$0,  $k_{2, d}^{\rm  BH}\equiv$0)  while all
others are the same as  those in the reference variant.  The resulting
best-fit  reduced $\chi^{2}$  increased by  as much  as 1.3,
which offers strong evidence for the requirement of redshift evolution
in the  SMBH energy fraction.  We then  ran a variant of  the model that
only  allows  monotonic  evolution  (i.e., $k_{1,  d}^{\rm  BH}$=free,
$k_{2,  d}^{\rm BH}\equiv$0).   The increase  in the  best-fit reduced
$\chi^{2}$ is smaller but still significant by 0.4.

\begin{figure}
\epsscale{0.8}
\plotone{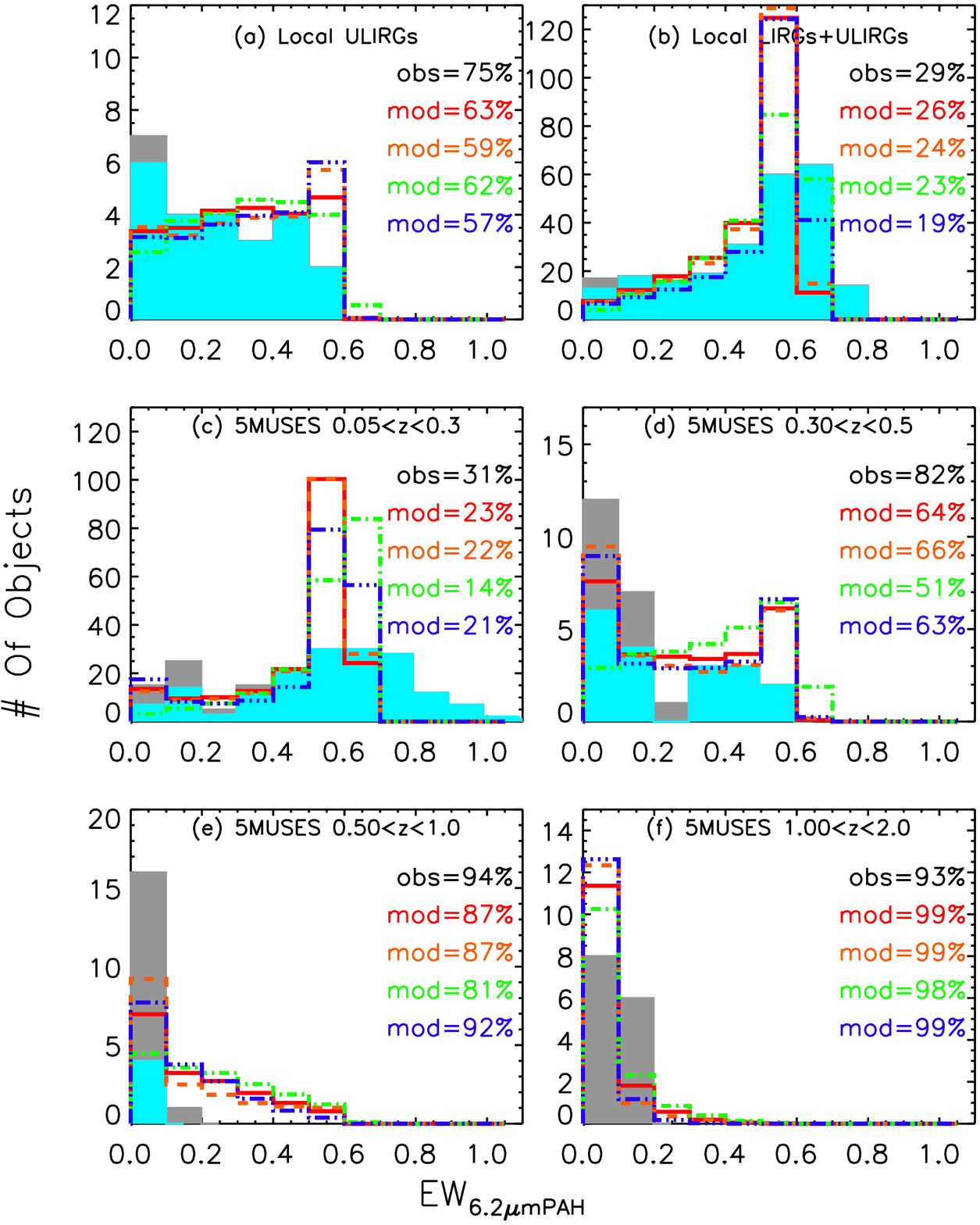}
\caption{\label{ew_dist} The 6.2 $\mu$m  EW distributions. The cyan and
grey   filled  histograms  represent   the  observed   detections  and
upperlimits,  respectively.   The  local  LIRGs and  ULIRGs  are  from
\citet{Stierwalt12}, while  those at higher redshifts  are from 5MUSES
\citep{Wu10}.   The  open  histograms  are  the  model's  predictions.
Different line styles  are for different variants of  the model: solid
line  for the reference  variant, dashed  line for  the fast\_evol\_SED\_SF
variant, dot-dashed  line for  the low\_IR2X\_BH  variant and
three-dot-dashed       line       for      high\_IR2X\_BH       (see
Table~\ref{table_list_model}). The listed percentages are the fractions of
objects with EW $<$ 0.4 for each distribution (``obs'' stands for the observed distribution while
``mod'' is for the modelled distributions). }
\end{figure}

In Figure~\ref{ew_dist}, we  compared the 6.2 $\mu$m EWs  in our model
to several observed distributions at different redshifts, which offers
a stringent check for our SMBH energy fraction measurements.  Here the
modelled samples  are defined in  similar ways to those  observed. All
EWs  are measured  through the  spline interpolations,  with  the pure
star-formation median  value around 0.6-0.7 $\mu$m.  Note  that our SF
templates do not have scatters in the EW, and thus the modelled EWs do
not have  values in bins higher  than EW=0.6-0.7 $\mu$m  and any value
below EW=0.6 $\mu$m is caused  by SMBH contributions in our model.  At
log$L_{\rm TIR}$=12.3 and $z$=0, the SMBH energy fraction in the model
is fixed to  that of 1-Jy ULIRGs, and is allowed  to be luminosity and
redshift dependent.   As shown in  Figure~\ref{ew_dist}(a) for ULIRGs,
the model produces  an almost flat trend, roughly  consistent with the
result  of GOALS's  ULIRGs \citep{Stierwalt12}  that  additionally show
excess objects in the the lowest EW bin above a flat trend.  For local
LIRGs+ULIRGs, the  distribution has a  peak at the  SF EW with  a tail
toward the  low EW  end, which is  roughly consistent with  the GOAL's
result  \citep{Stierwalt12}.  Beyond the  local universe,  we compared
the   model's   results  to   those   of   5MUSES  \citep{Wu10}.    At
0.05$<$$z$$<$0.3, the higher star-formation peak in the model compared
to  the observation  is  due to  zero  scatter in  our SED  templates.
Overall, the star-formation peak plus a low EW tail as produced by the
model is  consistent with  the observations.  At  0.3$<$$z$$<$0.5, the
model reproduces a bimodal EW distribution as observed, a SF peak plus
a peak at  the lowest bin with the latter higher  than the former. For
two  higher redshift bins  (0.5$<$$z$$<$1.0 and  1.0$<$$z$$<$2.0), the
model  reproduces a  peak at  the lowest  bin with  a tail  toward the
star-formation end as observed.  In Figure~\ref{ew_dist}, we also list
the fraction of objects with EW  $<$ 0.4 for each observed or modelled
distribution.   This  quantitative  comparison  further  supports  the
consistency  between model's  predictions and  observations,  with the
former is only slightly lower than the latter.

\subsection{HI Column Density Distribution And Compton-thick AGN Fraction}


\begin{figure}
\epsscale{1.0}
\plotone{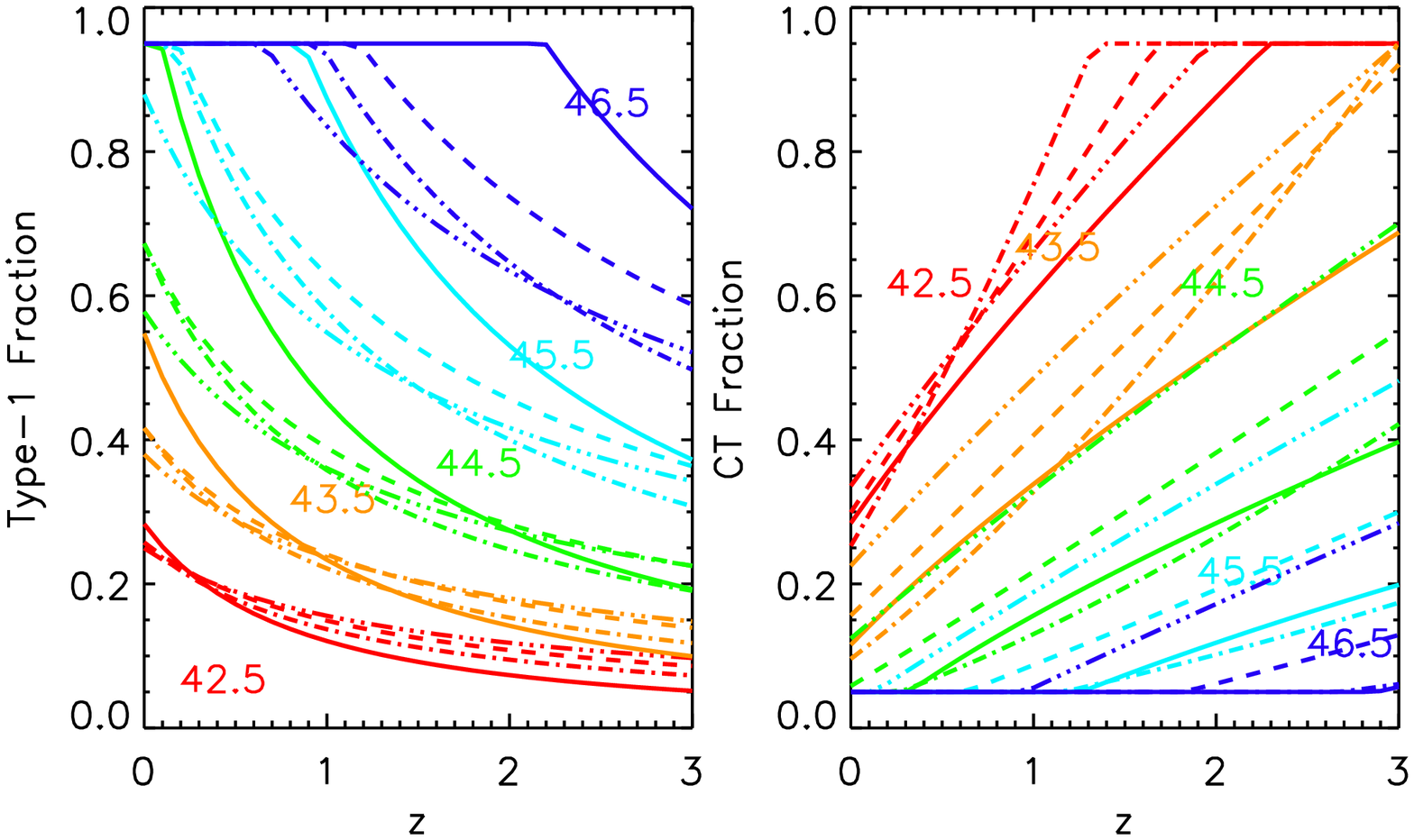}
\caption{\label{t1_ct_evl}  The  redshift   evolution  of  type-1  AGN
fraction  (left  panel) and  Compton-thick  AGN  fraction  in all  AGN
populations. Different  line styles are for different  variants of the
model:  solid line  for  the  reference variant,  dashed  line for  the
fast\_evol\_SED\_SF variant, dot-dashed line for the low\_IR2X\_BH
variant   and   three-dot-dashed   line  for   high\_IR2X\_BH   (see
Table~\ref{table_list_model}).  Different  colors  are  for  different
X-ray luminosities: blue  for log(L$_{\rm 2-10keV}$[erg/s])=42.5, cyan
for    log(L$_{2-10keV}$[erg/s])=43.5,     green    for    log(L$_{\rm
2-10keV}$[erg/s])=44.5, orange for log(L$_{\rm 2-10keV}$[erg/s])=45.5,
red for log(L$_{\rm 2-10keV}$[erg/s])=46.5.}
\end{figure}

The  model characterizes  the HI  column density  distribution through
three bins, the type-1 (log$N_{\rm HI}/$cm$^{2}$ $<$ 22), Compton-thin
type-2 (22 $<$ log$N_{\rm HI}/$cm$^{2}$  $<$ 24) and Compton thick AGN
(24  $<$  log$N_{\rm HI}/$cm$^{2}$  $<$  26).  Figure  \ref{t1_ct_evl}
shows the evolution  of the type-1 and Compton-thick  AGN fractions in
all AGN as  a function of the SMBH luminosity  and redshift.  All four
variants of the model require  that the  type-1  fraction  increases  with the  BH
luminosity but  decreases with redshift,  and the CT fraction  has the
opposite trend.

The  trend of  the  type-1 fraction  with  the SMBH  luminosity is  in
general   consistent  with  previous   studies  \citep[e.g.][]{Ueda03,
Hasinger04,  LaFranca05, Barger05, Treister06,  Ballantyne06, Gilli07,
Hasinger08}.   Although  previous CXB  models  can  explain the  X-ray
survey data either with the redshift dependence of the type-1 fraction
\citep{LaFranca05,   Ballantyne06}   or   without  it   \citep{Ueda03,
Treister05, Gilli07}, the requirement  of such a redshift evolution is
highly preferred by our model  that is constructed to explain both the
X-ray and IR data, consistent with recent studies of a large X-ray AGN
sample \citep{Hasinger08}.  Without redshift evolution ($\beta_{z, \rm
type-1}\equiv$0 \&  $\beta_{z, \rm CT}\equiv$0),  the best-fit reduced
$\chi^{2}$ rises by 0.5, which is essentially caused by the
increased  brightness  of high-z  AGN  at  both  X-ray and  24  $\mu$m
wavelengths.  In  the pure  CXB model that  only deals with  the X-ray
data, the  increase in the  apparent brightness of  individual objects
can be complemented  by the decrease in the  spatial number density in
order  to  keep the  number  of objects  above  a  given flux  density
unchanged.  This  is why  the CXB model  is not sensitive  to redshift
evolution  of the  type-1 fraction.   However, the  change of  the AGN
number  density in  our model  is additionally  constrained by  the IR
data.  The redshift  evolution of the break luminosity  and density of
the total IR LF is largely fixed through IR survey data.  Although the
change in the SMBH energy  fraction can vary the AGN number densities,
it also  modifies the  number density of  the star-formation-dominated
sources especially the far-IR/submm sources that lie at the bright end of the
IR  LF and  thus are  sensitive to  the SMBH  energy  fraction.  Without
redshift evolution of the HI  column density, the best-fit result show
that  the redshift distributions  at $z$  $>$ 1  of 24  $\mu$m sources
brighter  5  mJy  and  0.5-2  keV  sources  brighter  than  10$^{-14}$
erg/s/cm$^{2}$ are over-produced  by a factor of 10,  and the redshift
distribution of submm sources systematically  offsets by a factor of 3
along the Y-axis.

\begin{figure}
\epsscale{0.8}
\plotone{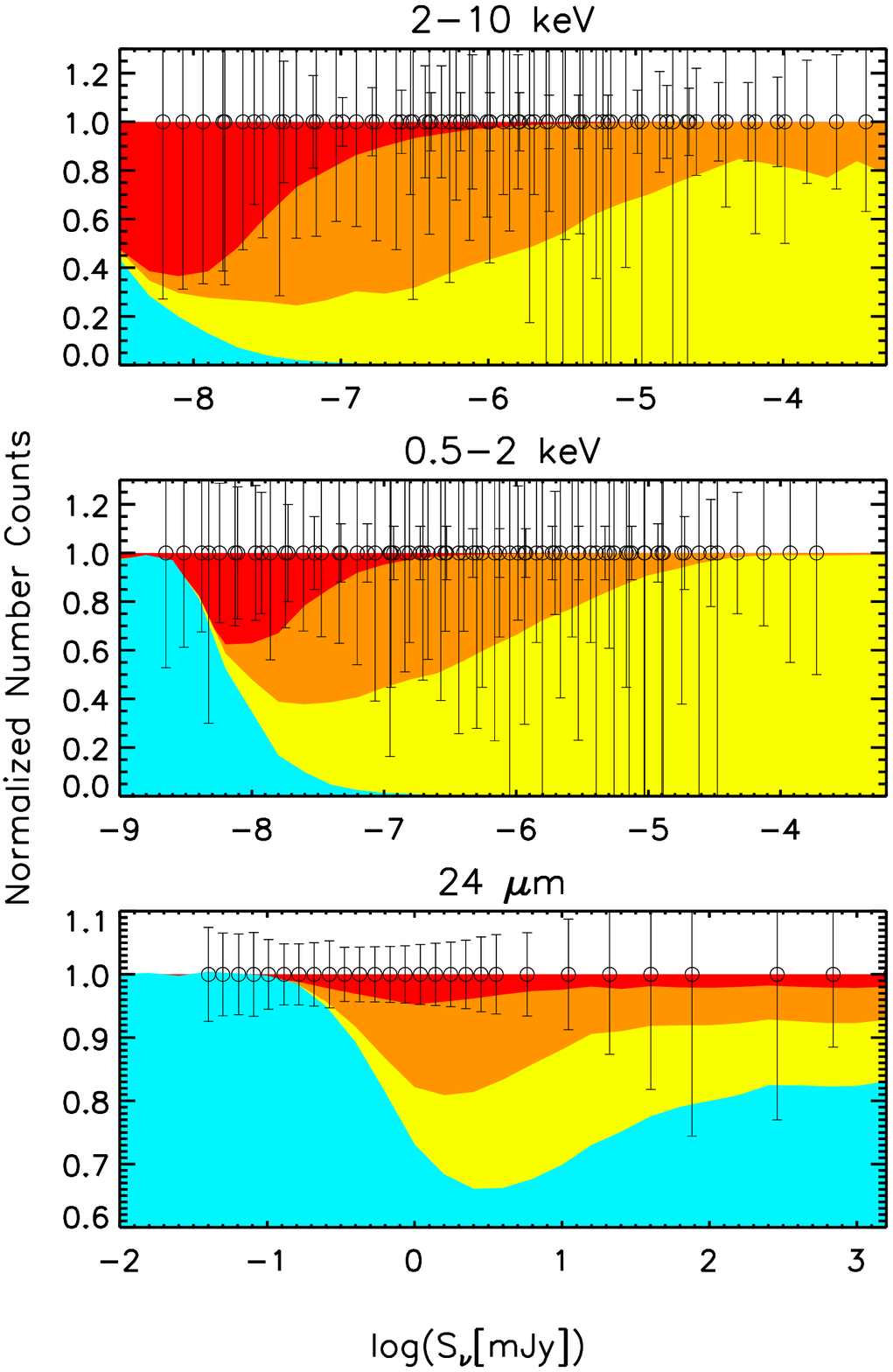}
\caption{\label{decompose_emission} The percentage energy contributions to the number counts at 2-10 keV,
0.5-2 keV and 24 $\mu$m, from different types of radiations in the reference model. 
Cyan for the pure star-formation emission, yellow for the pure SMBH emission of type-1, 
orange for the pure SMBH emission of Compton-thin type-2 and red for the pure SMBH emission of Compton-thick
AGN.  }
\end{figure}

Our model requires a large number  of the Compton-thick AGN. At $<$ 10
keV, Compton-thick AGN are  extremely faint. The traditional CXB model
first constrains the Compton-thin AGN through X-ray data at $<$ 10 keV
and then add the Compton-thick AGN  to match the CXB spectrum above 10
keV especially the 20-30  keV peak \citep[e.g.][]{Ueda03, Gilli07}. In
this paper, we  take advantage of the capability of  the mid-IR data in
probing these deeply buried AGN thanks to significantly lower apparent
gas-to-dust ratio. A test run of the model with zero Compton-thick AGN
abundance shows that the reduced $\chi^{2}$ would worsen by as much as
1.0.   The  main  reason   is  that  the  relative  fraction  between
Compton-thick and  Compton-thin type-2 AGN  as a function of  the flux
density is different at 24 $\mu$m from that at 0.5-2 and 2-10 keV, so
that it  is able  to break the  degeneracy between  Compton-thick and
Compton-thin  type-2  AGN  seen  in  pure CXB  models.   As  shown  in
Figure \ref{decompose_emission},   the   Compton-thick   AGN  are   only
important over  the faintest  1-2 orders of  magnitude flux  ranges in
X-ray bands, while Compton-thin type-2  AGN are important
over  a  much  larger  flux  ranges. The  resulting  Compton-thick  to
Compton-thin type-2  AGN fraction  increases with the  decreasing flux
density at both  bands.  Such similar behaviors at  both soft and hard
bands  imply that  if the  model  is forced  to replace  Compton-thick
objects with  Compton-thin type-2 AGN in  one band, the  result in the
other band  would not significantly violate  the observation. However,
at 24 $\mu$m, the Compton-thick contribution is noticeable over a flux
range as large as four orders  of magnitude, which is quite similar to
that of  the Compton-thin type-2  AGN. The relative ratio  between the
two populations is  not a strong function of the  flux density, with a
value  roughly around  1:3.  This means,  if  the model  is forced  to
replace  Compton-thick AGN  with Compton-thin  AGN at  24  $\mu$m, the
result in X-ray bands  would significantly over-produce objects at the
high  flux end  but under-produce  at the  faint flux  end.  This argument
was verified by running a test model without a Compton-thick AGN population and 
the expected behavior was observed in the model output.

\subsection{The Total IR LF}


\begin{figure*}
\epsscale{0.8}
\plotone{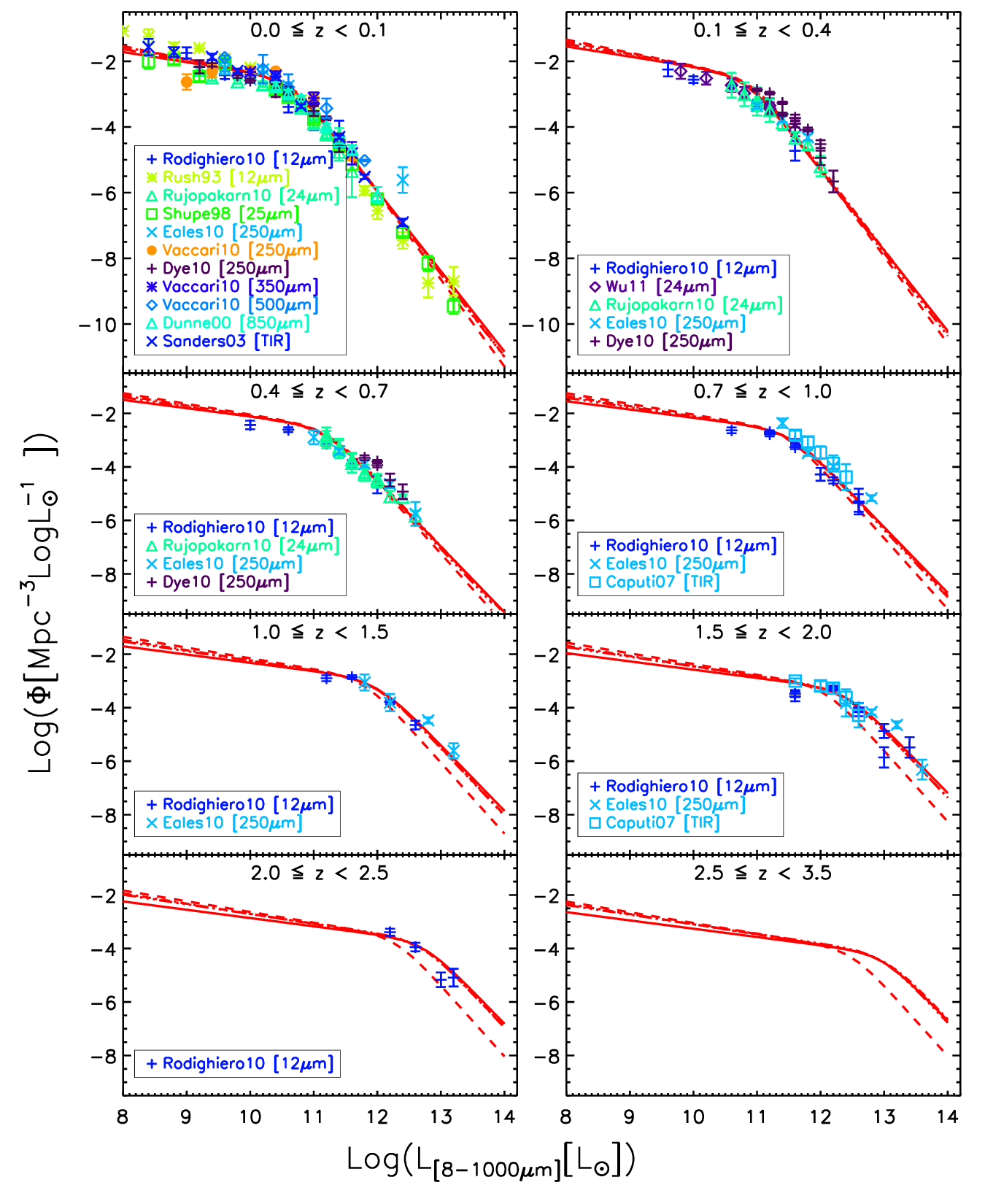}
\caption{\label{lf_tir}  The  luminosity  function  in  the  total  IR
(8-1000$\mu$m) at  different redshifts. Different line  styles are for
different variants of the model:  solid line for the reference variant,
dashed line for the fast\_evol\_SED\_SF variant, dot-dashed line for
the low\_IR2X\_BH variant   and   three-dot-dashed   line  for   high\_IR2X\_BH   (see
Table~\ref{table_list_model}). References: Caputi07 -- \citet{Caputi07};
Dunne00  --  \citet{Dunne00};   Dye10  --  \citet{Dye10};  Eales10  --
\citet{Eales10}; Rodighiero10 -- \citet{Rodighiero10}; Rujopakarn10 --
\citet{Rujopakarn10};   Rush93   --   \citet{Rush93};   Sanders03   --
\citet{Sanders03};   Shupe98    --   \citet{Shupe98};   Vaccari10   --
\citet{Vaccari10}; Wu11 -- \citet{Wu11}. }
\end{figure*}

\begin{figure*}
\epsscale{1.0}
\plotone{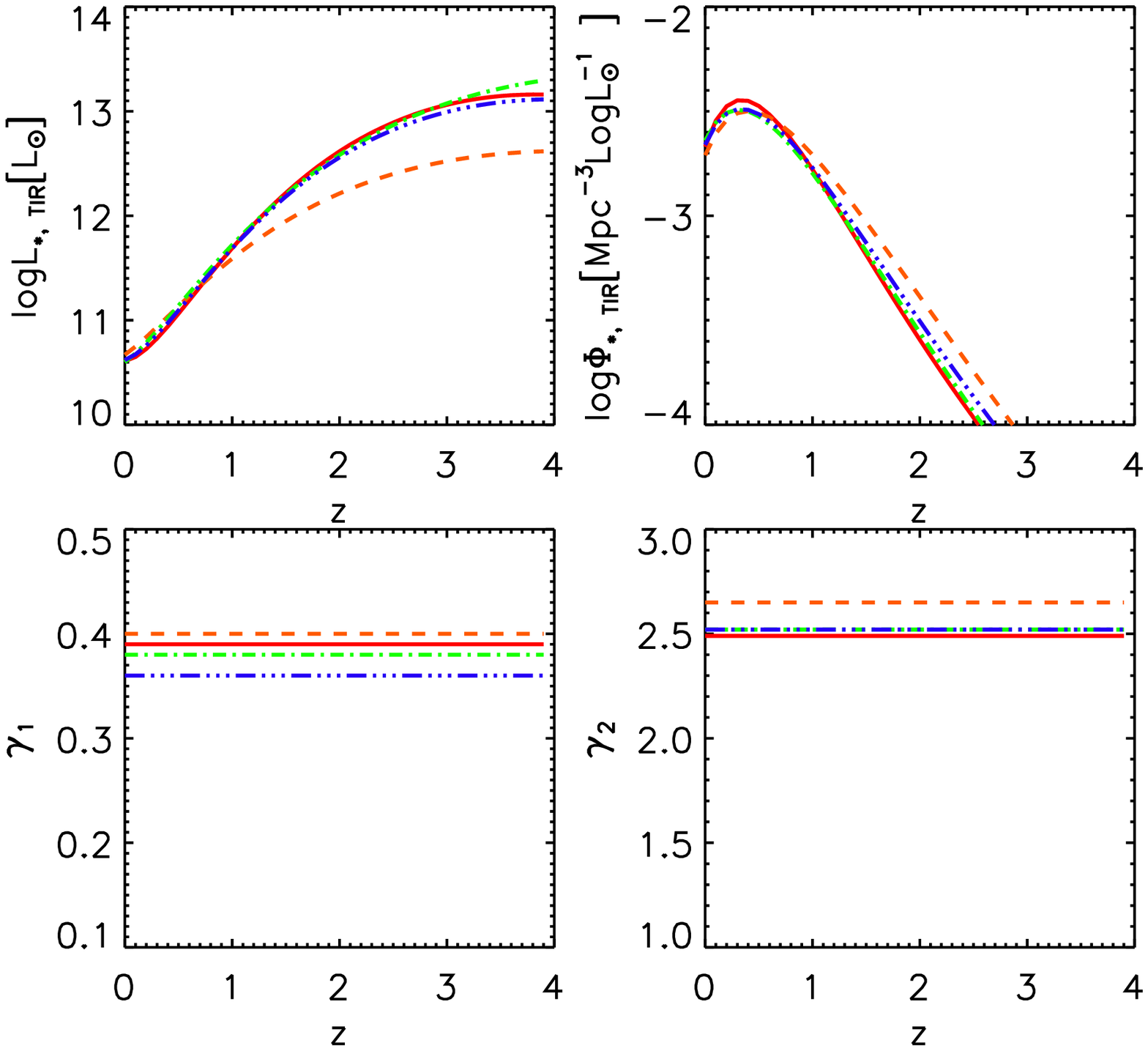}
\caption{\label{IRLF_par_evl}  Redshift evolution  of  four parameters
describing  the total  IR  LF including  the  break luminosity,  break
density, faint end and high  end slopes. Different line styles are for
different variants of the model:  solid line for the reference variant,
dashed line  for the fast\_evol\_SED\_SF variant, dot-dashed  line for the
low\_IR2X\_BH variant and three-dot-dashed line for high\_IR2X\_BH
(see Table~\ref{table_list_model}).}
\end{figure*}

As  shown  in Figure~\ref{lf_tir},  the  predicted  total  IR LFs  are
consistent with  the observations over redshift  and luminosity ranges
where  the  data are  available.   For  comparison,  we converted  the
observed LFs at given filters to  the total IR LFs using the reference
variant.    We   first   modelled   the  number   density   $\Phi_{\rm
model}$(filter, $z$, ${\nu}L_{\nu}$) at  the same filter, redshift bin
($z{\pm}{\Delta}z$)            and            luminosity           bin
(${\nu}L_{\nu}$${\pm}{\Delta}{\nu}L_{\nu}$)   as  the   observed  ones
($\Phi_{\rm  obs}$(filter, $z$, ${\nu}L_{\nu}$)).   The ${\nu}L_{\nu}$
is  then  corrected  to   the  total  IR  luminosity  with  luminosity
corrections  based on  the probability-weighted  average of  all model
SEDs              within             $z{\pm}{\Delta}z$             and
${\nu}L_{\nu}$${\pm}{\Delta}{\nu}L_{\nu}$.        This      $\Phi_{\rm
model}$/$\Phi_{\rm obs}$  is taken as  the difference of  the observed
and modelled  LFs in the  total IR band.  For the local LFs,  the IRAS
data offer the  largest dynamical range and are  thus the best dataset
to test the validity of the model's  LF.  The 12 and 25 $\mu$m LFs are
used   to   fit  the   model   and   the   result  is   discussed   in
\S~\ref{fit_result_local_lf}.  Here  we further compared  our model to
the  total   IR  LF   based  on  IRAS   60  $\mu$m   selected  sources
\citep{Sanders03},   and  found   a   good  match   between  the   two
(Figure~\ref{lf_tir}).  Consistency  is excellent with  other works of
significantly smaller fields that  cover the whole IR/submm wavelength
range.  Above $z$ $>$ 0.1 up to $z$ $\sim$ 3, current deep surveys are
able to  probe objects up  to one order  of magnitude below  the break
luminosity  as  shown in  Figure~\ref{lf_tir}.   At short  wavelengths
\citep{Wu11, Rodighiero10, Rujopakarn10}, no significant difference is
found between the model and  data up to $z$$\sim$2.5.  Some deviations
at the faint end are  most likely caused by incompleteness corrections
particularly associated with 1/$V_{\rm max}$ method \citep{Page00}. At
rest-frame  250   $\mu$m,  the  model  is  consistent   with  that  of
\citet{Eales10}  but  the result  of  \citet{Dye10} is  systematically
higher  at  0.1$<z<$0.   The   total  IR  LF  of  \citet{Caputi07}  is
systematically above our modelled LF; their work may over-estimate the
total    IR    luminosity    due    to    K-corrections    \citep[also
see][]{Magnelli11}.

Our model hypothesizes no evolution  in either the faint or bright end
slopes of the total IR LF.  As shown in Figure~\ref{lf_tir}, the fixed
bright end slope matches the  available data up to $z$$\sim$3. For the
faint  end slope,  we  did not  see  deviations from  the  data up  to
$z$$\sim$1 above that the available data are quite limited to justify.
With fixed faint  and bright end slopes, the total  IR LF as predicted
by   our   model   requires   evolution   in   both   luminosity   and
density. Although the detail  relies on the evolution formula adopted,
we found  in general that the break  luminosity (log$L_{*}$) increases
with  redshift but  the break  density (log$\Phi_{*}$)  decreases with
redshift.   As shown in  Figure~\ref{IRLF_par_evl}, for  the evolution
formula     given      in     Equations     \ref{eq_evl_lstar}     and
\ref{eq_evl_denstar},  the break  luminosity  increases rapidly  until
$z$=2 and  then gradually reaches the  peak at $z$=4,  while the break
density shows  a gradual increase  until z=0.4 and  then monotonically
decreases with redshift.  To test the redshift out to which luminosity
or density  evolution is  required, we have  carried out  several test
runs  of the  model by  modifying the  evolution formula  to  force no
evolution above  redshifts of  z=0.5, 1.0, 1.5,  2.0, 2.5 \&  3.0.  If
adopting ${\Delta}{\chi}^{2}$/d.o.f=1  as an indicator  of significant
worse fit,  we found that the  no-evolution scenario can  be ruled out
below  $z$ $\sim$  2.5.  Above  $z$$\sim$3, the  fit result  could not
differentiate  evolution vs.   no evolution,  which is  not surprising
since the  current constraint  on the $z$  $>$ $3$ infrared  and X-ray
populations  is still  of  low statistical  significance  as shown  in
Figure~\ref{red_dist}.   To   evaluate  the  degeneracy   between  the
luminosity  and density  evolution,  we carried  out  test runs  which
halted  density or  luminosity  evolution at  selected redshifts.  The
result shows that the data are not able to break the degeneracy at $z$
$\gtrsim$  1.5, which is  consistent with  Figure~\ref{lf_tir} showing
that  above   $z$  $\sim$  1.5   the  observed  LF  does   not  extend
significantly below  the break  luminosity.  However, no  matter which
evolution  formula is  adopted, the  robust result  is that  the break
luminosity  always increases  with redshift,  while the  break density
overall  decreases   with  redshift.   Our  result   of  the  positive
luminosity evolution  is consistent with  different literature studies
while the  negative trend  in the density  is not always  favored.  We
note  our result of  negative density  evolution is  inconsistent with
studies    in    \citet{LeFloch05},    \citet{Perez-Gonzalez05}    and
\citet{Rodighiero10}, but consistent with ones in \citet{Caputi07} and
\citet{Magnelli11}.

\section{Conclusions}\label{conclusions}

In this paper,  we have constructed a joint model of  the X-ray and IR
extragalactic background to account  simultaneously for both X-ray and
IR survey data.  The main conclusions are the following:

(1) The model successfully fits the deep survey data with the scenario
that individual galaxies are experiencing both star formation and SMBH
accretion  activities. The  model  with 19  free parameters  generates
number counts,  redshift distributions and  local luminosity functions
consistent with  617 data-points  over six decades  of electromagnetic
frequencies.  The  best-reduced $\chi^{2}$ reaches  2.7-2.9, with at
least  0.75   contributed  by  cosmic  variance,   photo-z  errors  or
limitations of the data sets.

(2) The  unique output  of this model,  compared to previous  ones for
either pure X-ray or IR data, is to constrain the SMBH energy fraction
in the total IR band.   The best-fit requires the SMBH energy fraction
to increase with the IR luminosity but decrease with the redshift back
to $z$$\sim$1.5.   The derived  trend is statistically  significant as
shown by the significantly increased best-fit $\chi^{2}$/d.o.f., i.e.,
$\Delta$$\chi^{2}$/d.o.f  $>$  1.0,  for  best  fits  where  the  SMBH
fraction  has either  no  luminosity or  no  redshift dependence.   An
independent  test comparing  the model's  predictions to  the observed
aromatic  feature  equivalent  width  distributions agrees  with  this
result.

(3) The  second important output of  the model is  the distribution of
the HI column density  obscuring the AGN, especially the Compton-thick
fraction.   A run  of the  model with  no Compton-thick  AGN  shows an
increase  in  the   $\chi^{2}$/d.o.f.   by  $\sim$1.0.   The  best-fit
parameters indicate that the Compton-thick AGN fraction decreases with
the SMBH luminosity but increases  with the redshift, while the type-1
AGN fraction has the inverse trend.

(4)  The  third output  is  the  total  IR luminosity  function.   The
best-fit   parameters  indicate  that   the  break   luminosity  rises
monotonically  with redshift while  the break  density has  an overall
inverse trend. The above result  is statistically important out to $z$
$\sim$  1.5, above  which the  degeneracy between  the  two parameters
cannot be resolved with available data.

\section{Acknowledgment}

We  thank the  anonymous  referee for  the  detailed and  constructive
comments.   We  thank  Matthieu  Bethermin  for  careful  reading  and
comments.   The work is  supported through  the Spitzer  5MUSES Legacy
Program 40539. The authors  acknowledge support by NASA through awards
issued by JPL/Caltech.

\clearpage


\begin{deluxetable}{llcccccccc}
\tabletypesize{\scriptsize}
\tablecolumns{5}
\tablecaption{\label{table_list_model}  A Summary of Different Variants of The Model }
\tablewidth{0pt}
\tablehead{ 
\colhead{Models} &  \colhead{description} &  \colhead{$p_{c}$}  & Log$L_{\rm 2-10keV}^{\rm BH}$ vs. LogL$_{12{\mu}m}^{\rm BH}$     \\
\colhead{[1]} & \colhead{[2]} & \colhead{[3]} & \colhead{[4]} 
}
\startdata
reference (solid line)               & reference-model            & 0.0 &  ${\rm log} \frac{L_{\rm 2-10keV}}{\rm 10^{43} erg/s} = 0.90{\rm log}\frac{L_{12.3{\mu}m}}{\rm 10^{43} erg/s} - 0.45$\\
fast\_evol\_SED\_SF (dashed line)        & strong evoluton of SF SED & 2.0 & ${\rm log} \frac{L_{\rm 2-10keV}}{\rm 10^{43} erg/s} = 0.90{\rm log}\frac{L_{12.3{\mu}m}}{\rm 10^{43} erg/s} - 0.45$\\
low\_IR2X\_BH (dot-dashed line)        & lower BH IR/X-ray ratio   & 0.0 & ${\rm log} \frac{L_{\rm 2-10keV}}{\rm 10^{43} erg/s} = 0.90{\rm log}\frac{L_{12.3{\mu}m}}{\rm 10^{43} erg/s} - 0.25$\\
high\_IR2X\_BH  (three-dot-dashed line)      & higher BH IR/X-ray ratio  & 0.0 &  ${\rm log} \frac{L_{\rm 2-10keV}}{\rm 10^{43} erg/s} = 0.90{\rm log}\frac{L_{12.3{\mu}m}}{\rm 10^{43} erg/s}- 0.65$\\
\enddata
\tablecomments{Col.(1): The names of  different variants of the model;
Col.(2):  the  description  to  the  specific variant  of  the  model;
Col.(3): the $p_{c}$ value (Equation~\ref{EV_T_LTIR}) used to quantify
the  star-forming  SED redshift  evolution.  Col.(4):  the SMBH  X-ray
vs. IR luminosity relationship used  to connect the AGN X-ray SED part
with   the  IR   part.\\  Compared   to  the   reference   model,  the
fast\_evol\_SED\_SF  variant  assumes  strong  redshift evolution  of  the
star-forming SED.   The low\_IR2X\_BH and  high\_IR2X\_BH variants
have  X-ray  luminosities  at  given  IR  luminosities  for  the  SMBH
radiation   0.2    dex   lower   and    higher,   respectively.    See
\S~\ref{IR_SED_HIGHZ_SF},         \S~\ref{Xray_SED_HI_AGN}         and
\S~\ref{sum_diff_mod}. The line style for each variant of the model is
used for all figures in this paper.}
\end{deluxetable}

\clearpage 
\LongTables
\begin{deluxetable}{lllllllllll}
\tabletypesize{\scriptsize}
\tablecolumns{15}
\tablecaption{\label{data_set} Data Sets To Derive The Best-Fit Parameters}
\tablewidth{0pt}
\tablehead{ \colhead{} & \colhead{} & \colhead{} & \colhead{} & \colhead{} }
\startdata
\multicolumn{5}{c}{differential number counts (371 data points)} \\ \hline
band             & ref                     & Fields        & Area[$\circ^{2}$] & Depth \\ \hline
17-60 keV        & \citet{Krivonos10}      & All-sky       & All-sky  & 7$\times$10$^{-12}$ erg/s/cm$^{2}$ [5$\sigma$]\\
15-55 keV        & \citet{Ajello12}        & All-sky       & All-sky  & 6$\times$10$^{-12}$ erg/s/cm$^{2}$ [5$\sigma$]\\
4-8 keV          & \citet{Lehmer12}        & CDFS          & 0.13     & 4.6$\times$10$^{-17}$ erg/s/cm [P$_{\rm det}$$>$0.004] \\
2-10  keV        &  \citet{Georgakakis08}  &  CDF-N        & 0.11     &  10$^{-16}$ erg/s/cm$^{2}$ [P$<$4$\times$10$^{-6}$]         \\
                 &                         &  CDF-S        & 0.06     & 2$\times$10$^{-16}$ erg/s/cm$^{2}$[P$<$4$\times$10$^{-6}$] \\
                 &                         &  ECDF-S       & 0.25     & 3$\times$10$^{-16}$ erg/s/cm$^{2}$[P$<$4$\times$10$^{-6}$] \\
                 &                         &  EGS          & 0.63     & 3$\times$10$^{-16}$ erg/s/cm$^{2}$[P$<$4$\times$10$^{-6}$] \\
                 &                         &  EN1          & 1.47     & 10$^{-15}$ erg/s/cm$^{2}$[P$<$4$\times$10$^{-6}$]          \\
                 &                         &  XBOOTES      & 9.24     & 6$\times$10$^{-16}$ erg/s/cm$^{2}$[P$<$4$\times$10$^{-6}$] \\
                 &  \citet{Ueda05}         &  AMSS         & 278      &  10$^{-13}$ erg/s/cm$^{2}$ [5$\sigma$]        \\
                 & \citet{Elvis09}         & COSMOS        & 0.5      & 7.3$\times$10$^{-16}$ erg/s/cm$^{2}$ [P$<$2$\times$10$^{-5}$] \\
                 & \citet{Lehmer12}        & CDFS          & 0.13     & 5.1$\times$10$^{-18}$ erg/s/cm$^{2}$ [P$_{\rm det}$$>$0.004] \\
0.5-2 keV        &  \citet{Georgakakis08}  &  CDF-N        & 0.11     &  10$^{-17}$ erg/s/cm$^{2}$[P$<$4$\times$10$^{-6}$]         \\
                 &                         &  CDF-S        & 0.06     &  2$\times$10$^{-17}$ erg/s/cm$^{2}$[P$<$4$\times$10$^{-6}$] \\
                 &                         &  ECDF-S       & 0.25     &  3$\times$10$^{-17}$ erg/s/cm$^{2}$[P$<$4$\times$10$^{-6}$] \\
                 &                         &  EGS          & 0.63     &  3$\times$10$^{-17}$ erg/s/cm$^{2}$[P$<$4$\times$10$^{-6}$] \\
                 &                         &  EN1          & 1.47     &  10$^{-16}$ erg/s/cm$^{2}$[P$<$4$\times$10$^{-6}$]          \\
                 &                         &  XBOOTES      & 9.24     &  6$\times$10$^{-17}$ erg/s/cm$^{2}$[P$<$4$\times$10$^{-6}$] \\
                 & \citet{Elvis09}         &  COSMOS       & 0.5      & 1.9$\times$10$^{-16}$ erg/s/cm$^{2}$  [P$<$2$\times$10$^{-5}$]\\
                 & \citet{Lehmer12}        & CDFS          & 0.13     & 5.1$\times$10$^{-18}$ erg/s/cm$^{2}$ [P$_{\rm det}$$>$0.004]  \\
Spitzer-24$\mu$m &   \citet{Bethermin10}   &  FIDEL ECDF-S & 0.23     &  60  $\mu$Jy [80\% complete]   \\
                 &                         &  FIDEL EGS    & 0.41     &  76  $\mu$Jy [80\% complete]   \\
                 &                         &  COSMOS       & 2.73     &  96  $\mu$Jy [80\% complete]   \\
                 &                         &  SWIRE LH     & 10.04    &  282 $\mu$Jy [80\% complete]   \\
                 &                         &  SWIRE EN1    & 9.98     &  261 $\mu$Jy [80\% complete]   \\
                 &                         &  SWIRE EN2    & 5.36     &  267 $\mu$Jy [80\% complete]   \\
                 &                         &  SWIRE ES1    & 7.45     &  411 $\mu$Jy [80\% complete]   \\
                 &                         &  SWIRE CDFS   & 8.42     &  281 $\mu$Jy [80\% complete]   \\
                 &                         &  SWIRE XMM    & 8.93     &  351 $\mu$Jy [80\% complete]   \\
IRAS-25$\mu$m    &   \citet{Hacking91}     &  All-sky      & All-sky  &  300  mJy [5$\sigma$]   \\
Spitzer-70$\mu$m &   \citet{Bethermin10}   &  FIDEL ECDF-S & 0.19     &  4.6  mJy [80\% complete]   \\
                 &                         &  COSMOS       & 2.41     &  7.9  mJy [80\% complete]   \\
                 &                         &  SWIRE LH     & 11.88    &  25.4 mJy [80\% complete]   \\
                 &                         &  SWIRE EN1    & 9.98     &  14.7 mJy [80\% complete]   \\
                 &                         &  SWIRE EN2    & 5.34     &  26.0 mJy [80\% complete]   \\
                 &                         &  SWIRE ES1    & 7.43     &  36.4 mJy [80\% complete]   \\
                 &                         &  SWIRE CDFS   & 8.28     &  24.7 mJy [80\% complete]   \\
Spitzer-160$\mu$m&   \citet{Bethermin10}   &  FIDEL EGS    & 0.38     &  45  mJy [80\% complete]   \\
                 &                         &  COSMOS       & 2.58     &  46  mJy [80\% complete]   \\
                 &                         &  SWIRE LH     & 11.10    &  92  mJy [80\% complete]   \\
                 &                         &  SWIRE EN1    & 9.30     &  94  mJy [80\% complete]   \\
                 &                         &  SWIRE EN2    & 4.98     &  90  mJy [80\% complete]   \\
                 &                         &  SWIRE ES1    & 6.71     &  130 mJy [80\% complete]   \\
                 &                         &  SWIRE CDFS   & 7.87     &  88  mJy [80\% complete]   \\
Herschel-100$\mu$m & \citet{Berta11}       &  GOODS-N      & 0.083    &  3.0 mJy [3$\sigma$]   \\
                   &                       &  GOODS-S      & 0.083    &  1.2 mJy [3$\sigma$]   \\
                   &                       &  LH           & 0.18     &  3.6 mJy [3$\sigma$]   \\
                   &                       &  COSMOS       & 2.04     &  9.0 mJy [3$\sigma$]   \\
Herschel-160$\mu$m & \citet{Berta11}       &  GOODS-N      & 0.083    &  5.7 mJy [3$\sigma$]   \\
                   &                       &  GOODS-S      & 0.083    &  2.4 mJy [3$\sigma$]   \\
                   &                       &  LH           & 0.18     &  7.5 mJy [3$\sigma$]   \\
                   &                       &  COSMOS       & 2.04     &  10.2 mJy [3$\sigma$]  \\
Herschel-250$\mu$m & \citet{Oliver10}      &  A2218        & 0.0225   &  13.8 mJy [50\% complete]  \\
                   &                       &  FLS          & 5.81     &  17.5 mJy [50\% complete]  \\
                   &                       &  Lockman-North& 0.34     &  13.7 mJy [50\% complete]  \\
                   &                       &  Lockman-SWIRE& 13.20    &  25.7 mJy [50\% complete]  \\
                   &                       &  GOODS-N      & 0.25     &  12.0 mJy [50\% complete]  \\
Herschel-350$\mu$m & \citet{Oliver10}      &  A2218        & 0.0225   &  16.0 mJy [50\% complete]  \\
                   &                       &  FLS          & 5.81     &  18.9 mJy [50\% complete]  \\
                   &                       &  Lockman-North& 0.34     &  16.5 mJy [50\% complete]  \\
                   &                       &  Lockman-SWIRE& 13.20    &  27.5 mJy [50\% complete]  \\
                   &                       &  GOODS-N      & 0.25     &  13.7 mJy [50\% complete]  \\
Herschel-500$\mu$m & \citet{Oliver10}      &  A2218        & 0.0225   &  15.1 mJy [50\% complete]  \\
                   &                       &  FLS          & 5.81     &  21.4 mJy [50\% complete]  \\
                   &                       &  Lockman-North& 0.34     &  16.0 mJy [50\% complete]  \\
                   &                       &  Lockman-SWIRE& 13.20    &  33.4 mJy [50\% complete]  \\
                   &                       &  GOODS-N      & 0.25     &  12.8 mJy [50\% complete]  \\
SCUBA-850$\mu$m    & \citet{Borys03}       &  HDF-N        & 0.046    & $\sim$ 3mJy [1$\sigma$]  \\
                   & \citet{Coppin06}      &  SHADES       & 0.20     & $\sim$ 2mJy [1$\sigma$]  \\
Aztec-1100$\mu$m   & \citet{Scott12}       & GOODS-N,LH,GOODS-S,ADF-S,SXDF & 1.6        & 0.4-1.7 mJy [1$\sigma$]                       \\
                   & \citet{Hatsukade11}   &  ADF-S        & 0.25     & $\sim$ 0.5mJy [1$\sigma$] \\
Mambo-1200$\mu$m   & \citet{Greve04}       &  EN2, LH      & 0.10     & 0.6 mJy [1$\sigma$]\\
                   & \citet{Lindner11}     &  LH-N         & 0.16     & 0.75 mJy [1$\sigma$]\\
\hline
\multicolumn{5}{c}{redshift distributions (195 data points)} \\ \hline
Bands                &   Fields &  spec-z   &  all-z  & ref  \\  \hline 
 2-10 keV            &   CDF-N  &  61\%     &  87\%   & \citet{Barger05}, \citet{Donley07} \\
                     &          &           &         & \citet{Trouille08} \\
                     &   COSMOS &  60\%     &  98\%   & \citet{Civano12}  \\
                     &   CDF-S  &  51\%     &  97\%   & \citet{Luo10}\\
                     &   ECDF-S &  64\%     &  95\%   & \citet{Silverman10} \\ 
                     &   XMS    &  87\%     &  87\%   & \citet{Barcons07} \\
0.5-2 keV            &   CDF-N  &  63\%     &  86\%   & \citet{Barger05}, \citet{Donley07} \\
                     &          &           &         & \citet{Trouille08} \\
                     &   COSMOS &  59\%     &  98\%   & \citet{Civano12}  \\
                     &   CDF-S  &  50\%     &  97\%   & \citet{Luo10}\\
                     & ROSAT-NEP    &  97\%     &  97\%   & \citet{Henry06} \\
                     & ROSAT-RIXOS  &  93\%     &  93\%   & \citet{Mason00} \\
Spitzer-24$\mu$m     & COSMOS       &  0\%      & 100\%   & \citet{LeFloch09} \\
                     & GOODS-S      &  90\%     & 90\%    & \citet{Barger08}  \\
                     & 5MUSES       &  98\%     & 98\%    & \citet{Wu10} \\
IRAS-60$\mu$m        & All Sky      & 100\%     & 100\%   & \citet{Sanders03}\\
SCUBA-850$\mu$m      & 7 fields     & 75\%      & 75\%    & \citet{Chapman05} \\
Aztec-1100$\mu$m     & COSMOS       & 41\%      & 100\%   & \citet{Smolcic12} \\  
\hline
\multicolumn{5}{c}{local luminosity functions (51 data points)} \\ \hline
Bands                & ref  \\  \hline 
15-55 keV            & \citet{Ajello12} \\
2-10 keV             & \citet{Ueda11} \\
IRAS-12$\mu$m        & \citet{Rush93} \\
IRAS-25$\mu$m        & \citet{Shupe98} \\
\enddata
\end{deluxetable}

\clearpage

\begin{deluxetable}{lccccccc}
\tabletypesize{\scriptsize}
\tablecolumns{25}
\tablecaption{\label{best_fit_parameter}  The Best-Fit Parameters }
\tablewidth{0pt}
\tablehead{ 
\colhead{Parameter} &  \colhead{reference} & \colhead{fast\_evol\_SED\_SF} & \colhead{low\_IR2X\_BH}  & \colhead{high\_IR2X\_BH}  \\
\colhead{[1]} & \colhead{[2]} & \colhead{[3]} & \colhead{[4]}   &  \colhead{[5]}
}
\startdata

\multicolumn{4}{c}{The total IR LF} \\ \hline
${\rm log}L_{*, 0}$[Log$L_{\odot}$]                    & 10.61$\pm$0.04    & 10.67$\pm$0.04    & 10.60$\pm$0.04    & 10.62$\pm$0.04    \\
$k_{1, l}$                                             & 0.35$\pm$0.40     & 1.65$\pm$0.37     & 1.65$\pm$0.36     & 0.81$\pm$0.39     \\
$k_{2, l}$                                             & 15.33$\pm$1.63    & 6.97$\pm$1.50     & 9.70$\pm$1.46     & 13.00$\pm$1.57    \\
$k_{3, l}$                                             & -15.19$\pm$1.76   & -7.64$\pm$1.63    & -9.32$\pm$1.60    & -12.95$\pm$1.70   \\
${\rm log}\Phi_{*, 0}$[Mpc$^{-3}$Log$L_{\odot}^{-1}$]  & -2.67$\pm$0.06    & -2.71$\pm$0.06    & -2.64$\pm$0.06    & -2.66$\pm$0.06    \\
$k_{1, d}$                                             & 3.68$\pm$0.57     & 2.96$\pm$0.57     & 2.57$\pm$0.53     & 2.80$\pm$0.56     \\
$k_{2, d}$                                             & -16.25$\pm$2.24   & -10.90$\pm$2.25   & -11.70$\pm$2.03   & -12.12$\pm$2.21   \\
$k_{3, d}$                                             & 9.39$\pm$2.36     & 3.61$\pm$2.42     & 4.77$\pm$2.12     & 5.33$\pm$2.34     \\
$\gamma_{1}$                                           & 0.39$\pm$0.03     & 0.40$\pm$0.03     & 0.38$\pm$0.03     & 0.36$\pm$0.03     \\
$\gamma_{2}$                                           & 2.49$\pm$0.04     & 2.65$\pm$0.05     & 2.52$\pm$0.05     & 2.52$\pm$0.04     \\
\multicolumn{4}{c}{BH Energy Fraction In The Total IR Band} \\ \hline
$\sigma_{0}^{\rm BH}$                                  & 0.64$\pm$0.03     & 0.75$\pm$0.05     & 0.55$\pm$0.02     & 0.63$\pm$0.02     \\
$k_{1, d}^{\rm BH}$                                    & -3.33$\pm$0.27    & -3.97$\pm$0.32    & -3.29$\pm$0.22    & -3.15$\pm$0.08    \\
$k_{2, d}^{\rm BH}$                                    & 3.05$\pm$0.44     & 6.02$\pm$0.52     & 2.89$\pm$0.40     & 2.87$\pm$0.11     \\
$p_{f^{\rm BH}}$                                       & 0.74$\pm$0.08     & 0.78$\pm$0.08     & 0.67$\pm$0.07     & 1.14$\pm$0.01     \\
$p_{\sigma}$                                           & -0.19$\pm$0.05    & -0.13$\pm$0.05    & -0.09$\pm$0.05    & -0.46$\pm$0.01    \\
\multicolumn{4}{c}{HI Column Density} \\ \hline
$\beta_{z, {\rm type-1}}$                              & 1.23$\pm$0.18     & 0.79$\pm$0.17     & 0.91$\pm$0.17     & 0.68$\pm$0.13     \\
$\beta_{l, {\rm type-1}}$                              & -0.66$\pm$0.06    & -0.48$\pm$0.06    & -0.48$\pm$0.07    & -0.42$\pm$0.05    \\
$\beta_{z, {\rm CT}}$                                  & -0.79$\pm$0.09    & -1.00$\pm$0.09    & -1.36$\pm$0.13    & -0.81$\pm$0.08    \\
$\beta_{l, {\rm CT}}$                                  & 0.44$\pm$0.08     & 0.41$\pm$0.07     & 0.73$\pm$0.12     & 0.21$\pm$0.05     \\
\hline
d.o.f.                                                 &          598      &          598      &          598      &          598      \\
$\chi^{2}$                                             & 1675.7            & 1752.8            & 1702.6            & 1608.1            \\

\enddata
\tablecomments{The best-fit parameters are listed for four variants of the model (see Table\ref{table_list_model}). }
\end{deluxetable}

\end{document}